\RequirePackage[l2tabu]{nag}		

\documentclass[a4paper,11pt,leqno,openbib,oldfontcommands]{memoir} 
\usepackage{datetime}
\usepackage{ifpdf}
\usepackage{physics}
\ifdraftdoc 
	\usepackage{draftwatermark}				
	\SetWatermarkScale{0.3}
	\SetWatermarkText{\bf Draft: \today}
\fi
\newsubfloat{figure}
\newsubfloat{table}
\settrimmedsize{297mm}{210mm}{*}
\settypeblocksize{634pt}{448.13pt}{*} 
\setulmargins{4cm}{*}{*} 

\setlrmargins{*}{*}{1} 

\setmarginnotes{17pt}{51pt}{\onelineskip} 
\setheadfoot{\onelineskip}{2\onelineskip} 
\setheaderspaces{*}{2\onelineskip}{*} 
\checkandfixthelayout
\usepackage{lmodern}
\usepackage{bbm}

\OnehalfSpacing 
\setsecnumdepth{subsection} 
\maxsecnumdepth{subsubsection}
\usepackage{calc,soul}
\makeatletter 
\newlength\dlf@normtxtw 
\newsavebox{\feline@chapter} 
\newcommand\feline@chapter@marker[1][4cm]{%
	\sbox\feline@chapter{%
		\resizebox{!}{#1}{\fboxsep=1pt%
			\colorbox{gray}{\color{white}\thechapter}%
		}}%
		\rotatebox{90}{%
			\resizebox{%
				\heightof{\usebox{\feline@chapter}}+\depthof{\usebox{\feline@chapter}}}%
			{!}{\scshape\so\@chapapp}}\quad%
		\raisebox{\depthof{\usebox{\feline@chapter}}}{\usebox{\feline@chapter}}%
} 
\newcommand\feline@chm[1][4cm]{%
	\sbox\feline@chapter{\feline@chapter@marker[#1]}%
	\makebox[0pt][c]{
		\makebox[1cm][r]{\usebox\feline@chapter}%
	}}
\makechapterstyle{daleifmodif}{

	\renewcommand\printchapternum{\null\hfill\feline@chm[2.5cm]\par}

} 
\makeatother 
\chapterstyle{daleifmodif}
\makepagestyle{myvf} 
\makeoddfoot{myvf}{}{\thepage}{} 
\makeevenfoot{myvf}{}{\thepage}{} 
\makeheadrule{myvf}{\textwidth}{\normalrulethickness} 
\makeevenhead{myvf}{\small\textsc{\leftmark}}{}{} 
\makeoddhead{myvf}{}{}{\small\textsc{\rightmark}}
\newcommand{\clearemptydoublepage}{\newpage{\thispagestyle{empty}\cleardoublepage}}
\makeindex
\usepackage{import}

\usepackage{lipsum}					
\usepackage{amsfonts} 					
\usepackage{physics}
\usepackage{mathtools} 
\usepackage{stmaryrd}					
\usepackage{amssymb}					
\usepackage{amsthm}					
\usepackage{newlfont}					
\usepackage{layouts}					
\usepackage{graphicx}					
\usepackage{longtable,rotating}			
\usepackage[utf8]{inputenc}			
\usepackage{colortbl}					
\usepackage{wasysym}					
\usepackage{mathrsfs}					
\usepackage{float}						
\usepackage{verbatim}					
\usepackage{upgreek }					
\usepackage{latexsym}					
\usepackage{booktabs}
\usepackage[square,numbers,
		     sort&compress]{natbib}		
\usepackage{url}						

\usepackage[english]{babel}		

\usepackage{color}                    				
\usepackage{hyperref}
\hypersetup{
  colorlinks   = true,    
  urlcolor     = blue,    
  linkcolor    = blue,    
  citecolor    = green      
}           
\usepackage{memhfixc}					
\usepackage{enumerate}					
\usepackage{footnote}					
\usepackage{microtype}					
\usepackage{rotfloat}					
\usepackage{alltt}						
\usepackage[version=0.96]{pgf}			
\usepackage{tikz}						
\usepackage{comment}
\newenvironment{dedication}
  {
   \thispagestyle{empty}
   \vspace*{\stretch{1}}
   \itshape             
   \centering          
  }
  {\par 
   \vspace{\stretch{3}} 
   \clearpage           
  }
\usepackage{calligra}

\usepackage{enumitem}

\newcommand{\setfont}[2]{{\fontfamily{#1}\selectfont #2}}
\usetikzlibrary{arrows,shapes,snakes,
		       automata,backgrounds,
		       petri,topaths}				
\newcommand{\pgftextcircled}[1]{                                                                    
    \setbox0=\hbox{#1}%
    \dimen0\wd0%
    \divide\dimen0 by 2%
    \begin{tikzpicture}[baseline=(a.base)]%
        \useasboundingbox (-\the\dimen0,0pt) rectangle (\the\dimen0,1pt);
        \node[circle,draw,outer sep=0pt,inner sep=0.1ex] (a) {#1};
    \end{tikzpicture}
}
\newcommand{\blackged}{\hfill$\blacksquare$}
\newcommand{\whiteged}{\hfill$\square$}
\newcounter{proofcount}

\let\oldsqrt\sqrt
\def\sqrt{\mathpalette\DHLhksqrt}
\def\DHLhksqrt#1#2{%
\setbox0=\hbox{$#1\oldsqrt{#2\,}$}\dimen0=\ht0
\advance\dimen0-0.2\ht0
\setbox2=\hbox{\vrule height\ht0 depth -\dimen0}%
{\box0\lower0.4pt\box2}}
\newcommand{\mycaption}[2][\@empty]{
	\captionnamefont{\scshape} 
	\changecaptionwidth
	\captionwidth{0.9\linewidth}
	\captiondelim{.\:} 
	\indentcaption{0.75cm}
	\captionstyle[\centering]{}
	\setlength{\belowcaptionskip}{10pt}
	\ifx \@empty#1 \caption{#2}\else \caption[#1]{#2}
}
\newcommand{\mysubcaption}[2][\@empty]{
	\subcaptionsize{\small}
	\hangsubcaption
	\subcaptionlabelfont{\rmfamily}
	\sidecapstyle{\raggedright}
	\setlength{\belowcaptionskip}{10pt}
	\ifx \@empty#1 \subcaption{#2}\else \subcaption[#1]{#2}
}
\usepackage{lettrine}
\newcommand{\initial}[1]{%
	\lettrine[lines=3,lhang=0.33,nindent=0em]{
		\color{gray}
     		{\textsc{#1}}}{}}
\theoremstyle{plain}

\theoremstyle{plain}

\theoremstyle{plain}
\theoremstyle{definition}

\theoremstyle{plain}

\theoremstyle{plain}

\theoremstyle{plain}

\usepackage{bm}
\usepackage{pdfpages}
\usepackage{enumitem,cleveref}

\usepackage{polski}
\usepackage[utf8]{inputenc}
\usepackage[T1]{fontenc}

\def\Tr{\operatorname{Tr}}

\def\sq{\operatorname{sq}}

\def\FS{\operatorname{FS}}
\def\GE{\operatorname{GE}}
\def\E{\operatorname{E}}
\def\BS{\operatorname{BS}}

\def\supp{\operatorname{supp}}
\def\LOCC{\operatorname{LOCC}}
\def\GHZ{\operatorname{GHZ}}

\def\id{\operatorname{id}}

\def\({\left(}
\def\){\right)}
\def\[{\left[}
\def\]{\right]}
\def\>{\rangle}
\def\<{\langle}

\newcommand{\mc}[1]{\mathcal{#1}}
\newcommand{\wt}[1]{\widetilde{#1}}

\newcommand{\tf}[1]{\textbf{#1}}

\newcommand{\msc}[1]{\mathscr{#1}}
\newcommand{\bbm}[1]{\mathbbm{#1}}

\usepackage[mathscr]{euscript}
\usepackage{esvect}

\newtheorem{definition}{Definition}

\newtheorem{remark}{Remark}

\usepackage{diagbox}
\usepackage{multirow}
\usepackage{adjustbox}

\begin{document}
%
%
%
%
%
\frontmatter
\pagenumbering{roman}
%
%
%
%
%
%
\begin{titlingpage}
\begin{SingleSpace}


\calccentering{\unitlength} 
\begin{adjustwidth*}{\unitlength}{-\unitlength}
\vspace*{24mm}
\begin{center}
\includegraphics[scale=0.19]{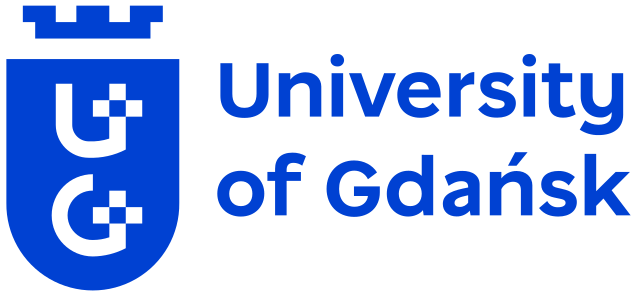}
\\ 
\vspace*{20mm}
\rule[0.5ex]{\linewidth}{2pt}\vspace*{-\baselineskip}\vspace*{3.2pt}
\rule[0.5ex]{\linewidth}{1pt}\\[\baselineskip]

{\HUGE  Fundamental Limitations within\\ 
the Selected Cryptographic Scenarios\\ 
and Supra-Quantum Theories\\[4mm]}
\rule[0.5ex]{\linewidth}{1pt}\vspace*{-\baselineskip}\vspace{3.2pt}
\rule[0.5ex]{\linewidth}{2pt}\\
\vspace{6.5mm}
{\large By}\\
\vspace{6.5mm}
{\large\textsc{M.Sc.Eng.~Marek Winczewski}}\\
\vspace{6.5mm}
Supervisor: {\large\textsc{dr hab. Karol Horodecki, prof. UG}}\\
\vspace{3.25mm}
{\large Institute of Theoretical Physics and Astrophysics,}
\\
{\large Faculty of Mathematics, Physics and Informatics}\\
\textsc{University of Gda\'nsk}
\\

\vspace{11mm}
\end{center}
\begin{flushright}
\end{flushright}
\end{adjustwidth*}
\end{SingleSpace}
\end{titlingpage}
\clearemptydoublepage
\chapter*{Acknowledgement}
\begin{SingleSpace}
\begin{quote}

\initial{T}his work would not have been possible without the mentorial and personal support of my supervisor, Karol Horodecki. In particular, I acknowledge and am grateful for support from the National Science Centre, Poland (NCN) grant Sonata Bis 5
(grant number: 2015/18/E/ST2/00327), and Foundation for Polish Science (IRAP project, ICTQT, contract No.~2018/MAB/5, co-financed by EU within Smart Growth Operational Programme).

I would also like to extend the deepest gratitude to Borhan Ahmadi, Stefan B{\"a}uml, Jakub Jan Borka{\l}a, Marek Czachor, Siddhartha Das, Micha{\l} Horodecki, Pawe{\l} Horodecki, Micha{\l} Kamo{\' n}, Ryszard P. Kostecki, Stanis{\l}aw Kryszewski, Marcin {\L}obejko, Antonio Mandarino, Gretchen Neuer, 
Adam Rutkowski, Omer Sakarya, Roberto Salazar, John H. Selby, and Bianka Wo{\l}oncewicz for many years of fruitful collaboration, support, and countless discussions. Last but certainly not least, I would like to especially thank Tamoghna Das for the long-lasting, close, and fruitful collaboration.

Finally, I would like to say thanks to all my friends and family for their support. In particular, I wish to thank members of ice hokey teams Marvin's Fighters and {{\' S}}lizg Gdynia for the time spent on the ice. Special thanks go to my friends Micha{\l} Miszczyszyn and Martyna Wygonna-Miszczyszyn for their unconditional love and support.

\end{quote}
\end{SingleSpace}
\clearpage
\frontmatter
\clearemptydoublepage
%
%
%

\chapter*{Dedication}
\begin{dedication}

\setfont{calligra}{\huge To my loving parents,} \\
    \vspace{\baselineskip}
    \setfont{calligra}{\huge Dorota Winczewska \& Piotr Winczewski}
\end{dedication}

\clearpage
\frontmatter
\clearemptydoublepage
%
%
%
%

\chapter*{Abstract}
\vspace{-0.3cm}
\begin{SingleSpace}

\initial{Q}uantum theory provides the ground for the most significant technological advances of the twenty-first century. The quantum revolution that is about to happen will bring us not only quantum computers but also quantum communication between quantum hardware. The global network of quantumly communicating devices will constitute the so-called quantum Internet. The promise quantum theory brings us about quantum communication is higher than the classical level of security and, in some scenarios, even unconditional security. Namely, there exist protocols for secret key distillation having entangled quantum states as a resource such that the security of the produced secret key is guaranteed by the laws of physics and are independent of the (possibly malicious) inner-working of the involved cryptographic devices. The more bits of the secure key can be distilled, given some entangled resources, the better the performance of the protocol. In this way, particular protocols constitute the lower bounds on the key rate achievable in the scenario. On the other hand, determining the optimal protocol or justification for pursuing it requires knowledge of the upper bounds on the achievable key rate. Finding the upper bounds on the secure key rate is, therefore, of crucial importance regarding the construction of quantum communication networks of the future. 

In the collection of articles constituting this dissertation, of which I am a coauthor, as our primary interest, we study the fundamental limitations within the selected quantum and supra-quantum cryptographic scenarios in the form of upper bounds on the achievable key rates. We investigate various security paradigms, bipartite and multipartite settings, as well as single-shot and asymptotic regimes. Specifically, our findings contribute to the secret key agreement scenario (SKA), device-dependent conference key agreement scenario (DD-CKA) both for quantum channels and quantum states in one-shot and asymptotic regimes, device-independent conference key-agreement scenario (DI-CKA) in both asymptotic and one-shot regime, and non-signaling device-independent secret key agreement scenario (NSDI). Our studies, however, extend beyond the derivations of the upper bounds on the secret key rates in the mentioned scenarios. In particular, we propose a novel type of rerouting attack on the quantum Internet for which we find a countermeasure and benchmark its efficiency. Furthermore, we propose several upper bounds on the performance of quantum (key) repeaters settings. We derive a lower bound on the secret key agreement capacity of a quantum network, which we tighten in an important case of a bidirectional quantum network. The squashed nonlocality derived here as an upper bound on the secret key rate is a novel non-faithful measure of nonlocality. Furthermore, the notion of the non-signaling complete extension arising from the complete extension postulate as a counterpart of purification of a quantum state allows us to study analogies between non-signaling and quantum key distribution (QKD) scenarios. 

From a larger perspective, our results are not only technical ones describing selected cryptographic tasks. In some cases, our findings confront fundamental questions regarding the foundations of quantum theory. The frameworks we develop allow to investigate this underlying subject insightfully both from the inside and outside of the quantum theory.

\end{SingleSpace}

\clearpage

\clearemptydoublepage
\chapter*{Abstrakt}

\begin{SingleSpace}


\initial{T}eoria kwantowa stanowi podwaliny dla najbardziej znacz{\k a}cych osi{\k a}gni{\k e}{\' c} technologicznych dwudziestego pierwszego wieku. Rewolucja kwantowa, kt{\' o}ra ma si{\k e} wkr{\' o}tce wydarzy{\' c}, przyniesie nam nie tylko komputery kwantowe, ale tak{\. z}e kwantow{\k a} komunikacj{\k e} mi{\k e}dzy urz{\k a}dzeniami nowego typu. Globalna sie{\' c} urz{\k a}dze{\' n} porozumiewaj{\k a}cych si{\k e} mi{\k e}dzy sob{\k a} z wykorzystaniem kwantowej komunikacji utworzy tzw. Internet kwantowy. Obietnica, jak{\k a} daje nam teoria kwantowa, w zakresie kwantowej komunikacji, gwarantuje poziom kt{\' o}ry jest wy{\. z}szy ni{\. z} klasyczny poziom bezpiecze{\' n}stwa, a w niekt{\' o}rych scenariuszach kryptograficznych nawet bezwarunkowe bezpiecze{\' n}stwo, niezale{\. z}ne od implementacji urz{\k a}dze{\' n} generuj{\k a}cych klucz kryptograficzny. Mianowicie, istniej{\k a} protoko{\l}y destylacji klucza kryptograficznego u{\. z}ywaj{\k a}ce jako zasobu spl{\k a}tanych stan{\' o}w kwantowych, takie {\. z}e bezpiecze{\' n}stwo wytworzonego klucza kryptograficznego gwarantowane jest prawami fizyki. Im wi{\k e}cej bit{\' o}w bezpiecznego klucza mo{\. z}na wydestylowa{\' c}, z okre{\' s}lonego stanu kwantowego, tym wy{\. z}sza wydajno{\' s}{\' c} protoko{\l}u. W ten spos{\' o}b poszczeg{\' o}lne protoko{\l}y wyznaczaj{\k a} granice dolne na ilo{\' s}{\' c} klucza mo{\. z}liwego do wydestylowania w danym scenariuszu kryptograficznym. Z drugiej strony okre{\' s}lenie optymalnego protoko{\l}u, lub uzasadnienie potrzeby dalszego go szukania, wymaga znajomo{\' s}ci ogranicze{\' n} g{\' o}rnych na mo{\. z}liw{\k a} do uzyskania ilo{\' s}{\' c} klucza. Znalezienie ogranicze{\' n} g{\' o}rnych na ilo{\' s}{\' c} bezpiecznego klucza kryptograficznego ma zatem istotne znaczenie dla budowy kwantowych sieci komunikacyjnych przysz{\l}o{\' s}ci.\\
\indent W zbiorze artyku{\l}{\' o}w sk{\l}adaj{\k a}cych si{\k e} na niniejsz{\k a} rozpraw{\k e} doktorsk{\k a}, kt{\' o}rych jestem wsp{\' o}{\l}autorem, naszym g{\l}{\' o}wnym celem jest okre{\' s}lenie fundamentalnych ogranicze{\' n} g{\' o}rnych na osi{\k a}galn{\k a} ilo{\' s}{\' c} bezpiecznego klucza kryptograficznego w wybranych kwantowych i supra-kwantowych scenariuszach kryptograficznych. Zosta{\l}y zbadane r{\' o}{\. z}ne paradygmaty bezpiecze{\' n}stwa, sytuacje dwu i wieloosobowe, a tak{\. z}e re{\. z}imy jednorazowe i asymptotyczne. W szczeg{\' o}lno{\' s}ci nasze rezultaty dotycz{\k a} scenariusza uzgadniania klucza sekretnego (ang. secret key agreement scenario, SKA), scenariusza  uzgadniania klucza konferencyjnego zale{\. z}nego od urz{\k a}dzenia (ang. device-dependent conference key agreement, DD-CKA) zar{\' o}wno dla stan{\' o}w kwantowych jak i kana{\l}{\' o}w kwantowych, w re{\. z}imach jednorazowych i asymptotycznych, scenariusza uzgadniania klucza konferencyjnego niezale{\. z}nego od urz{\k a}dzenia (ang. device-independent conference key agreement, DI-CKA) zar{\' o}wno w re{\. z}imie jednorazowym jak i asymptotycznym, oraz scenariusza uzgadniania klucza sekretnego w obecno{\' s}ci adwersarza ograniczonego jedynie wi{\k e}zami braku sygnalizacji (ang. device-independent non-signaling secret key agreement, NSDI). Nasze badania wykraczaj{\k a} jednak poza wyprowadzenie ogranicze{\' n} g{\' o}rnych na ilo{\' s}{\' c} klucza kryptograficznego osi{\k a}galnego we wspomnianych scenariuszach. W szczeg{\' o}lno{\' s}ci proponujemy nowy typ ataku przekierowuj{\k a}cego (ang. rerouting attack) na kwantowy Internet. Dla tego ataku znajdujemy {\' s}rodek zaradczy oraz kwantyfikujemy skuteczno{\' s}{\' c} naszego rozwi{\k a}zania. Proponujemy kilka ogranicze{\' n} g{\' o}rnych na wydajno{\' s}ci uk{\l}ad{\' o}w kwantowych powtarzaczy oraz powtarzaczy klucza kwantowego. Dodatkowo wyprowadzamy ograniczenie dolne na pojemno{\' s}{\' c} uzgadniania sekretnego klucza (ang. secret key agreement capacity) dla sieci kwantowej, kt{\' o}re to ograniczenie to ulepszamy w wa{\. z}nym przypadku sieci z{\l}o{\. z}onej z dwukierunkowych kana{\l}{\' o}w kwantowych (ang. bidirectional quantum channel). {\' S}ci{\' s}ni{\k e}ta nielokalno{\' s}{\' c} (ang. the squashed nonlocality) wyprowadzona jako ograniczenie g{\' o}rne na ilo{\' s}{\' c} klucza kryptograficznego jest w istocie now{\k a} miar{\k a} nielokalno{\' s}ci. Co wi{\k e}cej, poj{\k e}cie niesygnalizuj{\k a}cego kompletnego rozszerzenia (ang. non-signaling complete extension) wynikaj{\k a}ce z postulatu kompletnego rozszerzenia jako odpowiednik kwantowej puryfikacji pozwala nam na badanie analogii mi{\k e}dzy niesygnalizuj{\k a}cymi i kwantowymi scenariuszami uzgadniania klucza kryptograficznego. \\
\indent Patrz{\k a}c z szerszej perspektywy, nasze wyniki nie s{\k a} jedynie rezultatami technicznymi, opisuj{\k a}cymi wybrane zadania kryptograficzne. W niekt{\' o}rych przypadkach nasze odkrycia dotykaj{\k a} fundamentalnych kwesti dotycz{\k a}cych pryncypi{\' o}w teorii kwantowej. Ramy, kt{\' o}re opracowujemy, pozwalaj{\k a} wnikliwie bada{\' c} ten podstawowy temat zar{\' o}wno od wewn{\k a}trz, jak i od zewn{\k a}trz teorii kwantowej.

 \end{SingleSpace}
\clearpage
\clearemptydoublepage
%
%
%

\chapter*{Publications Included in the Dissertation}
\begin{SingleSpace}
\begin{enumerate}[label={[\Alph*]}]
    \item Omer Sakarya, \underline{Marek Winczewski}, Adam Rutkowski, Karol Horodecki. \emph{Hybrid quantum network design against unauthorized secret-key generation, and its memory cost}. Phys. Rev. Research 2, 043022 (2020). 
    \item Siddhartha Das, Stefan B{\"a}uml, \underline{Marek Winczewski}, Karol Horodecki. \emph{Universal Limitations on Quantum Key Distribution over a Network}. Phys. Rev. X 11, 041016 (2021). 
    \item Karol Horodecki, \underline{Marek Winczewski}, Siddhartha Das. \emph{Fundamental limitations on device-independent quantum conference key agreement}.  Phys.~Rev.~A~105, 022604 (2022).\\ With erratum in Phys.~Rev.~A 107, 029902 (2023).
    \item \underline{Marek Winczewski}, Tamoghna Das, Karol Horodecki. \emph{Limitations on device independent key secure against nonsignaling adversary via the squashed non-locality}. Phys. Rev. A 106, 052612 (2022).
    \item \underline{Marek Winczewski}, Tamoghna Das, John H. Selby, Karol Horodecki, Pawe{\l} Horodecki, {\L}ukasz Pankowski, Marco Piani, Ravishankar Ramanathan. \emph{Complete extension: the non-signaling analog of quantum purification}. Preprint arXiv:1810.02222. Published online 2018. Accepted for publication in Quantum 27.01.2023.
\end{enumerate}
\end{SingleSpace}
\clearpage
\clearemptydoublepage
\chapter*{Other articles}
\begin{SingleSpace}
\begin{enumerate}[label={[\Alph*]}]\setcounter{enumi}{4}
    \item \underline{Marek Winczewski}, Antonio Mandarino, Micha{\l} Horodecki, Robert Alicki. \emph{Bypassing the Intermediate Times Dilemma for Open Quantum System}. Preprint arXiv:2106.05776. Published online 2021.
    \item \underline{Marek Winczewski}, Robert Alicki. \emph{Renormalization in the Theory of Open Quantum Systems via the Self-Consistency Condition}. Preprint arXiv:2112.11962. Published online~2021.
    \item Marcin {\L}obejko, \underline{Marek Winczewski}, Gerardo Suárez, Robert Alicki, Micha{\l} Horodecki. \emph{Towards reconciliation of completely positive open system dynamics with the equilibration postulate}. Preprint arXiv:2204.00643. Published online 2022.
\end{enumerate}
\end{SingleSpace}
\clearpage
\clearemptydoublepage

\renewcommand{\contentsname}{Table of Contents}
\maxtocdepth{subsection}
\tableofcontents*
\addtocontents{toc}{\par\nobreak \mbox{}\hfill{\bf Page}\par\nobreak}
\clearemptydoublepage
%
%
%
%
\mainmatter
%
%

%
\let\textcircled=\pgftextcircled
\chapter{Introduction}\label{sec:Intro}


\initial{T}he desire to transfer messages in a confidential manner is probably as old as the history of humankind. It is known that ciphers were known already in ancient times. The~motivations to hide information from untrusted parties vary but were always the same, ranging from personal to military purposes. In particular, the Caesar cipher owes its name to the Roman general, statesman, and dictator Gaius Julius Caesar who used it in his military campaigns~\cite{Dooley2018,Bauer2021}. The development of methods on how to pass information secretly led to the establishment of a new area of science called cryptography. The distinctive feature of cryptography as a field of science and engineering is the fact that one has to confront an intelligent enemy called the eavesdropper and not only the laws of nature. Up to the beginning of the twentieth century, cryptography was in the hands of linguists~\cite{Bauer2021},~e.g., Auguste Kerckhoffs. Next, the engineers got involved, with Gilbert Vernam and his one-time pad encryption as a prominent example~\cite{Bauer2021}. In fact, one-time pad encryption was originally invented earlier by Frank Miller in 1882~\cite{miller1882telegraphic}. The later period belongs to mathematicians, who, amongst other successes, broke a German cipher device Enigma in the 1930s as the most famous achievement~\cite{Bauer2021}. In this way, the enormous efforts of Marian Rejewski, Jerzy R{\' o}{\. z}ycki, and Henryk Zygalski, continued by Alan Turing, contributed to the victory of the Allies in the  Second World War. Later, in the 1940s, Claude Shannon~\cite{Shannon1948} heavily developed the theory of information established by Harry Nyquist and Ralph Hartley in the 1920s~\cite{Nyquist1924,Hartley1928}, which is crucial for modern cryptographic purposes. Besides formalizing the concepts of processing and transmitting information and introducing the concept of entropy of information, Claude Shannon proved that the one-time pad encryption scheme is information-theoretically secure~\cite{Shannon1948}. The development of computers and communication between them entailed involvement in the cryptography of computer scientists in the middle 1970s~\cite{Bauer2021}. The~efforts of computer scientists allowed for the construction of a global network of communicating electronic devices called the Internet and gave it its present shape. Finally, in the 1980s, physicists realized that the quantum theory, studied from the beginning of the twentieth century, has the potential to be applied in cryptography. Namely, in 1983 Stephen Wiesner proposed the concept of quantum money resistant to forgery~\cite{Wiesner_1983}. In fact, he idea of Stephen Wiesner goes back to the 1970s, but it remained unpublished until 1983. Next, in 1984, Charles Bennett and Gilles Brassard developed the first quantum communication protocol named BB84 after the authors~\cite{BB84}, i.e., based on the formalism of quantum states and indeterminacy in their measurements. The BB84 protocol, under some assumptions, is provably secure as it produces a secure key ready for one-time pad encryption. Next, in 1991 Artur Ekert designed the E91 protocol~\cite{E91}, i.e., the first quantum cryptography protocol based on the entangled quantum states, that allows for the detection of the eavesdropper's presence. From these times, the combined power and knowledge of scientists from all of the aforementioned fields were used to develop a plethora of protocols, schemes, scenarios, and security paradigms for quantum, supra-quantum, and post-quantum cryptography. The ultimate goal of the efforts put into the development of the field is to establish communication schemes, the security of which is guaranteed by the laws of physics. In this way, quantum cryptography and recent progress in building quantum computers give a promise for constructing a quantumly secure Internet in the not-too-distant future~\cite{Dur_1999,Muralidharan_2016,Zwerger_2018,Wehner_2018}.

In the past several decades, four major security paradigms were developed for scenarios in which the cryptographic task of the honest parties is to distill the secret key in the presence of a malicious eavesdropper. Here, we list them with respect to the increasing power of the eavesdropper, which also reflects the level of professional paranoia in the number of incorporated assumptions. Furthermore, in this description, for the sake of its simplicity and conciseness of the notation, we restrict ourselves to the situation in which there are only two honest parties, usually called Alice and Bob, and the eavesdropping malicious Eve. The first one is the secret key agreement scenario (SKA)~\cite{CsisarKorner_key_agreement,Maurer93,Gisin-crypto,BB84}, in which the honest parties share $n$-copies of $P(AB)$ which is a marginal of a classical tripartite probability distribution $P(ABE)$, where $A$, $B$, and $E$ are the random variables describing the outputs of Alice, Bob and Eve, respectively. In order to distill the secret key, the honest parties process their data with so-called local operations and public communication (LOPC). At the same time, Eve listens to public communication and is in power to apply any stochastic map to her data. Furthermore, there are two distinct quantum key distribution (QKD) paradigms. The second paradigm to be described here is the quantum device-dependent (QDD)~\cite{BB84,E91,B92,AcinBBBMM2004-key} scenario introduced at the beginning of quantum cryptography. In this scenario, the honest parties share a marginal state $\rho_{AB}$ of (in the worst case from their perspective) a  tripartite pure quantum state $\ket{\psi_{ABE}}$, i.e., $\ket{\psi_{ABE}}$ is the purification of $\rho_{AB}$ to the eavesdropper's system $E$. Aiming to distill the secret key, the honest parties, Alice and Bob, process their subsystems using local quantum operations and classical communication (LOCC). Here, we remark that another definition based on LOPC is equivalent in the mentioned worst-case. On the other hand, the malicious Eve is assumed to receive any quantum system discarded by Alice and Bob, listen to the communication between the honest parties, and be able to perform a quantum operation on her subsystem. The problem with the QDD scenario is that Alice and Bob have to trust the inner workings of the device they operate on, used to transfer quantum states and to produce a classical output. Precisely speaking, the honest parties have to trust the dimensionality of the quantum state and measurements performed on it. A sophisticated solution to the drawback mentioned above is provided in the quantum device-independent (QDI)~\cite{E91,Bell-nonlocality,Mayers-Yao,AcinGM-bellqkd,acin-2007-98, Masanes2011, lit12} scenario. In the QDI scenario, Alice and Bob share an untrusted device, described with a conditional probability distribution $P(AB|XY)$ originating from local measurements $M_{A|X}$, $M_{B|Y}$ on a quantum state $\rho_{AB}$, i.e., $P(AB|XY) \equiv \Tr\left[M_{A|X} \otimes M_{B|Y} \rho_{AB}\right]$. Here, $X$ and $Y$ correspond to the choices of measurement settings (inputs) of Alice and Bob, respectively, and $A$ and $B$ to outcomes as before. The eavesdropping Eve is assumed to be restricted by the laws of quantum mechanics, and in particular, she may hold the purification $\ket{\psi_{ABE}}$ of the bipartite state $\rho_{AB}$. Therefore, the security in the QDI paradigm is based solely on the input-output statistic of the honest parties. In the last paradigm,  called the non-signaling device-independent secret key agreement (NSDI)~\cite{ Kent, Kent-Colbeck, Scarani2006, acin-2006-8,AcinGM-bellqkd, masanes-2009-102, hanggi-2009,masanes-2006} scenario, the assumptions on the power of the eavesdropper are even more relaxed than in the QDI scenario. Namely, the eavesdropper is restricted solely by the non-signaling condition that prevents Eve from changing the statistics of the honest parties and excludes instantaneous communication. Furthermore, the NSDI scenario allows the honest parties to share possibly supra-quantum (stronger than quantum) correlations constrained only by the no-signaling condition~\cite{PR}. Incidentally, the non-signaling conditional probability distributions constitute a generalized probabilistic theory (GPT) of non-signaling behaviors, wherein they have the role of states, see Ref.~\cite{InfoProcess} and references therein, and also Refs.~\cite{plavala2021general,muller2021probabilistic,lami2018non} for recent progress in this direction. Here, the device shared between parties is described as a tripartite conditional probability distribution $P(ABE|XYZ)$, with $E$ and $Z$ being the output and input of Eve, respectively, and $A$, $B$, $X$, $Y$ are as before. The honest parties choose their measurement settings (inputs) and process their output data with LOPC operations. The same as before, the device is assumed to be provided by Eve, who can listen to public communication and perform a certain class of operations on her subsystem. Moreover, the tripartite device $P(ABE|XYZ)$ is assumed to be the worst-case extension of the marginal $P(AB|XY) \equiv \sum_e P(ABe|XYz)$ shared by the honest parties. Finally, all of the mentioned scenarios have their extensions to the case of multiple honest parties. The secret key agreement between multiple honest parties is called the conference key agreement~\cite{Chen2005,AugusiakH2008-multi}. Similarly, QDD gives rise to the device-dependent conference key agreement (DD-CKA)~\cite{keyhuge,AugusiakH2008-multi}, QDI is lifted to the device-independent conference key agreement (DI-CKA)~\cite{RMW18}, and NSDI is naturally extended to the non-signaling conference key agreement (NS-CKA)~\cite{PKBW21}. Additionally, a modern research subject is to study secret (conference) key agreement capacities of quantum channels in scenarios that generalize the concept of the use of quantum states.

The most important property that quantifies the performance of the secret key distillation protocol is the number of bits of the secret key that is produced given some resource state (or device). The number of bits of the secret key produced in the protocol divided by the number of copies of the state (or device) used is called the protocol rate. In this way, the rates of particular protocols constitute the lower bounds on the maximal secret key rate achievable in the given scenario. The protocols of secret key distillation, and lower bounds on their rates, in the bipartite SKA~\cite{CsisarKorner_key_agreement,Maurer93}, QDD~~\cite{DevetakWinter-hash-prl, RGKinfo_sec_proof_short,pptkey,keyhuge}, QDI~\cite{Vidick-Vazirani,lit12}, NSDI~\cite{Kent,AcinGM-bellqkd, acin-2006-8, Scarani2006, masanes-2009-102, hanggi-2009, Kent-Colbeck, masanes-2006}, and multipartite CKA~\cite{multisquash}, DD-CKA~\cite{AugusiakH2008-multi}, and DI-CKA~\cite{RMW18} scenarios have been widely studied. The definition of the secret key rate involves a supremum over achievable rates by any theoretically possible protocols. Therefore, in a generic case, the maximal achievable secret key rate is unknown, as we usually do not know if the best known protocol is optimal. The complementary method to study, in a protocol-independent way, secret key rates achievable in cryptographic scenarios is to determine the upper bounds on them. Furthermore, the upper bounds can be determined both in the one-shot (single-run) and asymptotic regime. The one-shot regime refers to the key distillation from a single copy of a quantum state (or device), and the asymptotic regime considers a situation in which the honest parties have access to an unlimited number of copies of the resource state (or device). The upper bounds help to determine whether the known protocol is optimal, assess how much the protocols can be improved, or decide if the pursuit for a better protocol is still justified. Moreover, the upper bounds are also important for practical reasons. If an actually implemented cryptographic device produces more key than is determined by the upper bound, then the key provided by the device can not be secure. These reasons make finding the upper bounds an important direction of research. Therefore, not only the lower bounds but also the upper bounds on the secret key rate in the bipartite SKA~\cite{Intrinsic-Maurer,MauWol97c-intr,reduced-intrinsic,lit3}, QDI and  QDD~\cite{Christandl_2002,Christandl12,multisquash,Wil16,TGW14,Takeoka_2014,CFH21,AFL21,Farkas_2021}, NSDI~\cite{Kaur-Wilde} and multipartite CKA~\cite{secrecy-monotone}, DD-CKA~\cite{Carrara22,AugusiakH2008-multi,secrecy-monotone} scenarios have been studied. The same concerns multipartite DD-CKA scenario considering quantum channels~\cite{TGW14,Takeoka_2014,Pirandola2017}. However, apart from Ref.~\cite{acin-2006-8}, only recently the upper bounds in the DI-CKA~\cite{HWD22,PKBW21}, NSDI~\cite{Kaur-Wilde,WDH2021,CFH21,AFL21}, and NS-CKA~\cite{PKBW21} scenarios were studied. 

The vast development of cryptography based on the principles of non-classical physical theories rather than solely on mathematics in the form of number theory, combinatorics, and linear algebra, is due to the physical phenomena invisible in the classical description of the world stemming from intuitive perception. The advantage for the honest parties in secret key distillation tasks that comes with non-classical resources in the form of entanglement of quantum states or nonlocal correlations in devices comes with the flip side of the coin. Namely, the attack strategy of the eavesdropper possibly includes sharing of non-classical resources with the honest parties. Colloquially speaking, the classical eavesdropper is the weakest one (SKA), the quantum eavesdropper (QKD) is stronger than the classical, and the eavesdropper limited solely by the no-signaling principle (NS) is the most powerful one. Incidentally, the cryptography considering non-signaling adversary (NSDI) is still valid, even when some future theory will replace quantum theory as the most accurate description of reality, as long as it will be a non-signaling one, i.e., consistent with the special theory of relativity. The above hierarchy demonstrates, therefore, that the physical framework we adopt influences the notion of security and consequently determines the limitations of the physical theory concerning attainable quantities of secrecy. The converse statement is also true. The studies on the amounts of secret correlations achievable in different cryptographic paradigms can bring us new insights into quantum theory or non-signaling theories. In a similar manner, Peter Rastal~\cite{Rastall1985-RASLBT}, Sandu Popescu, and Daniel Rohrlich~\cite{PR} have shown that quantum theory is not maximally nonlocal, by inverting the logical order and making nonlocality an axiom. They demonstrated that there might exist theories that preserve relativistic causality but exhibit nonlocal correlations that violate some Bell inequalities stronger than quantum theory, for which the limit was proved by Boris Tsirelson~\cite{Tsirelson-bound}. Having the above past example of indirect investigation in mind, we believe that a research on relations between entanglement, nonlocality, secrecy, and their quantification can improve the understating of physical reality or at least develop frameworks to study it.

The articles included in this dissertation~\cite{Sakarya_2020,DBWH19,HWD22,WDH2021,CE}, contribute to almost all, except until recently untouched NS-CKA scenario, mentioned cryptographic paradigms. Our focus is directed mainly but not solely on the upper bounds on the maximal secret (conference) key rates (and capacities) achievable within the selected cryptographic scenarios, in some cases both single-shot and asymptotic regimes. The other topics of our concern are, among others, hacking attacks on the Quantum Internet~\cite{Makarov2009,Satoh_2021} and possible countermeasures to them, the upper bounds on the performance of quantum (key) repeater schemes~\cite{PhysRevLett.81.5932,Dur_1999,munro2015inside,CZC+21}, analogies between different cryptographic paradigms, the GPTs with a particular interest in the connection between the theory of non-signaling behaviors and NSDI cryptography, or the lower bounds on conference key agreement of selected quantum network schemes. We believe that the results presented in this dissertation display a considerably vast landscape of the fundamental limitations concerning the mentioned cryptographic scenarios, with some extra contributions. Here, we remark that by the fundamental, we mean limitations that stem solely from the formalism of quantum mechanics or the no-signaling principle. It is interesting, on its own, why Nature imposes constraints on the transmission and processing of information. The main results of the articles integrated into this dissertation are briefly described in the following paragraphs. We invoke the published articles \cite{Sakarya_2020,DBWH19,HWD22,WDH2021} in the chronological order of appearance, leaving the unpublished one, i.e., Ref.~\cite{CE} preprint, for the end. A detailed description is left for the next Section.

The first article \cite{Sakarya_2020} in this dissertation considers a quantum network under a threat of a new type of rerouting attack of our proposal~\cite{Satoh_2021}. The idea of the attack is the use of malicious software by dishonest end-users that performs entanglement swapping protocol in the hub node of a quantum network. This attack leads to an unauthorized consumption of the resources of the network in favor of the malicious end-users that gain shared entangled states. The example of a quantum network under study exhibits star-like topology with a single (central) hub node and multiple end-users. The end-users want to exchange only classical data with the central node, however, in a quantumly secure way. As the main result, we devise a countermeasure to the proposed attack based on special classes of quantum states that carry directly accessible (device-dependent) secret key but possess a low repeatable key rate~\cite{bauml2015limitations}. Precisely speaking, the countermeasure proposed by us is based on the fact that quantum security is not always transitive and employs either different types of the so-called private states~\cite{pptkey,keyhuge} or positive partial transpose (PPT) states that approximate the mentioned private stats. We benchmark our solution by quantifying the difference between the device-dependent secret key and repeatable key rates, i.e., the so-called gap of the scheme, as well as the amount of quantum memory spent solely for the security of the scheme, i.e., the so-called memory overhead of the scheme. This quantification is done by providing upper bounds on the repeatable key rate for the considered classes of states. At the same time, lower bounds on the gap of the scheme and memory overhead required us to identify both lower and upper bounds on the secret key rate. Moreover, we identify a low-dimensional example of a quantum state for which the scheme employing it exhibits a strict gap between secure and repeatable key rates~\cite{pptkey,Dobek_2011}. Our findings show that the protection of the scheme against the proposed attack has its cost in the amount of quantum memory required to be used in the scheme. In this way, our findings contribute to the development of the quantumly secure schemes of the Internet of the future.  

In the second article \cite{DBWH19}, we study the upper bounds on the achievable key rates in the device-dependent conference key agreement (DD-CKA) scenario. The conference key is the secret key shared by more than two parties. We do so by first introducing the concept of a multiplex quantum channel and subsequently determining general upper bounds on the device-dependent secret key agreement capacities of quantum channels in the multipartite scenario. In principle, the uses of a multiplex quantum channel interleaved with local operations and classical communication (LOCC) can simulate any conference key distribution protocol. Entanglement measures based on the notion of the generalized divergence provide case-specific upper bounds in terms of various relative entropy functions~\cite{Ume62,MDSFT13,D09,Dat09,MDSFT13,WWY14,BD10,WR12} on the plethora of secret-key distribution tasks both in asymptotic and one-shot (single-use) scenarios. In this way, the framework we developed has a unifying character as it provides a case-specific upper bound for a broad range of different scenarios within a common structure. We exemplify the use of our results in two specific cases, i.e., in measurement-device-independent quantum key distribution (MDI-QKD)~\cite{LCQ+12,braunstein2012side} scenario and the setup of quantum key repeaters~\cite{PhysRevLett.81.5932,Dur_1999,munro2015inside,CZC+21}. In the case of the dual-rail scheme~\cite{RP10} for the MDI-QKD scenario, our upper bound is tight and hence outperforms the so-called repeaterless bound (RB). For the setups of quantum repeaters, the upper bound provided by us is at least comparable with those in Refs.~\cite{bauml2015limitations,christandl2017private}. However, in our case, more parameters of the scheme can be incorporated. Additionally, we provide a non-trivial lower bound on the secret key agreement capacity for a bidirectional quantum network. Furthermore, we provide upper bounds on the achievable secret key rate in the scenario employing quantum states. Here, as a byproduct, we obtain an upper bound on Greenberger-Horne-Zeilinger (GHZ) states distillation in both single-shot and asymptotic regimes. This result is possible since GHZ states are an example of private states that are ideal final states of any key distillation LOCC protocol. In order to obtain upper bounds on the rate of distillation of GHZ states from noisy GHZ and W states, we conduct an extensive numerical investigation. One of the overcame difficulties is finding biseparable states close enough to GHZ and W or their noisy versions required for the numerical calculations. In summary, our work substantially contributes to the development of the field of DD-CKA cryptography, mainly but not only via providing a unified framework for deriving upper bounds on conference key rates and capacities for various DD-CKA scenarios.

The third article~\cite{HWD22} is devoted to upper bounds on the device-independent conference key agreement (DI-CKA) rates. We remark here that our paper is the first to consider upper bounds on DI-CKA rates, as before, only lower bounds on DI-CKA rates had been studied (see also recent work in Ref.~\cite{PKBW21}). We first define the reduced c-squashed entanglement as the multipartite generalization of cc-squashed entanglement \cite{AFL21,KHD21}. Next, we show that the reduced c-squashed entanglement upper bounds the device-independent conference key rate achievable by the standard protocols, i.e., the protocols that use a single input to generate the key~\cite{AFFRV18}. To be precise, we show the above with respect to two definitions of DI-CKA rates, i.e., the ``dev'' and the ``par'' definitions. We note that the independent and identically distributed (iid) setting is sufficient for our purposes. In the ``dev'' definition of DI-CKA secret key rate, the adversary is assumed to mimic the full statistics of the honestly implemented device. In contrast, in the ``par'' definition, the adversary has to mimic only some relevant parameters of the device, such as the level of violation of some Bell inequality or quantum bit error rate (QBER). Our goal is achieved by generalizing the upper bounds via intrinsic information studied in Ref.~\cite{Christandl12} to the case of an adversary which possibly can hold an infinite dimensional. We then compare one of our upper bounds with the lower bound on the DI-CKA secret key rate in Ref.~\cite{RMW18}. Next, we provide a non-trivial upper bound on the device-independent key rate in the parallel measurement scenario in which all parties simultaneously set all values of their inputs. Furthermore, we provide a multipartite generalization of the reduced bipartite entanglement measures~\cite{KHD21}. Consequently, we show that the reduced regularized relative entropy of genuine entanglement~\cite{DBWH19} is an upper bound on the DI-CKA secret key rate. This relation brings us to the discussion on genuine nonlocality~\cite{Bell-nonlocality} and genuine entanglement in the context of the DI-CKA scenario~\cite{Horodecki2009}. Finally, we provide proof of the existence of a strict gap between the rates of the DI-CKA secret key and the DD-CKA secret key. The proof is constructive, as the gap is inherited from the bipartite case in which an example of states that exhibits the gap is known~\cite{CFH21}. To conclude, we are the first to provide upper bounds on the secret key rates in the DI-CKA cryptographic paradigm. At the same time, we provide several upper bounds on different scenarios and regimes.

In the fourth, recently published article~\cite{WDH2021}, we initiate a systematic study of upper bounds on the secret key rate in the non-signaling device-independent secret key agreement (NSDI) scenario~\cite{Kent,Kent-Colbeck,Scarani2006,acin-2006-8,AcinGM-bellqkd,masanes-2009-102,hanggi-2009,masanes-2006}. Firstly, we show that the notion of security based on the so-called non-signaling norm (see also~\cite{ChristandlToner} in this topic), that we adopt, is equivalent to two security definitions present in the literature, i.e., to the one in Refs. \cite{masanes-2006,masanes-2009-102,Masanes2011} and to the second one in Refs. \cite{hanggi-2009,hanggi-2009b,Renner-Hanggi,Hanggi-phd}. To show the above-mentioned equivalence, we derive an explicit form of the non-signaling norm for the so-called $\textbf{c}$-d states. Next, we show our main result, i.e., the squashing procedure. Namely, we show that any secrecy quantifier that is an upper bound on the secret key rate in the SKA scenario~\cite{Maurer93,Intrinsic-Maurer,MauWol97c-intr,reduced-intrinsic,lit3} can be lifted to give an upper bound on the secret key rate in the NSDI scenario. The squashing procedure is based on the notion of the non-signaling complete extension \cite{CE} and on the suitable optimization over measurements performed by the honest parties and the eavesdropper. The lift-up of the formalism is possible since we were able to rephrase the definition of the secret key rate in the SKA scenario in the form known from quantum key distribution scenarios (QKD). Subsequently, we focus on one of the squashed functions, i.e., the non-signaling squashed intrinsic information, also called the squashed nonlocality. As we show, the squashed nonlocality is not only an upper bound on the NSDI secret key rate but also a novel and nonfaithful measure of nonlocality. Next, we develop the convexification technique that provides even tighter upper bounds. Namely, in the convexification technique we combine different convex upper bounds to produce lower convex hull of their plots. The mentioned lower convex hull provides then tighter upper bound than individual plots on their own. Finally, we perform a numerical investigation that allowed us to compare our upper bound via the squashed nonlocality with the lower bounds given by the rates of H{\"a}nggi-Renner-Wolf~\cite{hanggi-2009} and by Ac{\'{i}}n-Massar-Pironio~\cite{acin-2006-8} protocols. In summary, we devised a technique that allows constructing upper bounds on the secret key rate in the NSDI scenario from the upper bounds on the secret key rate in the SKA scenario. The upper bound via a nonfaithful measure of nonlocality shows that nonlocality is not always equivalent to secrecy. Our approach exposes analogies between different security paradigms. Therefore, our results contribute not only to the NSDI and SKA cryptographic paradigms but also to the fundamental topic of nonlocality.

In the fifth, recently accepted for publication manuscript ~\cite{CE}, we study the concept of the complete extension postulate (CEP). We do so in the framework of the so-called generalized probabilistic theories (GPTs)~\cite{InfoProcess,plavala2021general,muller2021probabilistic,lami2018non} that allows us to study candidates for the beyond-quantum theory. By beyond-quantum theory, we mean the future theory of physics with stronger explanatory power than the quantum theory. As we discuss, the explanatory power of the quantum theory is limited, Refs.~\cite{lee2018no,selbyReconstruction,Nakahira2020,westerbaan2022computer,hardy2005probability,Hardy_gravity,oreshkov2012quantum,chiribella2013quantum,araujo2017purification} in this context. The CEP is devised to be a relaxation of the purification postulate, i.e., one of the postulates in terms of which quantum theory can be reformulated~\cite{Chiribella2011}. In particular, contrary to the PP, the CEP does not require the existence, within the theory, of pure extensions for all systems and their states. Instead, the CEP requires the existence of extensions with the property of \textsc{generation}, i.e., extensions that can be transformed into any other extension, of the extended system, with dynamics available within the theory. In the manuscript in Ref.~\cite{CE}, we study the properties of GPTs in which the PP is replaced with the CEP. We first show that the PP can not hold in any discrete convex theory, as within such theories, there are not enough pure states. On the contrary, the CEP holds both in the quantum theory and the classical probability theory, while the latter does not satisfy the PP. Next, we show that the property of \textsc{generation} implies the property of \textsc{access}, i.e., the extending system of a generating extension has access to all statistical ensembles of the extended system via suitable choices of measurement settings. We show that replacing the PP with the CEP draws a demarcation line between the results that require the PP to hold and those for which the CEP is enough. In this context, we show that the impossibility of bit-commitment cryptographic task~\cite{May97, LC97a}, and its generalization to integer-commitment~\cite{sikora2018simple}, holds within GPTs which satisfy the CEP. More precisely, we show that the lower bound on the product of the cheating probabilities of two parties in the integer-commitment task derived within the framework of quantum theory still holds if we replace the PP with the CEP. On the other hand, we show that the proof for the no-go for hyperdecoherence, which relies on the PP, no longer holds when the PP is replaced with the CEP~\cite{lee2018no}. The above suggests that the CEP might be a valid postulate for theories that hyperdecohere to the quantum theory. In order to show that the CEP is not an empty postulate and that complete extensions can actually be constructed outside quantum and classical theory, as our case study, we choose the theory of non-signaling behaviors~\cite{PR,InfoProcess}. In this context, we first show that within the theory of non-signaling behaviors, the properties of \textsc{access} and \textsc{generation} are equivalent. We then define the non-signaling complete extension (NSCE) as an extension in which the extending system has direct access to all minimal ensembles upon suitable measurement choice. The minimal ensembles are simply ensembles that are not convex combinations of other ensembles. In fact, due to the dynamics available in the theory, access to minimal ensembles is equivalent to access to all ensembles by the extending system. Therefore, in analogy to quantum purifications, NSCEs satisfy both \textsc{access} and \textsc{generation}. We then explicitly construct NSCEs in some cases. In particular, we show that the NSCE of maximally mixed behavior with a single binary input and single binary output is the famous Popescu-Rohrlich (PR) box~\cite{PR}. We notice an analogy between NSCE described above and Bell's state, which is the purification of the maximally mixed state of a qubit. We see the above example as an independent derivation of the PR box in which we do not refer to any notion of Clauser-Horne-Shimony-Holt (CHSH) or other so-called Bell inequality. We also derive an upper bound on the dimension of the NSCE of an arbitrary behavior and show it has the lowest dimension amongst all extensions having the property of \textsc{acccess}. In this way, we show that the dimension of the NSCE of arbitrary behavior is always finite. This result is not only interesting on its own but also has consequences in the fields of device-independent cryptography against a non-signaling adversary~\cite{Kent, Masanes, hanggi-2009, Hanggi-phd} and private randomness \cite{Colbeck_randomness, Gallego_randomness, Mironowicz_randomness, Brandao_randomness, Ramanathan_randomness}. Namely, the adversary holding the extending system of NSCE possesses maximal operational power at the lowest memory cost required for it. To conclude, the CEP is an important relaxation of the PP in the context of studying possible beyond-quantum theories. Our findings reveal both the possibilities and the limitations of replacing the PP with the CEP. Our findings contribute not only to the field of GPTs but also to the field of device-independent cryptography against the non-signaling adversary.

%

%

\chapter{Summary of Dissertation}
\section{Preliminaries}\label{sec:Preliminaries}

In this Section, we would like to provide formal definitions of the selected mathematical objects appearing in this dissertation. We aim here to provide a possibly high level of completeness of the descriptions of articles in the forthcoming Sections. Because of its formal and supplementary character, this Section can be skipped during the first reading of the dissertation. The descriptions of the articles composed in the dissertation contain suitable references to specific notions described below.

\subsection{Private States and Positive Partial Transpose States}\label{subsec:PrivateAndPPT}

In Ref.~\cite{pptkey}, it was shown that the distillation of the secret key by two parties, $A$ and $B$, that want to communicate secretly is nonequivalent to the distillation of Einstein-Podolsky-Rosen~(EPR) states, i.e., states of the following form
\begin{align}
    \ket{\psi_+}_{AB} = \frac{1}{\sqrt{2}} \left(\ket{00}_{AB} + \ket{11}_{AB}\right). 
\end{align}
Namely, there exist the so-called bound entangled states that can be used to obtain the secret key. To create a bound entangled state \cite{Horodecki_1998}, the resource of pure entanglement is required. However, no pure entanglement can be distilled from bound entangled states. The general form of quantum states that contain ideal secrecy was found in Ref.~\cite{pptkey}. It was shown that a quantum state has an ideally secure key if and only if it has the form of the so-called private state $\gamma_{ABA^\prime B^\prime}$. Furthermore, private states were used in Ref.~\cite{keyhuge} to recast the theory of privacy, employing the notion of local operations and classical communication in terms of entanglement theory, where local operations and classical communication (LOCC) are a valid set of operations. The LOCC paradigm does not require explicitly considering the eavesdropper's presence in the cryptographic scenario. Furthermore, in Ref.~\cite{keyhuge}, equivalence between LOPC and LOCC secret key distillation paradigms was proved (see also Ref.~\cite{karol-PhD}). The above makes private states an important class of states, especially in the context of device-dependent quantum cryptography (DD-QKD) both in the bipartite and multipartite settings.  


\subsubsection{Bipartite Private States}\label{subsubsec:2Private}

In this Subsection, we provide the definitions of the bipartite private states and their special subclasses. In the definitions, we employ the notation used in Ref.~\cite{Sakarya_2020} (see Sec.~\ref{sec:A_Hybrid}) contained in this dissertation rather than the original one in Refs.~\cite{pptkey,keyhuge}.

\begin{definition}[Of a private state, cf. Ref.~\cite{keyhuge}]\label{def:PrivateStateBi}
A state $\rho_{AB A^\prime B^\prime}$ on a Hilbert space $\mathcal{C}^{d_A}\otimes\mathcal{C}^{d_B}\otimes\mathcal{C}^{d_{A^\prime}}\otimes\mathcal{C}^{d_{B^\prime}}$ with dimensions $d_A=d_b=d_k$, and $d_{A^\prime}=d_{B^\prime}=d_s$ of the form
\begin{align}
    \gamma_{d_k,d_s}:=\frac{1}{d_k} \sum_{i,j=0}^{d_k-1} \ketbra{ii}{jj}_{AB} \otimes X_{ij},~~X_{ij} := U_i \sigma_{A^\prime B^\prime} U_j^\dagger,
\end{align}
where the state $\sigma_{A^\prime B^\prime}$ is an arbitrary state of subsystem $A^\prime B^\prime$, $U_i$'s are arbitrary unitary transformations acting on $A^\prime B^\prime$ subsystem and $\mathcal{B}_A \equiv\{\ket{i}_A\}_{i=0}^{d_k-1}$, $\mathcal{B}_B \equiv\{\ket{j}_B\}_{j=0}^{d_k-1}$  are local (orthonormal) bases on  $A$ and $B$ respectively, is called a private state or pdit. In the case of $d_k=2$, the state $\gamma_{2,d_s}$ is called - a pbit. 
\end{definition}
In the above definition, we refer to the subsystem $AB$ as the key part because it contains directly accessible secret key via measurements in $\mathcal{B}_A$ and $\mathcal{B}_B$ bases. Similarly, the subsystem $A^\prime B^\prime$ is called the shielding system (the shield part). The shielding system has an important role in defining the LOCC secret key distillation paradigm. In the LOCC paradigm, it is assumed that no quantum system is discarded to outside of the laboratories of the honest parties. Instead, in the key distillation protocol, the shielding system serves as a dumpster, which is protected from the eavesdropper. The Definition~\ref{def:PrivateStateBi}, above, is constructed based on Definition 1 in Ref.~\cite{Horodecki_2018}. Yet, we employ the notation used in Ref.~\cite{Sakarya_2020} and restrict (with respect to Refs.~\cite{pptkey,keyhuge,Horodecki_2018}) to the case in which subsystems $A^\prime$ and $B^\prime$ are of the same dimension. See Sec.~\ref{subsubsec:NPrivate} for the most general definition of private states.

We now recall the definitions of special cases of private states. In the places where it does not lead to any confusion, we omit the subscript of $\gamma_{d_k,d_s}$ and write simply $\gamma$. In the following definitions $K_D(\rho)$ is the secret key rate of state $\rho$ (see Sec.~\ref{subsec:SKAtasks}).  
\newpage
\begin{definition}[Of irreducible private state~\cite{keyhuge}]\label{def:PSirreducible}
    Any pdit $\gamma$ (with $d_k$-dimensional key part) for which $K_D(\gamma)=\log_2 d_k$ is called irreducible.
\end{definition}
\begin{definition}[Of key-attacked private states~\cite{Horodecki_2018}]\label{def:PSattacked} We call states of the  following form key-attacked private states. 
\begin{align}
    \hat{\gamma}_{d_k,d_s}:=\frac{1}{ d_k} \sum_{i=0}^{d_k-1}|ii\>\<ii|\otimes X_{ii},
\end{align} 
where $X_{ii}=U_i\sigma U_i^\dagger$ for some state $\sigma$ of ${\cal C}^{d_s}\otimes {\cal C}^{d_s}$ and $U_i$ are some unitary transformations. They are private states which has been measured on their key part $AB$. We call $X_{ii}$ conditional states of the shield.
\end{definition}
\begin{definition}[Of strictly irreducible private state~\cite{christandl2017private,Horodecki_2018}]\label{def:PSstrictly} We say a private state $\gamma_{d_k,d_s}$ is strictly irreducible, if its key-attacked version $\hat{\gamma}_{d_k,d_s}$ consists of separable states with respect to the systems on the shield part of the diagonal, i.e., $U_i \sigma U_i^\dagger \in \mathcal{SEP}$. We denote such states $\gamma_{\langle d_k,d_s \rangle}$, and their attacked versions by~$\hat{\gamma}_{\langle d_k,d_s \rangle}$.
\end{definition}

We finish this Subsection with the following Remark.
\begin{remark}
It follows directly from the definitions that $K_D(\gamma_{d_k,d_s}) \ge \log_2 d_k$, $K_D(\hat{\gamma}_{d_k,d_s}) =0$, and $K_D(\gamma_{d_k,d_s}) = \log_2 d_k$ whenever $\gamma_{d_k,d_s}$ is irreducible.     
\end{remark}

\subsubsection{Positive Partial Transpose States}\label{subsubsec:PPT}

In this Subsection, we introduce the definitions of the operation of partial transposition and positive partial transpose states (PPT). It follows directly from the definition of private states that they have negative partial transpose (NPT). In Ref.~\cite{Horodecki2006} it was shown that all private states allow for distillation of pure entanglement. On the other hand, all PPT states are bound entangled~\cite{Horodecki_1998}. Furthermore, there exist PPT states that are arbitrarily close, in trace norm, to private states, and they contain almost perfect secret key~\cite{pptkey,keyhuge}. These feature of PPT states makes them an interesting class of states to study.

\begin{definition}[Of partial transpose]
Let $\rho_{AB}=\sum_{ijkl}\rho_{kl}^{ij} \ketbra{ik}{jl}_{AB}$ be a bipartite quantum state acting on Hilbert space $\mc{H}_A \otimes \mc{H}_B$. The operation of partial transpose $\Gamma_B$  with respect to the $B$ subsystem is defined as follows
\begin{align}
    \Gamma_B:~~\rho_{AB}^{\Gamma_B} \equiv \left(\mathbbm{1}\otimes \Gamma_B \right)\rho_{AB} = \sum_{ijkl}\rho_{kl}^{ij} \ketbra{il}{jk}_{AB}.
\end{align}
\end{definition}

\begin{definition}[Of positive partial transpose (PPT) states]
    Let $\rho_{AB}$ be a bipartite quantum state acting on Hilbert space $\mc{H}_A \otimes \mc{H}_B$. State $\rho_{AB}$ is called positive partial transpose (PPT) state, if and only if $\rho_{AB}^\Gamma$ has only non-negative eigenvalues, i.e., 
    \begin{align}
        \rho_{AB}^\Gamma \ge 0,
    \end{align}
    where $\Gamma$ is a partial transpose operation with respect to arbitrary subsystem. 
\end{definition}



\subsubsection{Multipartite Private States}\label{subsubsec:NPrivate}

Private states can also be defined in the multipartite setting and help to define the LOCC paradigm for conference key agreement (DD-CKA). In this Section, we assimilate the notation used in Ref.~\cite{DBWH19} (see Sec.~\ref{sec:B_Universal}). Let us consider $M$ parties $A\equiv A_1,\dots A_M$ with $\vv{K}\equiv K_1,\dots K_M$ key parts and $\vv{S} \equiv S_1 ,\dots S_M$ shielding systems, defined on $\mc{H} \equiv \mc{H}_1 \otimes \cdots \mc{H}_M$ and  $\mc{H}^\prime \equiv \mc{H}_1^\prime \otimes \cdots \otimes \mc{H}_M^\prime$, respectively.

In the following definition, $\Phi_{\vv{K}}^\mathrm{GHZ}$ denotes the projection on $M$-partite Greenberger-Horne-Zeilinger (GHZ) state, i.e., the state of the following form
\begin{align}
    &\vec{k} \equiv k_1,\dots,k_M:~\forall_{{i,j}\in[M]}~ k_i=k_j \in \mc{K}\label{GHZcondition},\\
    &\ket{\mathrm{GHZ}} := \frac{1}{K} \sum_{\eqref{GHZcondition}} \ket{\vec{k}}.
\end{align}
Here, $\mc{K}$ corresponds to the alphabet of outcomes for key parts of individual subsystems, and $K\equiv\abs{\mc{K}}=\abs{K_i}$ denotes the cardinality of $\mc{K}$. 

\begin{definition}[Of $M$-partite private state, cf. Refs.~\cite{AugusiakH2008-multi,DBWH19}]\label{def:PrivateStateMulti}
A state $\gamma_{\vv{SK}}$, with $\abs{K_i}=K$ for all $i\in[M]$, is called a ($M$-partite) private state if and only if
\begin{align}
    \gamma_{\vv{SK}} := U_{\vv{SK}}^\mathrm{tw} \left( \Phi_{\vv{K}}^\mathrm{GHZ} \otimes \omega_{\vv{S}}\right) \left(U_{\vv{SK}}^\mathrm{tw}\right)^\dagger,
\end{align}
where $U_{\vv{SK}}^\mathrm{tw}:=\sum_{\vec{k}\in \mathcal{K}^{\times M}} \ketbra{\vec{k}}{\vec{k}}_{\vv{K}}\otimes U_{\vec{S}}^{\vec{k}}$ is called a twisting unitary operator for some unitary operator $U_{\vec{S}}^{\vec{k}}$ and $\omega_{\vv{S}}$ is an arbitrary finite-dimensional density operator.
\end{definition}
The Definition~\ref{def:PrivateStateMulti}, above, is more general than Definition~\ref{def:PrivateStateBi}, but also because we do not assume that different shielding systems are of the same dimension. It follows directly from the definition that $K_D(\gamma_{\vv{SK}}) \ge \log_2 K$. We finish with the following remark.   
\begin{remark}
In this place, we remark that the definitions \ref{def:PSirreducible}, \ref{def:PSattacked} and \ref{def:PSstrictly}, defining different subclasses of private states, can be naturally extended to the multipartite case. 
\end{remark}

\subsection{Secret Key Agreement Tasks}\label{subsec:SKAtasks}

The secret key agreement is a cryptographic task in which two or more honest allies aim to distill the secure key that can be used for one-time pad encryption. A scenario that involves more than two parties is called a conference key agreement. More precisely, the task of the honest parties is to use resources and operations available in the scenario in order to obtain perfectly correlated bit strings which remain unknown to the eavesdropper. The sequence of operations performed by the honest parties, which may also contain communication via an authenticated but insecure classical channel, is called a protocol. Similarly, the action of the malicious eavesdropper is called the attack strategy. The particular resources and operations available in the cryptographic task depend on the considered cryptographic paradigm (see Chap.~\ref{sec:Intro}). The crucial mathematical objects quantifying the scenario are key rate and security conditions. The rate of a protocol is, roughly speaking, the number of bits of secret key distilled in protocol per one copy of the resource (state). The secret key rate is, therefore, the supremum over rates of all possible protocols. The secret key rate can be defined in both one-shot (single use) regimes, where the number of copies of the resource state is finite, as well as in the asymptotic regime, in which we assume that the number of copies of the resource state is arbitrarily large. The security conditions define what is understood by the secure key. The key distillation protocol is, therefore, called secure if the correlations shared by the honest parties at the end of the protocol satisfy the security conditions. The definitions of those are, again, scenario-dependent. In what follows, we assimilate the so-called ``optimistic'' definition of asymptotic key rate in which ``$\limsup_{n\to \infty}$'' is used rather than ``$\liminf_{n\to \infty}$'', where $n$ refers to the number of copies of the resource state, as it is done in, e.g., in Ref.~\cite{bauml2015limitations}. For the discussion about the ``optimistic'' and the ``pessimistic'' definitions of the asymptotic key rate, see Refs.~\cite{Ahlswede2006}. See also a different approach to ours in the case of entanglement theory, e.g., in \cite{Wilde_2021}.



\subsubsection{Secret Key Rate and Security}\label{subsubsec:Security}

We first describe the definition of secret key rate in the secret key agreement scenario (SKA)~\cite{Maurer93,MaurerWolf00CK}. As described in Chapter~\ref{sec:Intro}, the resource states in the SKA scenario are bipartite probability distributions $P(AB)$, and $A$ and $B$ are random variables in possession of respective parties. The honest parties, also called $A$ and $B$, process their marginal probability distributions using local operations in the form of local stochastic maps and public communication (LOPC). The eavesdropper $E$ is explicitly present in the scenario and is assumed to share with the honest parties a tripartite probability distribution $P(ABE)$ that is an extension of $P^{\otimes N}(AB)$, i.e., $\sum_e P(ABe)=P^{\otimes N}(AB)$. The secret key rate ${S}(A:B||Z)$ in the SKA scenario is defined as follows    
\begin{definition}[Of the SKA secret key rate, cf. Ref.~\cite{Maurer93,MaurerWolf00CK}]\label{def:Maurer93}
			The secret key rate of A and B with respect to E, denoted ${S}(A:B||Z)$, is the maximal $R \ge 0$ such that for every $\varepsilon >0$ and for all $N \ge N_0 (\varepsilon)$ there exists a protocol, using public communication over an insecure but an authenticated channel, such that Alice and Bob , who receive $A^N=[A_1,...,A_N]$ and $B^N=[B_1,...,B_N]$, can compute keys $S_A$ and $S_B$, respectively, with the following properties. First, $S_A=S_B$ hold with probability at least $1-\varepsilon$, and second,
			\begin{align}\label{eqn:M1}
			\frac 1N I(S_A:CE^N) \le \varepsilon~~~~\mathrm{and}~~~~ \frac 1N H(S_A) \ge R-\varepsilon
			\end{align} 
			hold. Here, C denotes the collection of messages sent over the insecure channel by Alice and Bob.
\end{definition}
In the Definition~\ref{def:Maurer93} above, $I(X:Y)$ is the classical mutual information function, and $H(X)$ is the Shannon entropy.

The SKA scenario can be naturally extended to the setting that involves more than two honest parties, i.e., conference key agreement CKA~\cite{multisquash}. Following the notation in Ref.~\cite{multisquash}, the object shared by the $m$ honest parties ($(m)A\equiv A_1 \dots A_m$) is $m$-partite probability distribution $P_{(m)A}$ defined on $\mc{X}_{(m)A} \equiv \mc{X}_{A_1} \times \cdots \times \mc{X}_{A_m}$. Here, $\mc{X}_{A_i}$ denotes the set values of random variable $A_i$ in possession of the $i$-th party. The n copies of such distribution belong then to $\mc{X}_{A_1}^n \times \cdots \times \mc{X}_{A_m}^n \equiv \mc{X}_{(m)A}^n$. In analogy to SKA scenario the extension $P^n_{(m)AE}$ of $P^{\otimes n}_{(m)A}$ to the eavesdropper system is defined as a probability distribution on  $\mc{X}_{(m)AE}^n \equiv \mc{X}_{A_1}^n \times \cdots \times \mc{X}_{A_m}^n \times \mc{X}_{E} $, where $\mc{X}_{E}$ is the set of values of random variable $E$ in possession of the eavesdropper. We first, provide the definition of the LOPC protocol in CKA scenario
\begin{definition}[Of CKA protocol, cf. Ref.~\cite{multisquash}] An LOPC protocol $\mathcal{P}$ is a family $\{\Lambda\}_n$ of classical channels $\Lambda_n:~\left( \mc{X}^n_{(m)AE}\right) \to \mc{X}^n_{(m)A^\prime E^\prime}$ which are a finite number of concatenation of local operations (local channels) and public communications steps (communication between honest parties via authorised but insecure  classical channel).
\end{definition}
The secret key rate $C_D^{(m)}$ in the $m$-partite CKA scenario is then defined as follows 
\begin{definition}[Of CKA secret key rate, cf. Ref.~\cite{multisquash}]\label{def:RateCKA} We say that LOPC protocol $\mathcal{P}$ is a classical key distillation protocol for a distribution $P_{(m)A} \in \mc{X}_{(m)A}$, if 
\begin{align}
    \lim_{n \to \infty} \norm{\Lambda^n \otimes \mathbbm{1}_E\left(P_{(m)AE}^{ n} \right) - K^{l_n}_{(m)AE}}_1 =0,
\end{align}
where $K^{l_n}_{(m)AE} = \frac{1}{l_n}(i_1\dots i_m) \delta_{i_1,\dots,i_m} \otimes P_E$ is the ideal key distribution on $\mc{X}_{(m)AE}$, for some distribution $P_E$ of the eavesdropper. The rate of the protocol if given by 
\begin{align}
    \mathcal{R}(\mc{P}) = \limsup_{n \to \infty} \frac{\log_2 l_n}{n}.
\end{align}
The classical distillable key of a distribution $P_{(m)A}$ is defined as supremum of the rates 
\begin{align}
    C_D^{(m)} (P_{(m)A}) = \sup_\mc{P} \mathcal{R}(\mc{P}).
\end{align}
\end{definition}

Definition~\ref{def:Maurer93} and Definition~\ref{def:RateCKA} are seemingly different. The first one refers to the notion of mutual information and Shannon Entropy, whereas the second uses trace norm distance~$\norm{\cdot}_1$ between the output state of the protocol and the ideal state. In Ref.~\cite{WDH2021} (see also Sec.~\ref{sec:D_Squashed}), the equivalence between the mentioned definitions was proved in the bipartite case. 

In the case of the device-dependent quantum key distribution (DD-QKD), the resources used in the cryptographic tasks are entangled quantum states. The set of operations available to the honest parties are, basically, transformations of quantum states and measurements on them. The secret key rate can then be defined both in LOPC and LOCC paradigms~\cite{keyhuge}. For the sake of this dissertation, we provide the definition of DD-CKA secret key rate referring to the LOCC paradigm that employs the notion of private states (see Sec.~\ref{subsec:PrivateAndPPT}). In the following definition of device-dependent conference key rate $K_D^{(m)}$, $\rho_{(m){A}}$ is a multipartite quantum state shared by $m$-honest parties ($(m){A}\equiv {A}_1 \dots {A}_m$), $n$ copies of which are shared by the honest parties at the beginning of the conference key distillation protocol.
\begin{definition}[Of DD-CKA protocol and conference key rate, cf. Ref.~\cite{AugusiakH2008-multi,multisquash}]\label{def:QCKArateAsym} For any given state $\rho_{(m){A}} \in \mc{B}\left(\mc{H}_{(m){A}}\right)$ let us consider a sequence $P_n$ of LOCC operations such that $P_n \left(\rho_{(m){A}}^{\otimes n}\right)=\sigma_n$. A set of operations $\mathcal{P}\equiv \bigcup_{n=1}^{\infty} \left\{P_n\right\}$ is called a pdit distillation protocol of state $\rho_{(m){A}}$ if there holds
\begin{align}
    \lim_{n \to \infty} \norm{\sigma_n-\gamma^{(m)}_{d_n}}_1 =0,
\end{align}
where $\gamma^{(m)}_{d_n}$ is a multipartite pdit whose key part is of dimension $d^{(1)}_n \times \cdots \times d^{(m)}_n$.
For a $\mathcal{P}$, its rate is given by 
\begin{align}
    \mc{R}(\mc{P})=\limsup_{n\to \infty} \frac{\log_2 d_n}{n}.
\end{align}
The distillable key of state $\rho_{(m){A}}$ is given by 
\begin{align}
    K_D^{(m)} \left(\rho_{(m){A}}\right)=\sup_{\mc{P}} \mc{R}(\mc{P}).
\end{align}
\end{definition}

Definitions of secret key rate \ref{def:Maurer93}, \ref{def:RateCKA} and \ref{def:QCKArateAsym} have an asymptotic character. Namely, they assume that the available number of copies of the resource state is not limited. Hence, the supremum over protocols in definitions of the rates also incorporates protocols that require an arbitrary large number of copies. On the other hand, in practical situations, the number of copies of the resource state is limited. Therefore, the number of bits of the secret key that can be distilled (per copy of resource state) from a finite number of copies is an important quantity to study. The following definition of one-shot (single run) secret key rate in the DD-CKA scenario corresponds to the one implicitly used in Ref.~\cite{DBWH19}.
\begin{definition}[One-shot secret key rate in DD-CKA scenario, cf. Ref~\cite{AugusiakH2008-multi,multisquash}]\label{def:QCKAsingleRate} For any given state $\rho_{(m){A}} \in \mc{B}\left(\mc{H}_{(m){A}}\right)$ let us consider a LOCC operation $\Lambda_n$  such that $\Lambda_n \left(\rho_{(m){A}}^{\otimes n} \right)=\sigma_n$. We call $\Lambda_n$ a $\varepsilon$-secure key distillation protocol from $n$ copies of state $\rho_{(m){A}}$ if there holds  
\begin{align}
    F\left( \sigma_n, \gamma_{d_n}^{(m)} \right) \ge 1 - \epsilon,
\end{align}
where $\gamma^{(m)}_{d_n}$ is a multipartite pdit whose key part is of dimension $d^{(1)}_n \times \cdots \times d^{(m)}_n$, and $F(\cdot,\cdot)$~denotes the fidelity between quantum states.
For a $\Lambda_n$, its rate is given by 
\begin{align}
    \kappa_n^{\varepsilon}\left(\Lambda_n (\rho_{(m){A}}^{\otimes n})\right) := \frac{\log_2 d_n}{n},
\end{align}
and the rate of $\varepsilon$-secure secure key distilled from $n$ copies of state $\rho_{(m){A}}$, by protocols satisfying the above security condition, is given by
\begin{align}
    K_D^{(n,\epsilon)}(\rho_{(m){A}}) := \sup_{\Lambda_n \in \mathrm{LOCC}} \kappa_n^{\varepsilon}\left(\Lambda_n (\rho_{(m){A}}^{\otimes n})\right). 
\end{align}
Finally, the one-shot $\varepsilon$-secure secret key is defined as 
\begin{align}
    K^{s,\varepsilon}_{D}(\rho_{(m){A}}) :=   K_D^{(1,\epsilon)} (\rho_{(m){A}}).
\end{align}
\end{definition}

Eventually, let us remark that the following relation holds between asymptotic and one-shot DD-CKA secret key rates, namely
\begin{align}
    K_D^{(m)} (\rho) = \inf_{\epsilon > 0} \limsup_{n \to \infty} K_D^{(n,\epsilon)}(\rho),\label{eqn:OStoAsym}
\end{align}
where $m$ denotes the number of honest parties. The Definition~\ref{def:QCKAsingleRate} is written in the LOCC paradigm, and the security condition is based on the fidelity between quantum states. On the contrary, the Definition~\ref{def:QCKArateAsym} is given in the LOPC paradigm where the security condition is based on the trace norm. The relation in Eq.~\eqref{eqn:OStoAsym} follows from the equivalence between LOCC and LOPC paradigms~\cite{keyhuge} and Fuchs-van de Graaf inequalities~\cite{Fuchs1999}.

In the device-independent conference key agreement scenario (DI-CKA), the $N$ honest parties are assumed to share an untrusted device, described with input-output statistics $\{p({\textbf a}\vert {\textbf x})\}_{{\textbf a}|{\textbf x}}$ originating from tensor product POVM (positive operator-valued measure) measurements $\mathcal{M}\equiv \{M^{x_1}_{a_1}\otimes M^{x_2}_{a_2}\otimes\ldots\otimes M^{x_N}_{a_N}\}_{\textbf{a}|\textbf{x}}$ on quantum state $\rho_{N(A)}$, i.e., a pair $(\rho_{N(A)},\mathcal{M})$. Here, $N(A)\equiv A_1\dots A_N$, where $A_i$ refers to the subsystems of the $i$-th party, and $ \textbf{x}\coloneqq (x_1,x_2,\ldots , x_N)$ and $\textbf{a}\coloneqq (a_1,a_2,\ldots ,a_N)$ refer to the inputs and outputs of the honest parties, respectively. The attack strategy of the adversary is to replace a device $(\rho_{N(A)},\mathcal{M})$ with another device $(\sigma_{N(A)},\mathcal{N})$ which yields the same attack statistics as the honest one. The power of the adversary is restricted then to the laws of quantum theory, and therefore, the adversary may hold the purifying system of $\sigma_{N(A)}$, i.e., the $E$ subsystem of the joint state $\ket{\psi_\sigma}_{N(A)E}$. We further omit writing subscripts like $N(A)$ in places in which it does not lead to any confusion.

Consider the following relations that are a basis for two different security conditions in the DI-CKA scenario, i.e., ``par'' and ``dev'' definitions.
\begin{align}
    (\rho,\mc{M}) &\approx_\varepsilon  (\sigma,\mc{N}) \label{eq:pre:k-1pre},\\
    \omega(\rho,\mc{M})& \approx_{\varepsilon} \omega(\sigma,\mc{N})\label{eq:pre:k-2pre},\\
    P_{err}(\rho,\mc{M})& \approx_{\varepsilon} P_{err}(\sigma,\mc{M})\label{eq:pre:k-3pre},
\end{align}
where $\omega(\rho,\mc{M})$ is the level of violation of some (chosen by honest parties) multipartite Bell inequality, and $P_{err}(\rho,\mc{M})$ is quantum bit error rate (QBER). Furthermore, relation in Eq.~\eqref{eq:pre:k-1pre} means that two input-output statistics $p$ and $p^\prime$ originating from devices $(\rho,\mc{M})$ and $(\sigma,\mc{N})$ respectively, satisfies
\begin{equation}
    d(p,p')=\sup_{{\textbf x}}\norm{p(\cdot\vert {\textbf x})-p'(\cdot \vert {\textbf x})}_1\leq \varepsilon.
\end{equation}

We restrict ourselves to the definitions of the DI-CKA secret key rates in the independent and identically distributed (iid) setting. See Definition \ref{def:QCKAsingleRate} for the meaning of $\kappa^\varepsilon_{n}$.
\begin{definition}[Of ``dev'' iid DI-CKA secret key rate, cf. Ref~\cite{CFH21}]\label{def:di-devPRE}
The (multipartite) device-independent quantum key distillation rate of a device $(\rho,\mc{M})$ with independent and identically distributed behavior is defined as
\begin{equation}
    K^{iid}_{DI,dev}(\rho,\mc{M})\coloneqq \inf_{\varepsilon>0}\limsup_{n\to \infty} \sup_{\hat{\mc{P}}} \inf_{\eqref{eq:pre:k-1pre}} \kappa^\varepsilon_{n} \left(\hat{\mc{P}}\left((\sigma,\mc{N})^{\otimes n}\right)\right),
\end{equation}
where $\kappa_n^\varepsilon$ is the rate of a key distillation protocol $\hat{\mc{P}}$ producing $\varepsilon$-secure output, acting on $n$ copies of the state $\sigma$, measured with $\mc{N}$. Here $\hat{\mc{P}}$ is a protocol composed of classical local operations and public (classical) communication (CLOPC) acting on $n$ identical copies of $(\sigma,\mc{N})$ which, composed with the measurement, results in a quantum local operations and public (classical) communication (QLOPC) protocol.
\end{definition}

\begin{definition}[Of ``par'' iid DI-CKA secret key rate, cf. Ref.~\cite{AFL21}]\label{def:di-parPRE}
The (multipartite) device-independent quantum key distillation rate of a device $(\rho,\mc{M})$ with independent and identically distributed behavior, Bell inequality violation $\omega(\rho,\mc{M})$, and QBER $P_{err}(\rho,\mc{M})$ is defined as
\begin{align}
    & K^{iid}_{DI,par}(\rho,\mc{M}) \quad\coloneqq \inf_{\varepsilon>0}\limsup_{n\to \infty} \sup_{\hat{\mc{P}}} \inf_{\eqref{eq:pre:k-2pre},\eqref{eq:pre:k-3pre}} \kappa^\varepsilon_{n}\left(\hat{\mc{P}}\left((\sigma,\mc{N})^{\otimes n}\right)\right).
\end{align}
\end{definition}

\begin{definition}[Of single-shot DI-CKA secret key rate, cf.~Ref.~\cite{HWD22}]
The single-shot device-independent quantum key distillation rate of a device $(\rho,\mc{M})$ with independent and identically distributed behavior is defined as
\begin{equation}
    K^\mathrm{single-shot}_{DI,dev}(\rho,\mc{M},\epsilon)\coloneqq \sup_{\hat{\mc{P}}} \inf_{\eqref{eq:pre:k-1pre}} \kappa^\varepsilon_{n} \left(\hat{\mc{P}}(\sigma,\mc{N})\right),
\end{equation}
where $\kappa_n^\varepsilon$ is the quantum key rate achieved for any security parameter $\varepsilon$ and measurements $\mc{N}$. Here $\hat{\mc{P}}$ is a protocol composed of classical local operations and public (classical) communication acting on a single copy of $(\sigma,\mc{N})$ which, composed with the measurement, result in local quantum operations and public (classical) communication protocol.
\end{definition}

\begin{definition}[Of the reduced DD-CKA secret key rate, cf.~\cite{CFH21,HWD22}]\label{def:ReducdDDpre}
The reduced device-dependent conference key rate of an $N$-partite state $\rho_{N(A)}$ reads
\begin{align}
K^{\downarrow}(\rho_{N(A)})\coloneqq \sup_{\cal M}\inf_{(\sigma_{N(A)}),{\cal L})=(\rho_{N(A)},{\cal M})} K_{DD}(\sigma_{N(A)}).
\end{align}
\end{definition}
In the Definition~\ref{def:ReducdDDpre} above, $K_{DD}$ correspond to $K_{D}^{(N)}$ in Definition~\ref{def:QCKArateAsym}. The reduced DD-CKA secret key rate is an important upper bound on the iid DI-CKA secret key rate \cite{CFH21}.



In the non-signaling device-independent secret key agreement scenario (NSDI), the two honest parties share $N$ copies of the non-signaling probability distribution (non-signaling device) $P(AB|XY)$, being the aforementioned resource, where $X$ and $Y$ denote the inputs and $A$ and $B$ correspond to outputs of the honest parties. In the NSDI scenario, the eavesdropper is limited solely with the no-signaling conditions~\cite{InfoProcess} and possesses an extending system of the device~$P(AB|XY)$. The device shared by all three parties is, therefore, described with tripartite conditional probability distribution $Q(ABE|XYZ)$, which has $N$ copies of $P(AB|XY)$ as its marginal state. Here, $E$ and $Z$ denote the output and input of the system held by the eavesdropper. Moreover, $Q(ABE|XYZ)$ can exhibit a supra-quantum correlation between its subsystems~\cite{PR}. In the considerations in Ref.~\cite{WDH2021}, we assume that the extension held by the eavesdropper is the non-signaling complete extension (NSCE) of $N$ independent and identically distributed (iid) copies of the device $P(AB|XY)$, i.e., $\mathcal{E}(P^{\otimes N})(ABE|XYZ)$~\cite{CE} (see also Sec.~\ref{sec:E_CE}). From the perspective of the honest parties, the NSCE is the worst-case extension the eavesdropper may use. This power is because NSCE allows access to all statistical ensembles of the extended system via suitable measurement and processing on the extending system~\cite{CE}.

We now provide the necessary definitions of MDLOPC operations, secret key distillation protocol and ideal state, its security condition, and the secret key rate in the iid NSDI scenario.
\begin{definition}[Of MDLOPC operations, cf. Ref.~\cite{WDH2021}]
    The MDLOPC operation consists of (i) direct measurement $\mathcal{M}^F_{x,y}$ on inputs $X$ and $Y$ of the honest parties, that changes the device into a probability distribution, followed by (ii) arbitrary bipartite LOPC operations. Direct measurement is a measurement that does not incorporate any external randomness on the measured input. 
\end{definition}

\begin{definition}[Of the MDLOPC protocol in the iid NSDI scenario, cf. Refs.~\cite{hanggi-2009,WDH2021}]\label{def:NSDIprotocol}
A~protocol of key distillation is a sequence of MDLOPC operations $\Lambda \equiv \left\{\Lambda_N\right\}$, performed by the honest parties on $N$ iid copies of the shared devices. Each of this $\Lambda_N$,  consists of a measurement stage $\{{\cal M}_N\}$, followed by post-processing $\{{\cal P}_N\}$, on $N$ iid copies of $P(AB|XY)$. Moreover, for each consecutive, complete extension of $N$ copies of shared devices $\mathcal{E}(P^{\otimes N})(\bm{AB}E|\bm{XY}Z)$, the protocol outputs a probability distribution in part of Alice and Bob and a device in part of Eve, which is arbitrarily close to an ideal distribution, and satisfies
\begin{align}
    &{||P_{\mathrm{out}}- P_\mathrm{ideal}^{(d_N)} ||}_\mathrm{NS} \le \varepsilon_N \stackrel{N\rightarrow \infty}{\longrightarrow} 0,\\
    &P_\mathrm{ideal}^{(d_N)} (s_A,s_B,q,e|z) := \frac{\delta_{s_A,s_B}}{|S_A|} \sum_{s_A^\prime,s_B^\prime} P_\mathrm{out} (s_A^\prime,s_B^\prime,q,e|z).
\end{align}
Here, $P_{\mathrm{out}} = \Lambda_N ~ \left({\mathcal{E}} \left(P^{\otimes N}\right)\right)$, and random variables $S_A$, $S_B$ and $Q$ correspond to the keys of the honest parties and communication respectively. Moreover, $\bm{A} = A_1A_2 \ldots A_N$, $\bm{B}$, $\bm{X}$ and $ \bm{Y}$ are similarly defined. 
\end{definition}
In the Definition~\ref{def:NSDIprotocol} above $\norm{\cdot}_\mathrm{NS}$ denotes the non-signaling norm described in Sec.~\ref{sec:D_Squashed} (see also Ref.~\cite{WDH2021}).

\begin{definition}[Of the secret key rate in the iid NSDI scenario, cf. Ref.~\cite{WDH2021}]\label{def:key_rate}
    Given a bipartite device $P\equiv P(AB|XY)$ the secret key rate of the protocol of key distillation  $\Lambda_N$, on $N$ iid copies of the device, denoted by $\mathcal{R}\left(\left.\Lambda\right|_P\right)$ is a number $\limsup_{N\rightarrow \infty} \frac{\log d_N}{
        N}
    $, where $\log d_N$ is the length of a secret key shared between Alice and Bob, with $d_N=\mathrm{dim}_\mathrm{A}\left(\Lambda_N \left({\cal E} \left(P^{\otimes N}\right)\right)\right) \equiv |S_A|$. The  device independent key rate of the {\it iid} scenario is given by
\begin{align}
    K_{DI}^{(iid)}(P)=\sup_{ \Lambda} {\cal R}\left(\left.\Lambda\right|_P\right),
\end{align}
where the supremum is taken over all MDLOPC  protocols~$\{\Lambda_N\}$.
\end{definition}

In Ref.~\cite{WDH2021}, we show that the Definition~\ref{def:key_rate} of the iid NSDI secret key rate is equivalent in terms of security
to the one present in the literature~\cite{masanes-2006,masanes-2009-102,Masanes2011,hanggi-2009b,Renner-Hanggi,hanggi-2009,Hanggi-phd}.\\
~~\\

\subsubsection{Key Repeater Rate}\label{subsubsec:RepeaterRate}

Quantum repeaters and quantum key repeaters ~\cite{PhysRevLett.81.5932,Dur_1999,munro2015inside,CZC+21} are 
devices that allow overcoming attenuation of (e.g., optical) signal, and decoherence of quantum states, in order to entangle quantum systems separated by arbitrarily large distances. The idea of quantum repeaters is based on the entanglement swapping protocol~\cite{PhysRevLett.81.5932}, which is performed inside a quantum repeater. In the simplest setup with one intermediate station denoted here $H$, the joint measurement in ``entangled'' basis performed in central station $H$ on subsystems $A^\prime E^\prime$ of two initially uncorrelated but entangled quantum states $\rho_{AA^\prime}$ and $\rho_{EE^\prime}$ can create entanglement between subsystems $A$ and $E$ that are spatially separated~\cite{PhysRevLett.81.5932,Bennett+DiVincenzo_1996}. The measurement outcome at hub station $H$ is then communicated to stations $A$ and $E$. Importantly, quantum repeaters can be combined in chains that have an arbitrary number of hub stations. Consequently, the distance at which entanglement can be created increases.  The idea of quantum key repeaters generalizes the concept of quantum repeaters. The task of a quantum key repeater is to distill private states (see Sec.~\ref{subsubsec:2Private}) between stations $A$ and $E$ using tripartite LOCC operations among nodes $A$, $E$, and $H$. The communication between the central station $H$ and stations $A$ and $E$ can be considered both one- and two-way. The rate of a quantum key repeater is defined as follows

\newpage
\begin{definition}[Of quantum key repeater rate, cf. Ref.~\cite{bauml2015limitations,christandl2017private}] The quantum key repeater rate with respect to arbitrary tripartite LOCC operations among $A$, $E$ and $H$ is defined as follows
 \begin{align}
 &R^{A \leftrightarrow H \leftrightarrow E}(\rho, \rho') := 
 \lim_{\epsilon \to 0} \limsup_{n \to \infty} \sup_{\Lambda_{n}^{LOCC},\gamma_{d_k,d_s}} \left\lbrace \frac{\log d_k}{n}: \tr_{H} \Lambda_n^\text{LOCC}((\rho \otimes \rho')^{\otimes n}) \approx_{\epsilon} \gamma_{d_k,d_s} \right\rbrace,
 \end{align}
 where Adam and Hub share state $\rho$ while Hub and Eve share $\rho'$. $\Lambda:=\left\{\Lambda_n^\text{LOCC}\right\}$ are tripartite LOCC protocols with two-way classical communication between nodes $H$, $A$, and $E$. In the case in which communication between central node and $A,E$ systems is restricted to one-way from $H$ to $A$ and $E$, we denote this rate with $R^{H \to A:E}$.
\end{definition}
In the above definition, $\gamma_{d_k,d_s}$ refers to a private state distilled in a quantum key repeater (see Sec.~\ref{subsubsec:2Private}).

\subsection{Quantum Channels}\label{subsec:Channels}

Quantum channels, also known as quantum operations, are the most general form of transformations between quantum states (see Ref.~\cite{Nielsen-Chuang} in this context). The (memoryless) quantum channel can be defined as follows
\begin{definition}[Of quantum operations, aka quantum channels, cf. Ref~\cite{Nielsen-Chuang}]
Let $\mathcal{D}(\mathcal{H})$ and $\mathcal{D}(\mathcal{H}^\prime)$ define the set of density operators acting on two arbitrary Hilbert spaces $\mathcal{H}$ and $\mathcal{H}^\prime$, respectively. Any completely positive and trace preserving (CPTP) linear map $\Lambda:~\mathcal{D}(\mathcal{H}) \to \mathcal{D}(\mathcal{H}^\prime)$ is called a quantum operation (quantum channel).
\end{definition}

In Ref.~\cite{DBWH19} (see also Sec.~\ref{sec:B_Universal}), we specify the use of quantum channels in situations concerning conference key agreement, in which the spatial distribution of the subsystems after the action of a quantum channel does not correspond to the initial distribution. We distinguish the roles of the senders and receivers $\tf{A}_a$, senders $\tf{B}_b$, and receivers $\tf{C}_c$. We call such types of quantum channels multiplex quantum channels. In the following definition, we use a notation in which $\vv{X} \equiv \tf{X}_1\dots \tf{X}_K$, where $K$ is the number parties of type $\tf{X}\ \in \{\tf{A},\tf{B},\tf{C}\}$.  
\begin{definition}[Of multiplex quantum channel, cf. Ref.~\cite{DBWH19}]
Consider multipartite quantum channel $\mc{N}_{\vv{A'}\vv{B}\to\vv{A}\vv{C}}$ where each pair $A_a', A_a$ is held by a respective party  $\tf{A}_a$ and each $B_b,C_c$ are held by parties $\tf{B}_b,\tf{C}_c$, respectively. While $\tf{A}_a$ is both sender and receiver to the channel, $\tf{B}_b$ is only a sender, and $\tf{C}_c$ is only a receiver to the channel. Such a quantum channel is referred to as the multiplex quantum channel. Any two different systems need not be of the same size in general. 
\end{definition}

\newpage
In principle, multiple uses of multiplex quantum channel interleaved by LOCC can simulate a wide range of QKD protocols, which are secure in the sense of the device-dependent paradigm. It is, therefore, useful to quantify conference key agreement rates achievable with particular multiplex quantum channels. The formal description of LOCC-assisted conference key agreement protocol over a multiplex quantum channel $\mc{N}_{\vv{A'}\vv{B}\to\vv{A}\vv{C}}$ and the definition of secret-key-agreement capacity $\hat{P}_{\LOCC}(\mc{N})$ of quantum channel $\mc{N}_{\vv{A'}\vv{B}\to\vv{A}\vv{C}}$ are provided in Sec.~\ref{sec:B_Universal} (see also Ref.~\cite{DBWH19}). The same concerns the definition of cppp-assisted (one-shot) secret-key-agreement capacities $\hat{P}_{\textnormal{cppp}}^{(1,\varepsilon)}$ of multiplex quantum channels. Here, cppp refers to classical preprocessing and postprocessing, which are operations employed in the protocol.


\subsection{Entanglement and Genuine Entanglement}\label{subsec:Entanglement}

The presence of entanglement in quantum theory is one of the basic features that differentiate quantum theory from the classical theory of Nature. It was already observed by A.~Einstein, B.~Podolsky, N.~Rosen~\cite{EPR35} and E. Schr\"odinger~\cite{Schrodinger} that systems described with entangled quantum states exhibit correlations that can not be explained with the classical theory of probability. The discussion about the meaning of quantum entanglement in modern physics is beyond the scope of this dissertation (see Ref.~\cite{Horodecki2009}). The entangled quantum states are defined as states that are not fully separable, i.e.,  
\begin{definition}[Of fully separable and entangled states, cf., e.g.,  Ref.~\cite{AugusiakH2008-multi}]\label{def:FS}
    If a $N$-partite quantum state $\rho_{N(A)}$, where $N(A)=A_1\dots A_N$, can be written as a convex combination of tensor product of states $\rho_{A_j}^{(i)}$ of individual subsystems
    \begin{align}
        \rho_{N(A)} = \sum_i p_i \rho_{A_1}^{(i)} \otimes \cdots \rho_{A_N}^{(i)},
    \end{align}
    then we call it a full separable state. All separable states constitute a set of fully separable states (FS). All other states are entangled.
\end{definition}
We remark here, that operationally entangled quantum states are quantum states that can not be papered with (multipartite) LOCC operations. In the bipartite setting the set of (fully) separable states is denoted $\mathcal{SEP}$, and called the set of separable states.

According to Definition~\ref{def:FS} a quantum state of the form $\rho_{A_1} \otimes \sigma_{A_2 A_3}$ is entangled, if and only if, $\sigma_{A_2 A_3}$ is entangled. However, system $A_1$ is not entangled with subsystems $A_2$ and $A_3$. Indeed, it is ``biseparable'' as it is of product form in (at least) one ``cut'' between its subsystems. It is, therefore, convenient, also from the cryptographic perspective, to define quantum states in which each subsystem $A_i$ is entangled with any other subsystem $A_{ \neq i}$ or collection of them $A_{ \neq i} \dots A^\prime_{ \neq i}$.
\newpage
\begin{definition}[Of biseparable state, cf. Ref.~\cite{Horodecki2009,DBWH19}]\label{def:BS}
    If a $N$-partite quantum state $\rho_{N(A)}$ can be written as a convex combination of $N$-partite states that are separable with respect to at least one partition of its subsystems 
    \begin{align}
        \rho_{N(A)} = \sum_i p_i \rho^{(i)}_{K_i(A)} \otimes \rho^{(i)}_{M_i(A)},~~K_i+M_i=N,
    \end{align}
    then we call it a biseparable state. All separable states constitute a set of biseparable states (BS). All other states are genuine entangled (GE).
\end{definition}
In the Definition~\ref{def:BS}, the word genuine emphasizes that the state therein, indeed, can not be prepared by LOCC without the resource of $N$-partite entangled pure states. In this place, we remark that the sets of fully separable (FS) and biseparable (BS) states are convex sets. However, the set of biseparable states is not closed under the tensor product. The subset of biseparable states that is closed under the tensor product is called the set of tensor-stable biseparable states and is defined as follows 
\begin{definition}[Of tensor-stable biseparable states, cf. Ref.~\cite{Palazuelos2022,DBWH19}] If any tensor power of some biseparable sate $\rho_{N(A)}$ is a biseparable state, i.e., 
\begin{align}
    \forall_{n \ge 1}~~ \rho_{N(A)}^{\otimes n} \in \mathrm{BS},
\end{align}
we call the state $\rho_{N(A)}$ tensor-stable biseparable state.
\end{definition}

\subsection{Nonlocality and Genuine Nonlocality}\label{subsec:Nonlocality}

Nonlocality is a notion that refers to the violation of Bell inequalities by quantum systems~\cite{Bell-nonlocality}. In analogy to the case of the theory of entanglement (see Sec.~\ref{subsec:Entanglement}), one can define the notions of local and genuine nonlocal quantum behaviors within the theory of nonlocality~\cite{Bell-nonlocality}. Quantum behaviors $P(\textbf{a}|\textbf{x})$ are conditional probability distributions that originate from measurement on quantum states (see Secs.~\ref{subsubsec:Security} and \ref{sec:C_Fundamenal} for more details).   
\begin{definition}[Of behavior local in a ``cut'', cf. Ref.~\cite{HWD22}]
We say that $N$-partite behavior $P(\textbf{a}|\textbf{x})$ is local in a cut $(A_{i_1}\ldots A_{i_k}):(A_{i_{k+1}}\ldots  A_{i_N})$ for some $k\in\{1,\dots,N\}$, if it can be written as a product of two behaviors on systems $A_{i_1}..A_{i_k}$ and $A_{i_{k+1}}\ldots  A_{i_N}$, respectively, i.e.,
\begin{align}
    P(\textbf{a}|\textbf{x}) = P_{1,k}(a_{i_1}\ldots a_{i_k}|x_{i_1}\ldots x_{i_k}) \otimes P_{k+1,N}(a_{i_{k+1}}\ldots  a_{i_N}|x_{i_{k+1}}\ldots  x_{i_N}),
\end{align}
for $(i_1,\dots,i_N)$ being some permutation of indices $(1,\dots,N)$
\end{definition}

There are various definitions of locality of behaviors in multipartite case~\cite{Bell-nonlocality}. In Ref.~\cite{HWD22} we use the following definition of locality.
\begin{definition}[Of locally quantum behaviors, cf. Ref.~\cite{Bell-nonlocality,HWD22}]\label{def:QL}
The locally quantum behaviors are convex mixtures of behaviors that are a  product in some cut and both behaviors in the product have quantum realization. 
\end{definition}
The class of locally quantum behaviors in Definition~\ref{def:QL} falls between the so-called TOBL and NSBL classes in Ref.~\cite{TOBL} (see also Ref.~\cite{Svetlichny1,Bell-nonlocality}). Definition~\ref{def:QL} of locally quantum behaviors induces the following definition of genuinely nonlocal quantum behaviors
\begin{definition}[Of genuinely nonlocal quantum behaviors, cf. Ref.~\cite{HWD22}]
The behavior $P(\textbf{a}|\textbf{x})$ is  genuinely nonlocal if and only if it is not locally quantum, i.e., it is not 
a mixture of behaviors that are a
product in at least one cut and have quantum realizations. 
\end{definition}


\subsection{Generalized Probabilistic Theories}\label{subsec:GPTs}

The framework of generalized probabilistic theories (GPTs)~\cite{hardy2001quantum,InfoProcess} is a formalism that describes the operational predictions of essentially arbitrary conceivable physical theories. In particular, the quantum theory, the classical probabilistic theory, and the theory of non-signaling behaviors (see Sec.~\ref{subsubsec:NSthoery}) are known examples of theories that can be described and systematically studied within the framework of GPTs. Each generalized probabilistic theory (GPT) is defined by some set of postulates, the implications of which can be methodically investigated within the framework of GPTs. As discussed in Sec.~\ref{sec:E_CE} here, the framework of GPTs can help in finding the future theory of Nature that would be more fundamental and have greater explanatory power than the quantum theory. See Refs.~\cite{plavala2021general,muller2021probabilistic,lami2018non} for a more insightful introduction to the framework of GPTs.

\subsubsection{Generalized Probabilistic Theories in a Nutshell}\label{subsubsec:GPTsInNutshell}

The primitive building blocks of any GPT, $\mathcal{G}$ are the systems, denoted $A, B,...\in \textsf{Syst}[\mathcal{G}]$, that are equipped with an associative bilinear composition rule $\otimes:\textsf{Syst}[\mathcal{G}]\times \textsf{Syst}[\mathcal{G}] \to \textsf{Syst}[\mathcal{G}]$ which allow constructing composite systems out of the systems. Each system $A$ is described as a vector in finite-dimensional real vector space $V_A$. In this place, we note that $V_{A\otimes B}$ is not necessarily equal to $V_A\otimes V_B$. The equality holds only under the assumption of tomographic locality \cite{hardy2001quantum}. The convex, compact, and closed set $\Omega_A \subset V_A$ describes the state space for the system $A$. Moreover, $\Omega_A $ must have an affine dimension at least one less than the linear dimension of the vector space such that the affine span of $\Omega_A$ does not intersect with the origin. One can also define the convex state cone $K_A:=\{r s | r\in \mathbbm{R}^+, s\in \Omega_A\}$ which includes also subnormalised ($r<1$), normalized ($r=1$) and supernormalised ($r>1$) states. 

Let $V_A^*$ be the dual vector space to $V_A$, i.e.,  the vector space of all linear functionals on $V_A$. A specified proper convex subset $\mathcal{E}_A$ of $V_A^*$, i.e., $\mathcal{E}_A \subset V_A^*$, describes the effect space for the system $A$. The subset $\mathcal{E}_A$ is assumed to be full dimensional (with respect to $V_A^*$), compact and closed. The effects $e\in\mathcal{E}_A$ assign probabilities to measurement outcomes whenever the measurement is performed on an arbitrary state $s \in\Omega_A$, they, therefore, must satisfy  $e(s) \in [0,1]$. The effect space $\mathcal{E}_A$ contains also the unique unit effect $u_A \in \mathcal{E}_A$ which is defined as $u_A(s)=1$ for all $s \in \Omega_A$. The uniqueness of the unit effect is built into the setup of the GPTs framework (see Ref.~\cite{Chiribella2010} for an alternative approach).

In analogy to the case of GPT systems, both state and effect spaces are equipped with an associative composition rule, $\Omega_A\otimes \Omega_B := \Omega_{A\otimes B}$, $\mathcal{E}_A\otimes \mathcal{E}_B:=\mathcal{E_{A\otimes B}}$ and $u_A\otimes u_B := u_{A\otimes B}$. This composition is assumed to be bilinear, and satisfy $e\otimes f(s\otimes t)= e(s) f(t)$ for all $e\in \mathcal{E}_A$, $f\in \mathcal{E}_B$, $s\in\Omega_A$ and $t\in\Omega_B$. The conditions above allow unambiguously defining a kind of ``partial trace''. Consequently, they ensure that a GPT satisfies the no-signaling principle~\cite{Chiribella2010,coecke2014terminality,kissinger2017equivalence} (see also Sec.~\ref{subsubsec:NSthoery}). We remark here that, in general, $\otimes$ is not the tensor product of vector spaces. The symbol ``$\otimes$'' is used in analogy to the notation of the tensor product in composing quantum systems.

Given any two GPT systems $A$ and $B$, there is a space of transformations $\mathcal{T}_A^B$ from system $A$ to system $B$. The space of transformations forms a closed compact and convex set. There are two associative composition rules for transformations that are relevant (i) parallel composition, $\mathcal{T}_A^B \otimes \mathcal{T}_C^D := \mathcal{T}_{A\otimes C}^{B\otimes D}$, and (ii) sequential composition, $\mathcal{T}_B^C \circ \mathcal{T}_A^B:= \mathcal{T}_A^C$. Both composition rules are bilinear and must satisfy the condition that $(T_1 \otimes T_2) \circ (T_3\otimes T_4) = (T_1\circ T_3) \otimes (T_2\circ T_4)$. States can be viewed as transformations with the trivial input system, denoted $\star$, and effect as transformations with trivial output. The trivial system $\star$ is a unit of system compositions, i.e.,  $A\otimes \star = A = \star \otimes A$. Consequently, one can write that $\Omega_A = \mathcal{T}_\star^A$ and $\mathcal{E}_A=\mathcal{T}_A^\star$. Finally, there is also  an identity transformation $\mathbbm{1}_A$ which  for every $T:A\to B$ must satisfy $\mathbbm{1_B}\circ T = T = T\circ \mathbbm{1}_A$.

Withing the framework of GPTs, it is convenient to work with the convention in which every GPT contains classical systems, denoted $\Delta_I$. These systems are used as registers in which measurement outcomes can be stored, and which control variables of ``experiments'' can be encoded~\cite{gogioso2017categorical,selbyReconstruction}. Let $\Delta_I$ be a classical system which corresponds to outcome degree of freedom on some measurement device where $I$ labels the set of possible outcomes. Then, vector space $V_{\Delta_I}$ corresponds to the real vector space $\mathbbm{R}^I$ of real-valued functions from $I\to \mathbbm{R}$. Consequently, the state space $\Omega_{\Delta_I}$ is the space of probability distributions over $I$, i.e., real-valued functions $p:I\to \mathbbm{R}$ such that $p(i) \in [0,1]$ and $\sum_i p(i)=1$ for all $i\in I$.
In~fact, geometrically $\Omega_{\Delta_I}$ is a simplex in which the elements of $I$ label the vertices. Furthermore, the vertices correspond to delta function probability distributions $\delta_i$. The effect space $\mathcal{E}_{\Delta_I}$ is the vector space dual to $V_{\Delta_I}$. However, due to the Riesz representation theorem via the inner product $\sum_{i\in I} f(i) g(i)$, the effects can be seen as the elements of $\mathbbm{R}^I$. In this representation, the effects correspond to functions $e:I\to \mathbbm{R}$ such that $e(i) \in [0,1]$. Geometrically, the set of such functions forms a hypercube that contains the simplex of states. In Ref.~\cite{CE}, we denote the vertices of the simplex  (when interpreted as effects via the Riesz representation theorem) with $\epsilon_i$, then $\epsilon_i(\delta_j) = \sum_{k\in I} \delta_i(k)\delta_j(k) = \delta_{ij}$. The classical systems compose, therefore, via $\Delta_I\otimes \Delta_J := \Delta_{I\times J}$. In consequence, the transformations between classical systems are stochastic linear maps. Moreover, an important property of classical theory is that identity transformations $\mathbbm{1}_{\Delta_I}$ can be expanded as $\sum_{i\in I} \delta_i \circ \epsilon_i$. Consequently, any measurement $M:A\to\Delta_I$ must satisfy $M = \mathbbm{1}_{\Delta_I} \circ M = \sum_{i\in I}  \delta_i \circ \epsilon_i \circ M$. Now, because, $e_i:=\epsilon_i \circ M \in \mathcal{T}_A^\star = \mathcal{E}_A$ one can reexpress the above $\mathbbm{1}_{\Delta_I}$ as $\sum_{i\in I} \delta_i \circ e_i$. Therefore, it is possible to construct an isomorphism between (i) measurements as a collection of effects and (ii) measurements as transformations to a classical system.

\subsubsection{The Theory of Non-Signaling Behaviors}\label{subsubsec:NSthoery}

The theory of non-signaling behaviors (aka the Box-world theory) is an instance of a generalized probabilistic theory (GPT) based on the no-signaling principle wherein the states are multipartite conditional probability distributions $P_{\textbf{A}|\textbf{X}}$ called behaviors, which satisfy the so-called no-signaling conditions~~\cite{InfoProcess,PR}. Here,  $\textbf{X} \equiv X_1X_2\dots X_N$ and $\textbf{A} \equiv A_1A_2\dots A_N$ refer to the inputs (measurement choices) and outputs (measurement outcomes) of $N$ parties, respectively. Importantly, not all non-signaling behaviors have quantum realizations. The theory of non-signaling behaviors exhibits stronger correlations between subsystems than quantum theory and saturates the algebraic maximum of the Clauser-Horne-Shimony-Holt (CHSH) inequality~\cite{PR}. The no-signaling principle that prohibits instantaneous communication (faster-than-light communication, colloquially)  between spatially separated subsystems yields the no-signaling conditions on the behaviors $P_{\textbf{A}|\textbf{X}}$. In what follows, we refer to the systems described with the theory of non-signaling behaviors as to the non-signaling devices (see also Sec.~\ref{sec:D_Squashed}). In the bipartite setting, the no-signaling conditions and non-signaling behaviors are defined as follows
\begin{definition}[Of bipartite non-signaling behaviors, cf. Ref.~\cite{InfoProcess,Bell-nonlocality}]\label{def:NS2behavior}
Let $P_{AB|XY}$ be a behavior describing a bipartite device. If the following so-called no-signaling conditions hold    
\begin{align}
    & \forall_{a,x, y, y'}~~P_{A|X}(a|x) = \sum_b P_{AB|XY}(ab|xy) = \sum_b P_{AB|XY}(ab|xy'), \label{eq:nons:A}\\
	&\forall_{b,x, x', y}~~ P_{B|Y}(b|y) = \sum_a P_{AB|XY}(ab|xy) = \sum_a P_{AB|XY}(ab|x'y),
\end{align}
we call $P_{AB|XY}$ a non-signaling behavior, and $P_{A|X}(a|x)$ , $P_{B|Y}(b|y)$ the marginal distributions of $A$ and $B$ respectively.
\end{definition}
The Definition~\ref{def:NS2behavior} of the no-signaling conditions and non-signaling behaviors can be readily extended to the multipartite setting.
~~\\
\newpage
\begin{definition}[Of $N$-partite non-signaling behaviors, cf. Ref.~\cite{hanggi-2009,AxiomContext}]\label{def:NSmultiBehavior}
Let $P_{\textbf{A}|\textbf{X}}$ be a behavior describing a $N$-partite device, where $\textbf{A} \equiv A_1A_2\dots A_N$ and $\textbf{X} \equiv X_1X_2\dots X_N$ are the outputs and inputs of $N$ parties, respectively. If the following, condition hold 
\begin{align}
    \forall_{1\le i \le N, \textbf{a}^{\neq i},\textbf{x}^{\neq i}} \forall_{x_i,X_i^\prime} ~~ P_{\textbf{A}^{\neq i}|\textbf{X}^{\neq i}}(\textbf{a}^{\neq i}|\textbf{x}^{\neq i}) =\sum_{a_i}  P_{\textbf{A}|\textbf{X}} (\textbf{a}|\textbf{x}^{\neq i},x_i) = \sum_{a_i}  P_{\textbf{A}|\textbf{X}} (\textbf{a}|\textbf{x}^{\neq i},x_i^\prime),
\end{align}
where $\textbf{a}^{\neq i}\equiv a_1 \dots a_{i-1}a_{i+1}\dots a_N$ and mutatis mutandis, for $\textbf{A}^{\neq i}$, $x^{\neq i}$ and $X^{\neq i}$. We call $P_{\textbf{A}|\textbf{X}}$ a $N$-partite non-signaling behavior, and $P_{\textbf{A}^{\neq i}|\textbf{X}^{\neq i}}(\textbf{a}^{\neq i}|\textbf{x}^{\neq i})$ the marginal distribution of $\textbf{A}^{\neq i}$ subsystem.
\end{definition}
The no-signaling condition in Definition~\ref{def:NSmultiBehavior} above implies that the no-signaling principle is satisfied between any two subsets of parties sharing a non-signaling device~\cite{hanggi-2009}.

The state spaces in the theory of non-signaling behaviors are non-signaling polytopes~\cite{InfoProcess}. A polytope is a multidimensional (to dimensions greater than 3) generalization of the notion of polyhedron. A convex polytope is a convex hull of a finite number of points in some real vector space. The set of all $N$-partite non-signaling behaviors $\{P_{\textbf{A}|\textbf{X}}\}$ with $m_i$ being the number of inputs of $i$-th party, and $v_{ij}$ being the number of outputs for the $j$-th input, is a proper subspace of $\mathbb{R}^t$, where $t=\prod_{i=1}^{N} \sum_{j=1}^{m_i} v_{ij}$ is the total number of outputs in the behavior~\cite{Pironio2005lift}. Generally, the polytope of $N$-partite non-signaling behaviors is the set of all behaviors that satisfy the no-signaling conditions in Definition~\ref{def:NSmultiBehavior}. For the sake of Sec.~\ref{sec:E_CE} we provide a more insightful definition of non-signaling polytopes.
\begin{definition}[Of non-signaling polytope, cf. Ref.~\cite{Pironio2005lift,Vertices,CE}]\label{def:NSpolytope}
Let $x \equiv (p_{ijk})_{i=1,j=1,k=1}^{i=N,j=m_i,k=v_{ij}}$ be a vector in a real vector space $\mathbb{R}^t$ corresponding to the outputs of some $N$-partite behavior $P_{\textbf{A}|\textbf{X}}$, where $m_i$ is the number of inputs of $i$-th party, and $v_{ij}$ is the number of outputs for the $j$-th input, and $t=\prod_{i=1}^{N} \sum_{j=1}^{m_i} v_{ij}$. The non-signaling polytope $\mathcal{B}$ is then a convex region within $\mathbb{R}^t$, i.e., 
\begin{align}\label{def:polyhedron}
    \mathcal{B}=\left\{x \in \mathbb{R}^t: Ax \le b\right\},
\end{align} 
where A is an $h \times t$ matrix of reals and let $b \in \mathbb{R}^m$ be a real vector, if and only if $A$ and $b$ encode (i) probabilistic constraints, i.e., $0 \le p_{ijk} \le 1$, (ii) normalization constraints, i.e., $\sum_k^{v_{ij}} p_{ijk}=1$, (iii) non-signaling constraints (see Definition~\ref{def:NSmultiBehavior}), that all together form $h$ linearly independent constraints. 
The dimension $\dim \mathcal{B}$ of the polytope $\mathcal{B}$ is given by~\cite{Pironio2005lift} 
\begin{align}\label{eq:dim_def}
    \dim \mathcal{B} = \prod_{i=1}^{n} \left(\sum_{j=1}^{m_i}(v_{ij}-1)+1\right)-1.
\end{align}
\end{definition}


As it was noted in Ref.~\cite{PR} (see also Refs.~\cite{Rastall1985-RASLBT}) the quantum theory is nonlocal but not maximally. Namely, there exist non-signaling behaviors that exhibits stronger violations of CHSH inequalities than allowed by quantum theory~~\cite{Tsirelson-bound}.
\begin{definition}[Of Popescu-Rohrlich (PR) boxes, see Ref.~\cite{PR}]\label{def:OfPRbox}
The Popescu-Rohrlich (PR) box or PR behavior, and anti-PR behavior $\overline{\mathrm{PR}}$ are bipartite conditional probability distributions defined as follows
\begin{align} 
    &\mathrm{PR}_\mathrm{AB|XY}\left(ab|xy\right) = \left\{ \begin{array}{ll}
	1/2 & \mbox{if $a\oplus b = xy$} \\
	0 & \mbox{otherwise},\end{array} \right., \\
    &\overline{\mathrm{PR}}_\mathrm{AB|XY}\left(ab|xy\right) = \left\{ \begin{array}{ll}
	1/2 & \mbox{if $a\oplus b = xy \oplus 1$} \\
	0 & \mbox{otherwise},\end{array} \right.
\end{align}
		Alternatively, in form of input-output probability tables, the $\mathrm{PR}$ and $\overline{\mathrm{PR}}$ are explicitly expressed as follows
\begin{align}
		&\mathrm{PR}_\mathrm{AB|XY}\left(ab|xy\right)=
		\begin{array}{cc|cc|cc}
			&\multicolumn{1}{c}{x} & \multicolumn{2}{c}{0}	& \multicolumn{2}{c}{1} \\
			y &$\diagbox[width=2.4em, height=2.4em, innerrightsep=0pt]{$b$}{$a$}$~~  & 0	& 1	&0  & 1 \\
			\hline \\ [-0.9em]
			\multirow{2}{*}{0 }  & 0 & \frac 12	& 0	& \frac 12	& 0\\[0.3em]
			& 1	& 0& \frac{1}{2}	& 0	& \frac 12  \\ [0.2em]
			\hline  \\ [-0.9em]
			\multirow{2}{*}{1 }   &0 & \frac 12	   &  0  &   0 & \frac 12 \\ [0.3em]
			& 1 & 0	&       \frac 1 2 &  \frac 12 	& 0
		\end{array},\\
		&\overline{\mathrm{PR}}_\mathrm{AB|XY}\left(ab|xy\right)=
		\begin{array}{cc|cc|cc}
			&\multicolumn{1}{c}{x} & \multicolumn{2}{c}{0}	& \multicolumn{2}{c}{1} \\
			y &$\diagbox[width=2.4em, height=2.4em, innerrightsep=0pt]{$b$}{$a$}$~~  & 0	& 1	&0  & 1 \\
			\hline \\ [-0.9em]
			\multirow{2}{*}{0 }  & 0 & 0	& \frac 12	& 0	& \frac 12\\[0.3em]
			& 1	& \frac 12& 0	& \frac 12	& 0  \\ [0.2em]
			\hline  \\ [-0.9em]
			\multirow{2}{*}{1 }   &0 & 0	   &  \frac 12  &   \frac 12 & 0 \\ [0.3em]
			& 1 & \frac 12	&       0 &  0 	& \frac 12
		\end{array}. 
\end{align}
\end{definition}
The PR and anti-PR boxes in Definition~\ref{def:OfPRbox} are extreme points in the polytope of bipartite devices with two binary inputs and two binary outputs (for each input) behaviors, i.e., (2,2,2,2) behaviors. Furthermore, they reach the algebraic maximum for the CHSH inequality and consequently do not have quantum realizations~\cite{Tsirelson-bound}. In Ref.~\cite{WDH2021}, we refer to the convex mixtures of the PR and anti-PR behaviors as to isotropic behaviors.
\begin{definition}[Of isotropic non-signaling behaviors, cf. Ref.~\cite{Bell-nonlocality,WDH2021}]
The isotropic behaviors $\mathrm{P}_\mathrm{iso}(\varepsilon)$ are defined as probabilistic mixtures, i.e., convex combinations, of $\mathrm{PR}$ and $\overline{\mathrm{PR}}$ behaviors
\begin{align}
    \mathrm{P}_\mathrm{iso}(\varepsilon) := (1-\varepsilon) \mathrm{PR}+ \varepsilon \overline{\mathrm{PR}},
\end{align}
for any $\varepsilon \in \left[0,1\right]$.
\end{definition}

Finally, we remark that the (2,2,2,2) polytope $\mathrm{P_{HRW}}$ of bipartite binary input-output devices~\cite{hanggi-2009,Hanggi-phd} studied by us in Ref.~\cite{WDH2021} (see also Sec.~\ref{sec:E_CE}) consists of $24$ extremal behaviors \cite{Barret-Roberts}.  Among extremal behaviors, $16$  are local or deterministic behaviors, and the remaining $8$ are nonlocal. The local behaviors are given by
\begin{align}
    \mathrm{L}_{\alpha\beta\gamma\sigma}(ab|xy) = \left\{ \begin{array}{ll}
	1 & \mbox{if $a= \alpha x\oplus \beta$, $b = \gamma y \oplus \sigma$} \\
	0 & \mbox{otherwise}.\end{array} \right.
\end{align}
		where $\alpha,\beta,\gamma,\sigma \in \{0,1\}$. And the nonlocal devices are
\begin{align} 
    \mathrm{B}_{rst}(ab|xy) = \left\{ \begin{array}{ll}
	1/2 & \mbox{if $a\oplus b = xy\oplus rx \oplus sy \oplus t$} \\
	0 & \mbox{otherwise},\end{array} \right.
\end{align}
where $r,s,t \in \left\{0,1\right\}$.

\newpage
\section{Hybrid Quantum Network Design Against Unauthorized Secret-Key Generation, and Its Memory Cost [A]}\label{sec:A_Hybrid}

In the first article \cite{Sakarya_2020}, we notice a possible vulnerability of the future Quantum Internet designs \cite{Dur_1999,Muralidharan_2016,Zwerger_2018,Wehner_2018} and propose a countermeasure to it. In this way, we first design a rerouting attack on the classical parts (classical computers) of the devices that would constitute the Quantum Internet and discuss the possible consequences for the Quantum Internet service provider \cite{Satoh_2021}. The introduced countermeasure is based on the hybrid quantum network design, which employs the fact that quantum security is not always transitive, i.e., there exist so-called nonrepeatable secure states \cite{bauml2015limitations,christandl2017private}. Namely, our secure network scheme employs quantum states that possess directly accessible secret key but small amount of repeatable key, i.e., specific kinds of private and positive partial transpose (PPT) states (see Sec.~\ref{subsec:PrivateAndPPT}). We refer to the difference between the secret and repeatable keys of a state as to the gap of the scheme for which we find lower bounds. Moreover, we show an example of a quantum state for which the gap of the scheme is strictly larger than zero. Application of the countermeasure brings an expense in terms of the required amount of quantum memory, i.e., the memory overhead. Consequently, we study the performance of our solution quantitatively in terms of lower bound on the amount of quantum memory needed to apply the proposed scheme.

In parallel to the advances in quantum computing that is about to enter the so-called NISQ (noisy intermediate scale quantum) era \cite{NISQ_2018}, a great effort has been put into the development of the so-called Quantum Internet \cite{Dur_1999,Muralidharan_2016,Zwerger_2018,Wehner_2018} in which qubits rather than the classical bits would be exchanged between NISQ processing units. The main advantage of the future Quantum Internet would be its inherent security of the sent signals guaranteed by the laws of physics \cite{Wiesner_1983,BB84}. In the first generation of the Quantum Internet \cite{Muralidharan_2016}, the security would be based on the quantum correlations called entanglement and their beneficial property of transitivity. Namely, two disconnected nodes of the network can obtain a mutual, unconditionally secure connection if only they share maximally entangled states with a common node. This outcome is achieved via performing the so-called entanglement swapping protocol \cite{Zukowski_1993,Bennett_1996}, which allows for the initially disconnected nodes to share the maximally entangled state after the protocol is executed. On the other hand, the monogamy of quantum correlations reflected in the no-cloning theorem \cite{Wootters_1982} guarantees the secrecy of the newly established link. By these means, an extensive network of nodes that can perform the entanglement swapping protocol (quantum repeaters \cite{Dur_1999}, see also Sec.~\ref{subsubsec:RepeaterRate}) connecting the end-nodes (end-users) is about to establish the first generation of the Quantum Internet in the near future \cite{Wehner_2018}.

\begin{figure}
    \centering
    \includegraphics[width=0.7\linewidth]{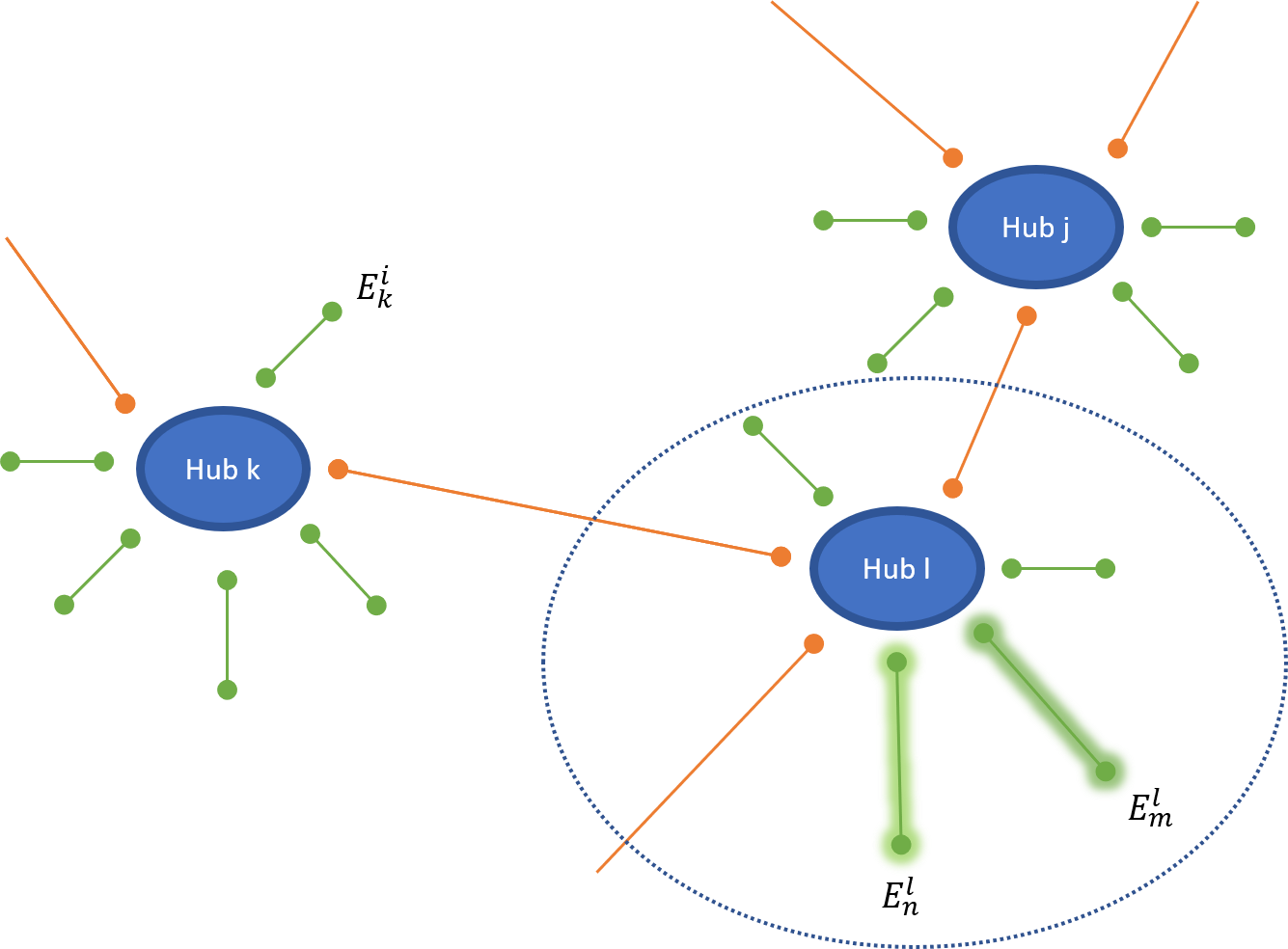}
    \caption{\label{fig:NetworkLarge} Design of the hybrid quantum network \cite{Sakarya_2020}. Thin green lines represent the connections between end-users and the hubs. In these, only classical data can be transferred. Thick orange lines represent the connection between the hubs being the routing nodes. In these lines also, quantum states can be transmitted. Shaded green lines connect the hub node with two end-users communicating classical data with the former. The selected region is the star-shaped network we focus on in our considerations.   } 
\end{figure}

\begin{figure}
    \centering
    \includegraphics[width=0.5\linewidth]{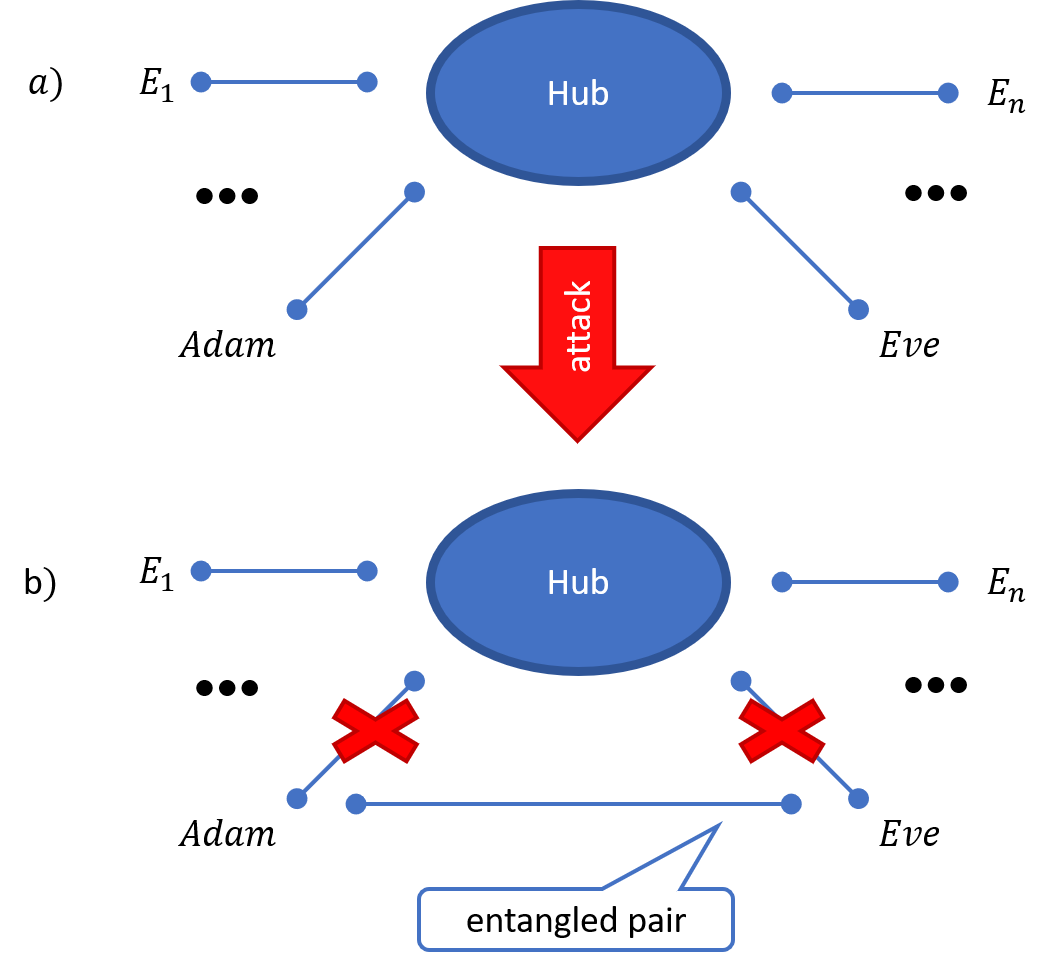}
    \caption{\label{fig:Attack} The main idea of the proposed attack \cite{Sakarya_2020}: (a) the hub shares pure entangled states with end users $E_i$, Eve and Adam, amongst others. Adam and Eve can attack the hub via malware which makes the hub perform the entanglement swapping protocol in their favor. (b) The malicious users share an entangled state after a successful attack. }   
\end{figure}

\newpage
As we discuss in the article, the new quantum technology that would come in the form of the Quantum Internet would not only allow for new possibilities but would also open new threats unknown from the era of the classical Internet~\cite{Satoh_2021}. In order to exemplify the possible danger, we focus on the quantum network with a star-shaped topology, i.e., a single, centrally placed quantum repeater (the hub node) connected to multiple end-users (end nodes) $E_i$ with $i \in [1,2,...,N]$ (see Fig.~\ref{fig:NetworkLarge}). The secure connections in the network are established due to entangled states shared between the hub and each end-user. Moreover, the hub is a processing unit with a classical and possibly a quantum computer inside. The critical observation is that if the network is based on pure entanglement (e.g., maximally entangled states), then the topology of the network can be altered by performing the entanglement swapping protocol incorporating the hub and two end-users (see Fig.~\ref{fig:Attack}). The described situation realizes the so-called rerouting attack~\cite{Satoh_2021}. In this manner, we consider a situation in which two malicious end-users, call them Adam and Eve, perform a Trojan horse attack on the classical computer of the hub by installing malware (malicious software) on it. In this way, Adam and Eve take control over the functions of the central (hijacked) node of the network, and, via the entanglement swapping protocol, they change the topology of the network into a (locally) disconnected graph. Therefore, Adam and Eve gain access to unconditionally secure communication with each other illegally, i.e., without permission and at the cost of the owner of the hub. Another possibility is when the dishonest administrator of the hub sells the connections by themselves on the expense of the owner. In both situations, any physical resources needed for the functioning of the hub (e.g., energy) would be stolen. In the article, we focus on the former situation of the Trojan horse attack and study cases in which the hijacked node can perform the three-party classical communication but also when the communication is limited to one-way from the central node to the hackers. 

\begin{figure}
    \centering
    \includegraphics[width=0.64\linewidth]{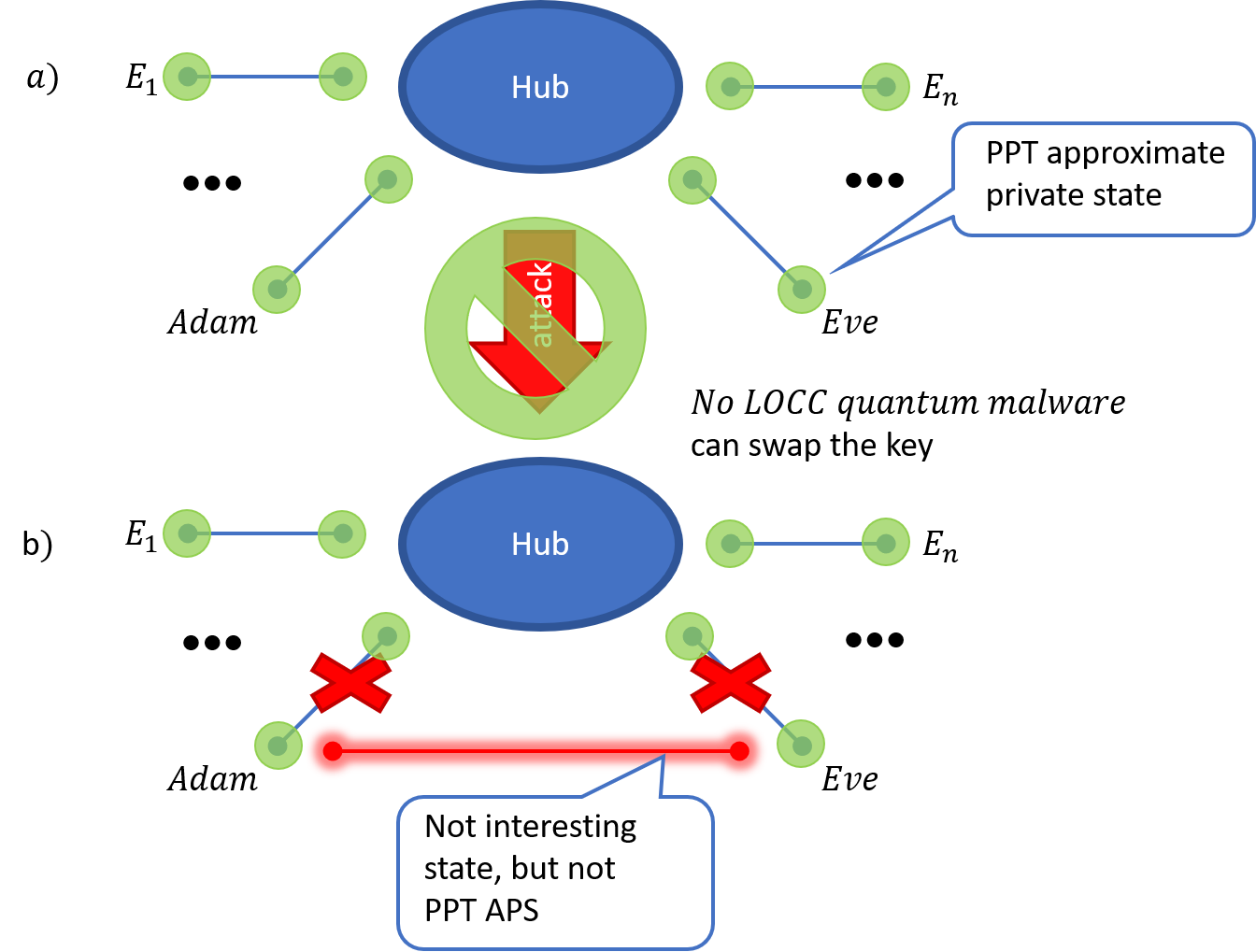}
    \caption{\label{fig:Countermeasure} The main idea of the proposed countermeasure: (a) the connection between the end-users $E_i$ and the hub is realized with bound entangled states (each having nearly 1 bit of secret-key at least), in particular, with Eve and Adam (shaded lines). No malware based on LOCC can efficiently swap the key. (b) Eve and Adam can not share a state with a non-negligible amount of the key (shaded red line).  }   
\end{figure}

We further propose a countermeasure to the above threat based on the recently proved no-go theorem (impossibility) \cite{bauml2015limitations}. Namely, \emph{there exist quantum states that allow for point-to-point security of classical data against a quantum adversary, and in spite of this fact, they can not be effectively used in quantum key repeaters} \cite{bauml2015limitations}. The results in Ref. \cite{bauml2015limitations} show that quantum security is not always transitive, i.e., for certain states, call them nonrepeatable secure quantum states when an agent in node A has a secure link (via state $\rho$) with an agent in node B, and B has a secure link with an agent in node C, there is no possibility to (efficiently) create a link with non-negligible security between A and C with the help of B via three-partite local quantum operations and classical communication (3-LOCC), that is protected against B as well (see Fig.~\ref{fig:Countermeasure}). The above feature can be realized with certain bound entangled states \cite{Horodecki_1998} (from which pure entanglement can not be distilled via local operations and classical communication \cite{Bennett+DiVincenzo_1996}). Moreover, highly noisy private states \cite{pptkey} also fit the above scheme in the case of 3-way and one-way classical communication for node  B to nodes A and C (with nodes A and C communicating freely)~\cite{bauml2015limitations,christandl2017private}. In that way, we propose a specially designed hybrid quantum network that is more robust against the aforementioned kind of attacks than the original design of a quantum network using solely quantum repeaters. The hybrid design is based on both quantum repeaters, and special relay stations called here hubs. Our approach applies in the scenario in which 
\begin{enumerate}[label=(\roman*)]
    \item the hub nodes can be connected via quantum repeaters,
    \item only classical data is transferred between the end nodes and the hub node,
    \item the distance between the end nodes and the hub node is up to the distance achievable for repeaterless quantum networks \cite{Takeoka_2014,Pirandola2017,Tamaki_2018},
    \item a single instance of the attack incorporates a single hub node and its two adjacent nodes
    \item hackers perform the ``honest but curious'' attack, which means that only the functioning of the classical processor is altered by the malware, while the sensitive data remain unread.
\end{enumerate}
The above scheme fits the realistic use of quantum-secure Internet, in which the end-users, as in the traditional Internet, want to exchange anonymously classical data rather than quantum states, but in a quantum-secure way. In this scheme, the role of the hub nodes is to provide the registered users (end-nodes) access to the online services rather than to provide a connection between the users. The examples of the possible application of the scheme considered in the paper are online baking, data clouds, access to medical data from medical laboratories, online shopping, and possibly many other applications in which the hub node takes a role similar to the server in the traditional Internet. 

As it is usually in real life, any good also comes indispensably with a price. In the above scheme that implements the proposed countermeasure, the price is the number of qubits needed to be stored and processed in the quantum memory of the nodes. Because, in the NISQ era, there is no technology to store qubits coherently for a long time, the amount of quantum memory used in the quantum network architecture is of primary importance. Therefore, in the paper, we study the lower bounds on the memory cost of the implementation of the hybrid quantum network in which every link is represented by the same quantum state $\rho$. Moreover, we show that it can be realized with quite modest memory requirements. We quantify the memory needed to realize the proposed scheme $S_\rho$ in terms of the memory overhead 
\begin{equation}
    V(S_\rho) \coloneqq M(\rho)(1-\mathcal{D}(\rho)),
\end{equation}
where $M(\rho)$ is the total memory of the scheme based on quantum states $\rho$, and $\mathcal{D}(\rho)$ is the density of the secure-key (cf.~Refs.~\cite{bauml2015limitations,keyhuge,Badziag_2014}), i.e., the ratio between the distillable key of $\rho$ and the number of qubits (of the memory at the hub-node) needed to store the state $\rho$. The efficiency of the scheme is quantified by the difference between the amount of the key that can be repeated $R$ and the initial key in the link $K_D$. The repeatable key in our approach is the hackable one, so we want to diminish its amount while keeping the distillable key $K_D$ possibly large.  We call the difference between these two quantities ($K_D-R \ge 0$) the gap of the scheme. Furthermore, we say that a scheme is $(\theta,\eta)$-good when $K_D \ge \eta$ but $R \le \theta$ (assuming $\eta>0$). The above notions were used for the assessment of the performance of the scheme in terms of the number of qubits (or their percentage) intended in use solely for the security of the scheme against the considered attack.

In the paper, we recognize that states especially useful for the realization of the scheme proposed by us are, among others, the so-called private states \cite{pptkey,keyhuge,Horodecki_2018}. These states are equipped with secret-key directly accessible via measurement on the key part. Moreover, private states can also possess a shielding part (shield) that is responsible for their property of having a low repeatable key \cite{christandl2017private} (see also Secs.~\ref{subsec:PrivateAndPPT} and \ref{subsubsec:RepeaterRate}). The shield protects the key but costs the quantum memory needed to implement it. Additionally, if the shielding system of a private state does not possess any secrecy content we call such a private state irreducible private state (denoted $\gamma_{\langle d_k,d_s \rangle}$, see also Sec.~\ref{subsubsec:2Private}). Moreover, as we show in the paper, the states having positive partial transpose (PPT states, see also Sec.~\ref{subsubsec:PPT}) that approximate private states prove their usefulness in the proposed anti-malware solution.

\newpage
For the quantitative results, we firstly derive a lower bound on the overhead of the $(\theta,\log_2 d_k)$-good secure network scheme $S_{\gamma_{d_k,d_s}}^\rightarrow$ employing irreducible private states $\gamma_{d_k,d_s}$, and one-way classical communication from the hub to the attacking end-nodes. The lower bound reads (Theorem~1 in Ref.~\cite{Sakarya_2020})
\begin{align}
    V\left(S_{\gamma_{d_k,d_s}}^\rightarrow\right) \ge \Delta \log_2 (d_k d_s) \left(1-\frac{1}{2-\frac{\theta}{\log_2 k_k}}\right) \approx_{\theta \approx 0} \frac{1}{2} M (\gamma_{d_k,d_s}),
\end{align}
where $\Delta$ is the degree of the central node (hub) equal to the number of connections between the hub and the end nodes. We observe that the requirement for the rate of the repeatable key to be approximately zero implies that at least half of the memory must be spent on the shielding system that protects the scheme. A similar result is proved for any quantum states (Theorem~2 in Ref. \cite{Sakarya_2020}). In the latter case, we observe a general property of secure schemes in which $\theta \approx 0$, i.e., at least half of the memory, must be spent to protect the scheme and does not store the secure-key $K_D$.  

In the further part, we develop a lower bound on the memory overhead in the case of a secure scheme that employs strictly irreducible private states $\gamma_{\langle d_k,d_s \rangle}$ hardly distinguishable from their attacked versions $\hat{\gamma}_{\langle d_k,d_s \rangle}$~\cite{Horodecki_2018} (see Sec.~\ref{subsubsec:2Private}). The degree to which a strictly irreducible private state differs from its attacked version is quantified with the aid of the trace distance for which we assume the following
\begin{align}
    \norm{\gamma_{\langle d_k,d_s \rangle}^\Gamma -\hat{\gamma}_{\langle d_k,d_s \rangle}^\Gamma }_1 \le \epsilon, 
    \label{eqn:TD_HD}
\end{align}
where $\Gamma$ stands for partial transpose. In this way, we show that special private states that satisfy the above, and for which conditional shield states $X_{ii}$ are positive partial transpose operators, i.e.,  $X_{ii}^\Gamma \ge 0$, satisfy $d_s \ge \frac{d_k-1}{\epsilon}$ (Lemma~1 in Ref.~\cite{Sakarya_2020}). Finally, we find a lower bound on the overhead of the $(\theta,\eta)$-good scheme (with one-way classical communication) which employs strictly irreducible private states (Theorem~3 in Ref.~\cite{Sakarya_2020}) which satisfy the condition in Eq.~\eqref{eqn:TD_HD} with separable conditional shield states ($X_{ii} \in \mathrm{SEP}$)
\begin{align}
    V\left(S_{\gamma_{\langle d_k,d_s \rangle}}^\rightarrow\right) \ge M \left( \gamma_{\langle d_k,d_s \rangle} \right) \left(1-\frac{\log_2 d_k}{\log_2 d_k+\log_2 \frac{d_k-1}{\epsilon}}\right) \approx_{\epsilon \to 0}  M \left( \gamma_{\langle d_k,d_s \rangle} \right),
    \label{eqn:Thm3_OH_LB}
\end{align}
for $\theta = 2 \log_2 (1+\epsilon) \approx_{\epsilon \ll 1} \frac{2}{\ln 2} \epsilon$  and $\eta=\log_2 d_k$. Here, the upper bound $\theta$ on the repeatable key rate is due to Observation~2 in Ref.~\cite{Sakarya_2020}, and the lower bound $\eta$ (saturated here) is due to the properties of the strictly irreducible private states.

We observe that the parameter $\epsilon$ appears both in the formula for the lower bound on the overhead and the upper bound on the repeater rate (see Eq.~\eqref{eqn:Thm3_OH_LB} and lines below it). Therefore, one can not diminish the repeater rate to zero as desired while keeping the memory overhead at a considerably low level. Instead, one should decide on an acceptable level of the repeater rate for which the memory overhead is still reasonable. In this way, we identify low-dimensional examples of states for which such a tradeoff is controllable \cite{pptkey,Dobek_2011}. The block matrix representation of these states reads
\begin{align}
\Omega_{d_s}= \frac{1}{2}
    \begin{bmatrix}
    \frac{I}{d_s^2}        & 0 & 0 & \frac{\mathrm{F}}{d_s^2} \\
    0       & 0 & 0 & 0  \\
    0       & 0 & 0 & 0  \\
    \frac{\mathrm{F}}{d_s^2}       & 0 & 0  & \frac{I}{d_s^2} 
    \end{bmatrix},
\end{align} 
where $F=\sum_{i,j=0}^{d_s-1}\ketbra{ij}{ji}$ is a matrix of swap quantum logic gate of dimension $d_s^2$. In this way, we find the first non-trivial case in which a secure network scheme has an advantage over malicious parties. Namely, we show that already for $d_s=3$ case of $\Omega_{d_s}$, the gap between the secure-key and repeatable key is strictly larger than zero. 

The later results concern secure network schemes that employ positive partial transpose (PPT) states $\rho$ approximating private states and two-way classical communication. In Theorem~4 of Ref.~\cite{Sakarya_2020}, we derive a lower bound on the memory overhead in the case of PPT states approximating strictly irreducible private bit (pbits, $d_k=2$). Without loss of generality, we assume PPT states, of dimension $d_k^2 \times d_s^2$, are of the following form $ \rho = \sum_{i,j,k,l=0}^{d_k-1} \ket{ij}\bra{kl} \otimes A_{ij,kl}$, where $A_{ij,kl}$ are blocks of dimension $d_s^2$. We obtain even more interesting results in the case of the higher dimensions. Firstly, we show the following implication (Lemma~2 in Ref.~\cite{Sakarya_2020})
\begin{align}
    \norm{\rho-\gamma_{d_k,d_s}}_1 \le \epsilon ~~\implies~~ d_s \ge \left(\frac{d_k-1}{\epsilon}\right)(1-\epsilon d_k).
\end{align}
The above implication has an important corollary, namely
\begin{align}
    \norm{\rho-\gamma_{d_k,d_s}}_1 \ge \frac{d_k-1}{d_s+d_k(d_k-1)}.
\end{align}
\begin{figure}[t]
    \centering
    \includegraphics[trim=1.3cm 0cm 0cm 0.5cm,clip,width=0.49\linewidth]{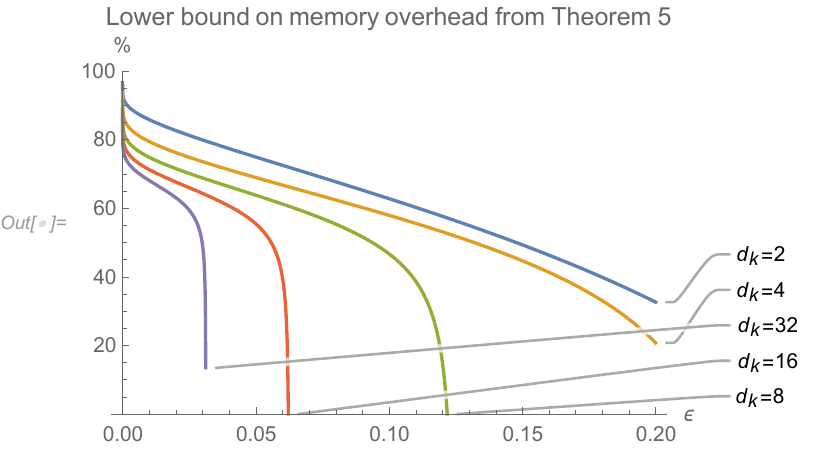}
    \includegraphics[trim=0cm 0cm 0cm 0.5cm,clip,width=0.49\linewidth]{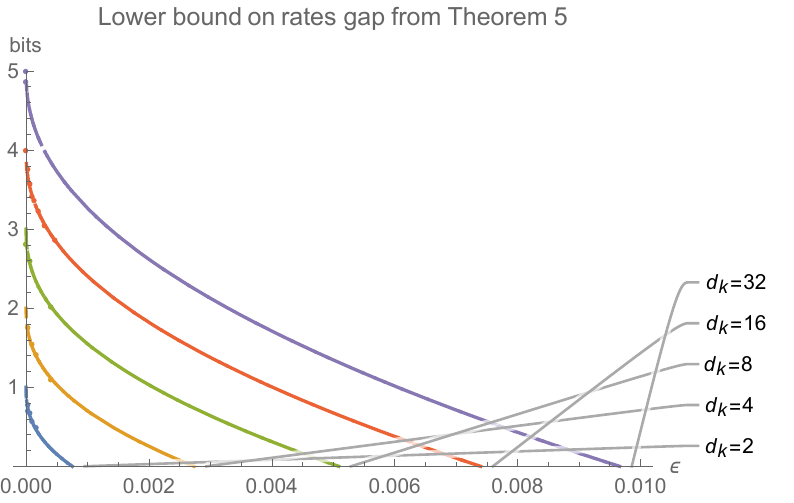}
    \caption{\label{fig:plots:PPT} Sample results from Theorem~5. in Ref.~\cite{Sakarya_2020}. Plots of lower bounds on the memory overhead~(left) and the gap of the $(\theta,\eta)$-good secure scheme~(right) for $\rho \in \mathrm{PPT}$ approximating strictly irreducible private dit (pdit) $\gamma_{\langle d_k,d_s \rangle}$, such that $\norm{\rho - \gamma_{\langle d_k,d_s \rangle}}_1 \le \epsilon$ for $\frac{d_k-1}{d_s+d_k(d_k-1)}\le \epsilon < \frac {1}{d_k}$, $\sum_{i \neq j} \norm{A_{ij,ji}^\Gamma}_1 \le \epsilon$, for different values of the key part, i.e., $d_k$.}   
\end{figure}
The above inequality not only holds for any dimension but, in the case of $d_k=2$ is tighter than the known results \cite{Badziag_2014,Dobek_2011}. The above findings are the basis for our further developments. In Proposition~2 of Ref.~\cite{Sakarya_2020}, we show (under some mathematical conditions) an upper bound on the two-way repeater rate for PPT states approximating strictly irreducible private dit (pdit, $d_k >2$). Finally, we derive a lower bound on the memory overhead for a $(\theta, \eta)$-good secure scheme involving two-way classical communication, and  $\rho \in \mathrm{PPT}$, such that $||\rho - \gamma_{\left<d_k,d_s\right>}||\leq \epsilon$ for $\frac{d_k-1}{d_s+d_k \left(d_k-1\right)} \le \epsilon < \frac{1}{d_k}$, $\sum_{i\neq j} \norm{A_{ij,ji}^\Gamma} \le \epsilon$, and its conditional shield states are separable (Theorem~5 in Ref.~\cite{Sakarya_2020})
\begin{align}
    &V(S_{\rho})  \ge M(\rho) \left(1-\frac{\epsilon}{2}-f(d_k,\epsilon)\right),\label{eqn:Thm5_OH}\\
    &f(d_k,\epsilon) :=\frac{ \log{d_k} +(1 + \frac{\epsilon}{2})h(\frac{\frac{\epsilon}{2}}{1+\frac{\epsilon}{2}})}{\log d_k + \log  \left( \frac{d_k-1}{\epsilon} \right)+\log (1-\epsilon d_k)},
\end{align}
with $\eta = \log d_k-8\epsilon \log d_k  - 4 h(\epsilon) $ (where $h(.)$ is the binary Shannon entropy), and $\theta =   2\left(\sqrt{\epsilon}+ \epsilon\right) \log \dim_H (\rho) + \left(1 + 2\sqrt{\epsilon}+2 \epsilon\right)h(\frac{\sqrt{\epsilon}+ \epsilon}{\frac 12 +\sqrt{\epsilon}+\epsilon})$. In Fig. \ref{fig:plots:PPT}, we illustrate the performance of the lower bounds in Theorem~5 in Ref.~\cite{Sakarya_2020} (here, Eq. \eqref{eqn:Thm5_OH} and lines below it).

All of the above results contribute either directly or indirectly to the analysis of the efficiency of the hybrid quantum network design, as a countermeasure to the proposed rerouting attack, in terms of its memory cost. Finally, we note that we close the problem of the threat in the form of the malware attack proposed by us. Our findings fit the novel field of attacks on quantum internet \cite{Satoh_2021} (Ref.~\cite{Sakarya_2020} was already cited therein). We believe our results can help to design and construct the future Quantum-secure Internet that will meet the expectations of the rapidly developing information-era society.

\newpage
\section{Universal Limitations on Quantum Key Distribution over a Network [B]}\label{sec:B_Universal}

In Ref.~\cite{DBWH19}, we consider the distribution of secret key over a network both in bipartite and multipartite (conference) settings. Our primary achievement is establishing a unified framework capable of providing case-specific upper bounds on the achievable rates in the asymptotic and single-shot (single-use) regimes of device-dependent conference key agreement (QDD-CKA). We aim to describe a general network setting. We do this by employing the idea and framework of quantum channels. In a nutshell, any operation that transforms one quantum state into another can be considered a quantum channel (see Sec. \ref{subsec:Channels} for more details), with unitary evolution or decoherence as prominent examples. In this way, we introduce the notion of a multiplex quantum channel, which links an arbitrary finite number of honest parties. In a multiplex quantum channel, each party can have the role of the sender to a channel, a receiver from a channel, or both sender and receiver. Within this scheme, we define asymptotic and single-shot (single-use) local operations and classical communication-assisted (LOCC-assisted) conference key agreement (CKA) secret capacities for multiplex quantum channels. The results of Ref.~\cite{DBWH19} consist of weak and strong converse (upper) bounds on the LOCC-assisted SKA capacities of quantum channels. To achieve our goal, we first show that any output state of a multipartite protocol distilling secret key amongst the honest parties must be genuinely multipartite entangled. The protocols we consider manifest an adaptive strategy to the secret key and entanglement distillation (in the form of private states and GHZ states, respectively) over an arbitrary multiplex quantum channel. Therefore, the structure of the protocols we consider is generic. In particular, our approach allows us to study the performance of quantum key repeaters, measurement-device-independent quantum key distribution (MDI-QKD) setups, or teleportation-covariant multiplex quantum channels. For the latter upper bounds on the SKA, capacities are given in terms of the entanglement measures of their Choi states. The bounds we provide are given in terms of the generalized relative entropies, the form of which has a topology-dependent character. Moreover, our approach allows us to obtain upper bounds on the rates at which tripartite Greenberger-Horne-Zeilinger (GHZ) states~ \cite{GHZ89} can be distilled from a limited number of copies of an arbitrary multipartite quantum state. Additionally, we provide lower bounds on the SKA rate of a multiplex channel achievable in the sense of \cite{DevetakWinter-hash} by classical preprocessing and postprocessing~(cppp) as a generalization of \cite{pirandola2009direct} and show a non-trivial lower bound for a specific setup of a bidirectional quantum network. 
\begin{figure}[t]
\centering
 \includegraphics[trim=0.0cm 0.0cm 8.0cm 0.0cm,width=11.5cm,clip]{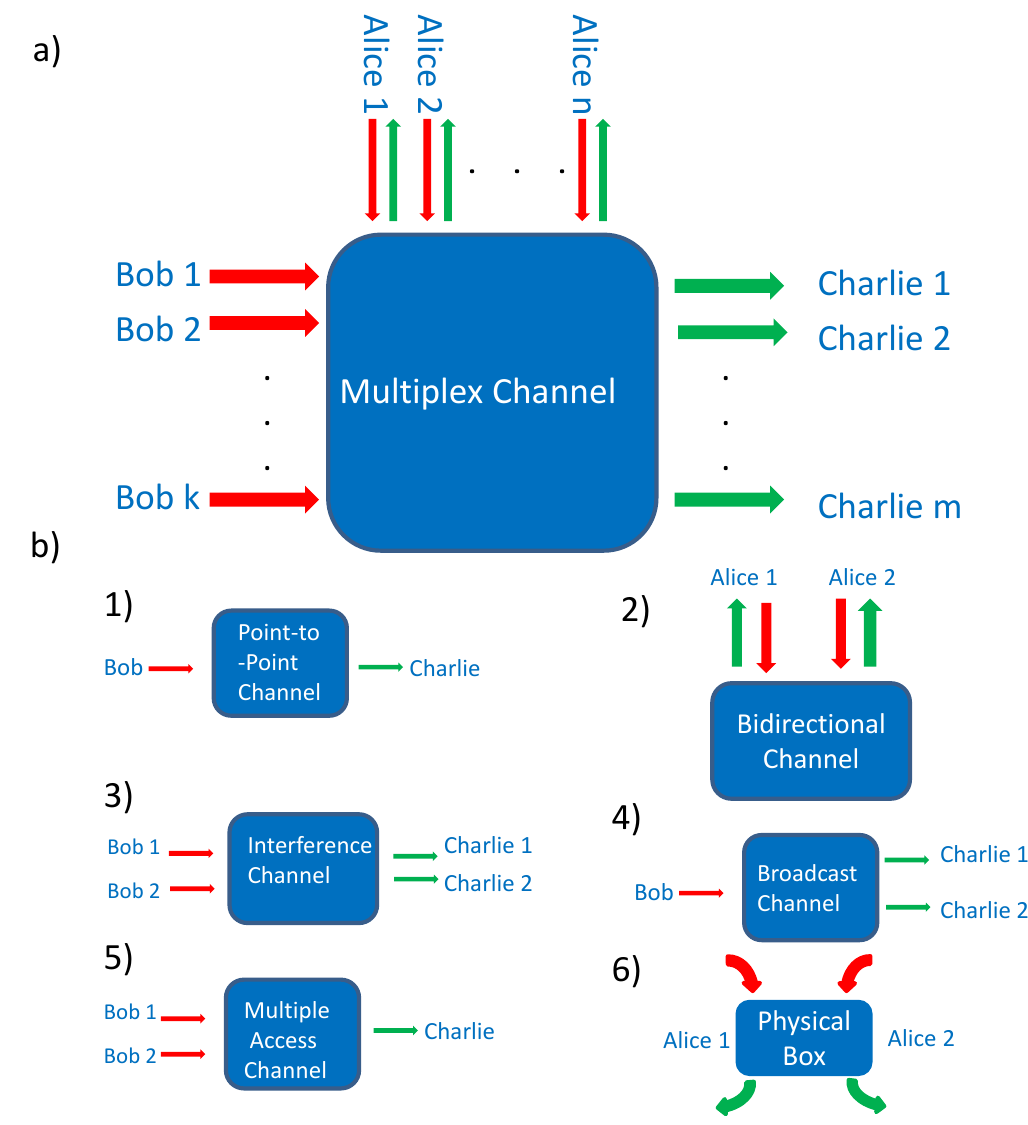}
        \caption{Pictorial illustration of multiplex quantum channel in Ref.~\cite{DBWH19}. The illustration depicts the universal nature of multiplex quantum channels from which all other (memoryless) quantum network channels arise. Here, red arrows represent the inputs, and green arrows represent the outputs to channels. See section IIIB of Ref.~\cite{DBWH19} for more details.}\label{fig:UNMQC} 
\end{figure}

Quantumly secure communication is one of the greatest promises of the Quantum Internet that is currently in the design phase~\cite{Dur_1999,Muralidharan_2016,Zwerger_2018,Wehner_2018,DM03,Kim08,Wehner_2018}. The formidable task of quantum communication over a network raises both fundamental and application-focused questions~\cite{BB84,E91,DM03,RennerThesis,CLL+09,VBD+15,ZXC+18}. The development of the technology~\cite{HBD+15,Mst19,BRA+19,CZC+21} together with the concerns about privacy~\cite{Sho94,ZXC+18} create a need for designing protocols and determining criteria for secure communication between multiple (trusted) parties in the network~\cite{Chen2005,AugusiakH2008-multi}. A complex structure of a quantum network consists of various quantum channels, with possibly complex spatial alignment, due to environmental conditions. Moreover, the exponential attenuation of the signal, which can not be amplified by cloning or broadcasting, along an optical fibre~\cite{ATL15} and the difficulties in preserving entanglement for a long time due to the interaction with the environment~\cite{BRA+19} make the task of building a network even more complicated. On the other hand, the advancements in the technology of quantum repeaters~\cite{PhysRevLett.81.5932, Dur_1999,munro2015inside,CZC+21} give hope to overcome these issues. Given broad interest in the application of technologies such as quantum networks, quantum repeaters, and measurement-device-independent QKD (MDI-QKD) protocol \cite{LCQ+12,braunstein2012side} makes the understanding of the fundamental limitations on the achievable key rates in the mentioned scenarios an important task. The seminal papers~\cite{pptkey,christandl2004squashed} on secret key distillation from states together with results in Refs.~\cite{Bennett+DiVincenzo_1996,VPRK97,VP98,HHH99,Dat09} open a pathway to studies in the mentioned direction, in the case of point-to-point LOCC-assisted quantum channels~\cite{TGW14,Pirandola2017,WTB16,CM17}. Consequently, further progress has been made in the restricted network settings, for example, the case of quantum repeaters~\cite{bauml2015limitations,CM17}, between two parties over bidirectional~\cite{DBW17,BDW18,D18thesis}, multiple access, and interference quantum channels~\cite{LP17}, and networks consisting of point-to-point~\cite{AML16,RKB+17,pirandola2019capacities} or broadcast channels~\cite{bauml2017fundamental}.

In Ref.~\cite{DBWH19}, we focus on providing a unifying framework for obtaining upper bounds on the conference key rates achievable in the general network setting also in the one-shot scenario. To achieve this task, we introduce a multiplex quantum channel (see Fig.~\ref{fig:UNMQC}) as the most general form of memoryless transformation between multipartite quantum states, where each of $M$ honest parties can have a role of the sender (Bob), receiver (Charlie) or both sender and receiver (Alice). We introduce a concise notation in which systems that are input to multiplex quantum channel of multiple Alice parties are denoted $\vv{A'}\coloneqq \{A_a'\}_{a\in\msc{A}}$, and similarly outputs of multiple Alice parties are denoted by $\vv{A}\coloneqq \{A_a\}_{a\in\msc{A}}$. The systems sent to multiplex quantum channel by multiple Bob parties are denoted by $\vv{B}=\{B_b\}_{b\in\msc{B}}$, and systems received from the channel by multiple Charlie parties are denoted $\vv{C}=\{C_c\}_{c\in\msc{C}}$. The reference systems kept by Alice, Bob, and Charlie parties are denoted $\vv{L}$, $\vv{R}$, and $\vv{P}$, respectively. Furthermore, we use $:\vv{A}:$ to denote the partition with respect to all systems in the set $\vv{A}$ as spatially separated parties keep them. For example 
$:\!\vv{LA}\!:\!\vv{RB}:$ denotes systems of spatially separated Alice and Bob parties, for which system $L_a$ are not separated from $A_a$, and similarly $R_b$ are not separated from $B_b$. Additionally, $\vv{K}=\{K_i\}_{i=1}^M$ denotes the systems holding the secret key at the end of a protocol, and $\vv{S}=\{S_i\}_{i=1}^M$ the shielding (see Sec.~\ref{subsec:PrivateAndPPT} for the definition) systems ($\vv{SK}$ altogether). Furthermore, we introduce secret key agreement LOCC-assisted protocols over multiplex quantum channels that provide a unifying framework for many seemingly different QKD scenarios. The unification is achieved by constructing a multiplex quantum channel in a way that its uses interleaved by LOCC simulates the protocol. Our upper bounds are based on entanglement measures called sandwiched R\'enyi relative entropies~\cite{WWY14,MDSFT13}, for which relative entropy is a special case. One of the sandwiched R\'enyi relative entropies, i.e., the relative entropy of entanglement and its regularisation, have already proved their usefulness in providing upper bounds on the secret key rate in the bipartite case~\cite{pptkey}. As we show, these entanglement measures are topology-dependent, as the upper bound depends on the partition of roles between the honest parties.
\begin{figure}[t]
\centering
 \includegraphics[trim=0.0cm 3.0cm 0.0cm 1.0cm,width=15cm,clip]{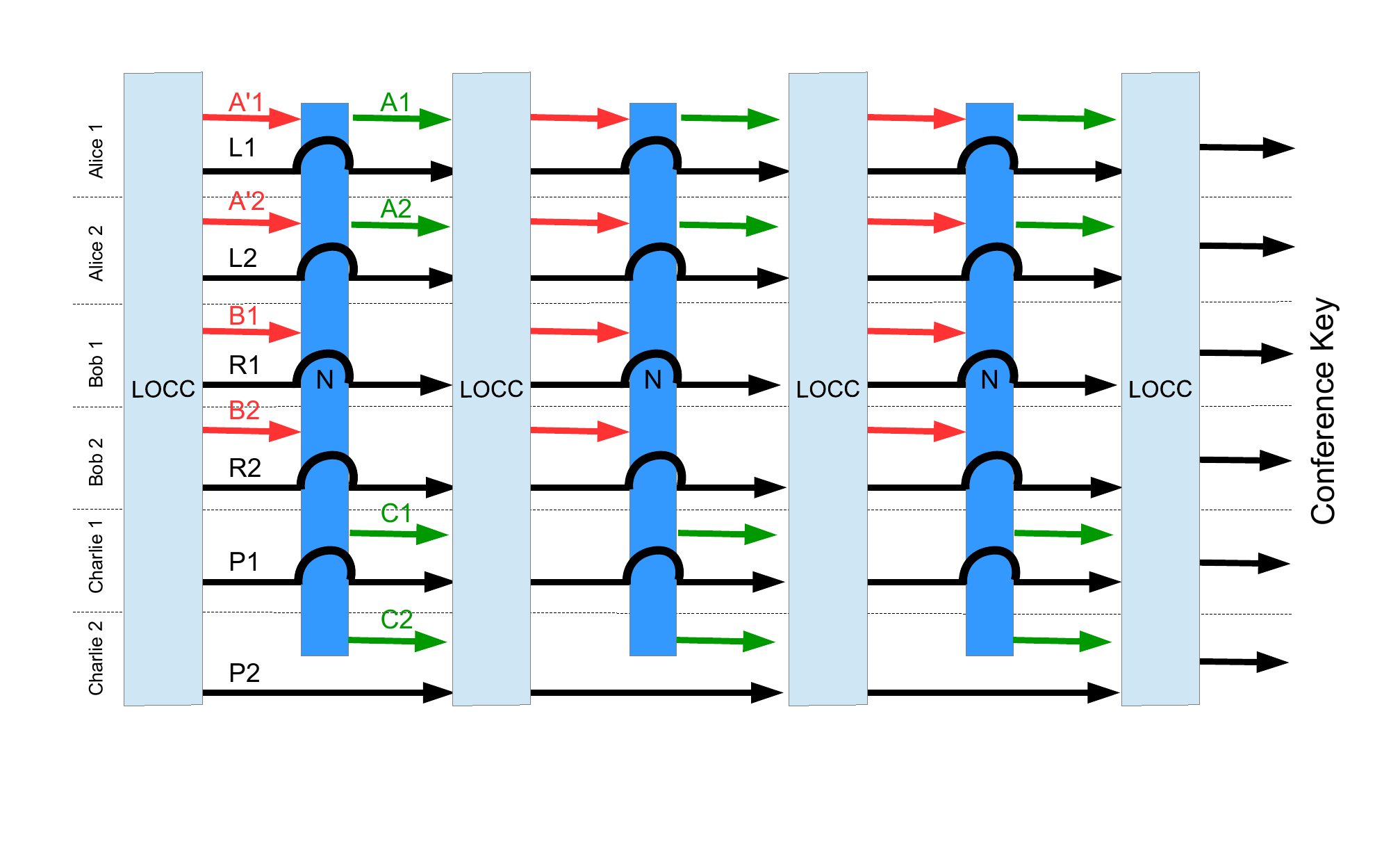}
        \caption{Illustration in Ref.~\cite{DBWH19} of a sample LOCC-assisted conference key agreement protocol incorporating six parties. Inputs to quantum channel $\mathcal{N}$ are represented by red arrows, outputs are represented by green arrows, and reference systems by black arrows. Alice1 and Alice2 both send and receive quantum systems from channel $\mathcal{N}$, Bob1 and Bob2 only send their systems, and Charlie1 and Charlie2 only receive systems as an output of channel $\mathcal{N}$. In the end, all parties share a six-partite conference key. }\label{fig:EQKDC} 
\end{figure}

In Ref.~\cite{DBWH19} we consider LOCC-assisted conference key agreement protocol amongst $M$ honest parties employing multiplex quantum channel $\mc{N}_{\vv{A'}\vv{B}\to\vv{A}\vv{C}}$ (see Fig. \ref{fig:EQKDC}). We assume that environmental part $E$ of the isometric extension $U^\mc{N}_{\vv{A'}\vv{B}\to\vv{A}\vv{C}E}$ of the quantum channel~$\mc{N}$ is accessible to the eavesdropper together with all classical information communicated between the honest parties while performing LOCC. All other quantum systems that are locally available to the honest parties are assumed to be secure against the eavesdropper. Fortunately, in the paradigm based on the private states, we do not need to consider the eavesdropper explicitly~\cite{pptkey}.  In a LOCC-assisted CKA-DD, the uses of a multiplex quantum channel $\mc{N}$ are interleaved with the uses of LOCC channels. In the first round, the honest parties perform LOCC channel $\mathcal{L}^1$ to generate a fully separable state $\rho_1\in\FS(:\!\vv{L^{(1)}A^{(1)'}}:\!\vv{R^{(1)}B^{(1)}}\!:\!\vv{P^{(1)}})$ that subsequently enter multiplex quantum channel $\mc{N}^1_{\vv{A^{(1)'}}\vv{B^{(1)}}\to \vv{A^{(1)}}\vv{C}^{(1)}}$ to yield output state $\tau_1\coloneqq \mc{N}^1(\rho)$. Next, the LOCC channel $\mathcal{L}^2$ acts on $\tau_1$, which enters the multiplex quantum channel again. After the final ($n$-th) round, the decoding $\mathcal{L}^{n+1}$ LOCC channel generates the final state $\omega_{\vv{SK}}$, where $K_i$ and $S_i$ denote the key and shielding systems of the honest parties respectively. It is assumed that the eavesdropper has access to purification $\omega_{\vv{SK}Y^{n+1}E^n}$ of $\omega_{\vv{SK}}$ as it possesses all environmental systems from the isometric extension $U^\mathcal{N}$ and copies of classical data $Y$ exchanged among the honest parties. The protocol for which $F(\gamma_{\vv{SK}},\omega_{\vv{SK}})\geq 1-\varepsilon$ is called $(n,K,\varepsilon)$ LOCC-assisted secret key agreement protocol (see Sec. \ref{subsubsec:NPrivate} for more details upon multipartite private states $\gamma_{\vv{SK}}$). The rate $P$ of the protocol is equal to the number of the conference (secret) bit per use of multiplex quantum channel, i.e., $P\coloneqq\frac 1n \log_2 K$ (here, $K=\dim \mathcal{H}_{K_i}$). A rate P is (weak converse) achievable if for $\varepsilon\in(0,1),\delta>0$, and $n$ sufficiently large, there exists an $(n,2^{n(P-\delta)},\varepsilon)$ LOCC-assisted secret key agreement protocol. The supremum over all achievable rates is called the LOCC-assisted secret-key-agreement capacity $\hat{P}_{\LOCC}(\mc{N})$ of a multiplex quantum channel $\mc{N}$. Furthermore, a rate $P$ is called a strong converse rate for LOCC-assisted secret key agreement if for all $\varepsilon\in[0,1), \delta>0$, and $n$ sufficiently large, there does not exist a $(n, 2^{n(P+\delta)},\varepsilon)$ LOCC-assisted secret key agreement protocol. The strong converse LOCC-assisted secret-key-agreement capacity of quantum channel  $\widetilde{P}_{\LOCC}(\mc{N})$ is defined as the infimum of all strong converse rates. It follows directly from the definitions that
\begin{equation}
\hat{P}_{\LOCC}(\mc{N})\leq \widetilde{P}_{\LOCC}(\mc{N}).
\end{equation}
Moreover, we call a protocol in which the honest parties are allowed only for classical preprocessing and postprocessing (cppp) communication, i.e., one use of LOCC channel for encoding and a second for decoding, a cppp-assisted secret key agreement protocol over multiplex quantum $\mc{N}$. The $(1,K,\varepsilon)$ LOCC-assisted secret key agreement protocol is identical to $(1,K,\varepsilon)$ cppp-assisted secret key agreement protocol. 
Moreover, the LOCC-assisted secret-key-agreement capacity $\hat{P}_{\LOCC}$ of the channel $\mc{N}$ is always greater or equal to $\hat{P}_{\text{cppp}}$
\begin{equation}
\hat{P}_{\text{cppp}}(\mc{N})\leq \hat{P}_{\LOCC}(\mc{N}). 
\end{equation}
Finally, let $\hat{P}^{\mc{N}}_{\textnormal{cppp}}(n,\varepsilon)$ denote the maximum rate such that $(n,2^{nP},\varepsilon)$ cppp-assisted secret key agreement is achievable, for any given multiplex quantum channel $\mc{N}$.

Our first main result has a technical significance and constitutes a background for further investigation. Here, we focus on the multipartite private states~\cite{AugusiakH2008-multi} (see Sec.~\ref{subsubsec:NPrivate}) that constitute the most general class of states that provide quantum conference key directly by local measurements with no further distillation required 
\begin{equation}
\gamma_{\vv{SK}}\coloneqq U^{\text{tw}}_{\vv{SK}} (\Phi^{\GHZ}_{\vv{K}}\otimes\omega_{\vv{S}}) (U^{\text{tw}}_{\vv{SK}})^\dag.
\end{equation}
Here, $\vv{K}=K_1, ..., K_N$ denotes the key part, that is, the systems that the $N$ honest parties have to measure in order to obtain the conference key, and $\vv{S}=S_1, ..., S_N$ denotes the so-called shield that has to be kept secure from the eavesdropper. Furthermore, each multipartite private state $\gamma_{\vv{SK}}$ has at least $\log_2 K$ secret bits (key)~\cite{keyhuge}, where $K=\dim\left(\mathcal{H}_i\right)$, for all $i \in \{1, \dots, N\}$ . We prove that multipartite private states are necessarily genuine multipartite entangled. To show this we construct a multipartite privacy test ($\gamma$-privacy test), defined as a $\{\Pi^\gamma,\bbm{1}-\Pi^\gamma\}$ such that any $\epsilon$-approximate multipartite private state $\rho$ with fidelity $F(\rho,\gamma)\geq1-\epsilon$ passes the test with success probability $\Tr[\Pi^\gamma\rho]\geq 1-\epsilon$. Consequently, in Theorem 1 of Ref.~\cite{DBWH19} we show that any biseprable state $\sigma_{\vv{SK}}\in\BS(:\!\vv{SK}\!:)$ can not pass any multipartite privacy test with probability greater than $1/K$, i.e., 
\begin{align}
    \Tr[\Pi^{\gamma}_{\vv{SK}}\sigma_{\vv{SK}}]\leq \frac{1}{K}.
\end{align}
The above result is our starting point for showing how achievable rates of quantum channels capacities can be upper-bounded with various divergence measures. 

We observe that interleaving uses of a multiplex quantum channel with LOCC amongst the honest parties provide the general framework to describe a number of different conference key agreement protocols. The critical point is the idea that the protocol can be simulated by a suitably constructed multiplex channel laced with LOCC operations. In this manner, we generalize the results for point-to-point~\cite{Pirandola2017,WTB16,CM17} and bidirectional~\cite{D18thesis,DBW17,BDW18} channels. Namely, we show that secret-key-agreement capacities of multipartite generalizations of the mentioned channels are upper bounded with divergence-based measures of entangling abilities of multiplex quantum channels. The measures of entanglement and genuine entanglement (see Sec.~\ref{subsec:Entanglement} for details) we provide are of the form 
\begin{equation}\label{eqn:DivergenceForm}
\tf{E}_{r}(\mc{N})\coloneqq \sup_{\tau\in\FS(:\vv{LA'}:\vv{RB}:)}\tf{E}_{r}(:\!\vv{LA}\!:\!\vv{R}\!:\!\vv{C}\!:)_{\mc{N}(\tau)}.
\end{equation} 
Here, $r=\text{E}$ or $r=\text{GE}$ (E and GE denote entanglement and genuine entanglement, respectively), and $\FS$ denotes the set of fully separable states (see Sec.~\ref{subsec:Entanglement} for more details). Furthermore, we define the divergence $\tf{E}_{r}$ form the convex sets $\tf{S}_E$ and $\tf{S}_{GE}$ of separable and biseparable states respectively for any partition $:\vv{X}:$, measured by some generalized divergence $\tf{D}$
\begin{equation}\label{eqn:DivergenceForm2}
\tf{E}_{r}(:\vv{X}:)_{\rho}\coloneqq \inf_{\sigma\in\tf{S}_r(:\vv{X}:)}\tf{D}(\rho\Vert\sigma).
\end{equation} 
We call a quantity a generalized divergence if it satisfies the data processing inequality. Namely, for any channel $\mathcal{N}$ the following inequality must hold
\begin{equation}\label{eq:gen-div-mono}
\mathbf{D}(\rho\Vert \sigma)\geq \mathbf{D}(\mathcal{N}(\rho)\Vert \mc{N}(\sigma)).
\end{equation}
The examples of generalized divergence are therefore quantum relative entropy $D(\rho\Vert\sigma)$~\cite{Ume62}, the max-relative entropy $D_{\max}(\rho\Vert\sigma)$~\cite{MDSFT13,D09,Dat09},  the sandwiched R\'enyi relative entropy $\wt{D}_\alpha(\rho\Vert \sigma)$~\cite{MDSFT13,WWY14}, the $\varepsilon$-hypothesis-testing divergence $D^\varepsilon_h(\rho\Vert\sigma)$~\cite{BD10,WR12}, trace distance $\norm{\rho-\sigma}_{1}$ and negative of fidelity of quantum states $-F(\rho,\sigma)$.

The main results of Ref.~\cite{DBWH19} are upper bounds on the secret-key-agreement capacities of multiplex quantum channels. The upper bounds on these capacities are, therefore, upper bounds on the maximum rates at which multipartite private states can be distributed amongst the honest parties by using the multiplex quantum channel as well as free operations of classical preprocessing and postprocessing (cppp). In the one-shot (single use) case of multiplex quantum channel $\mathcal{N}$ use, for any fixed $\varepsilon\in(0,1)$, in Theorem 2 of Ref.~\cite{DBWH19}, we have the following weak converse bound on the achievable region of cppp-assisted secret-key-agreement capacity $\hat{P}_{\textnormal{cppp}}^{(1,\varepsilon)}(\mc{N})$ of $\mathcal{N}$
\begin{equation}
\hat{P}_{\textnormal{cppp}}^{(1,\varepsilon)}(\mc{N})\leq E^\varepsilon_{h,\GE}(\mc{N}).
\end{equation} 
Here, $E^\varepsilon_{h,\GE}(\mc{N})$ denotes the $\varepsilon$-hypothesis testing relative entropy of genuine multipartite entanglement for the multiplex quantum channel. $E^\varepsilon_{h,\GE}(\mc{N})$ is based on the $\varepsilon$-hypothesis testing divergence \cite{BD10}, see Eqs.~\eqref{eqn:DivergenceForm} and~\eqref{eqn:DivergenceForm2}. Further, in the case of multiple uses of multiplex quantum channel interleaved with LOCC operations in Corollary 3 of Ref.~\cite{DBWH19}  (see also Theorem 3 therein), we show the following strong converse bound on asymptotic capacity ${P}_{\LOCC}(\mc{N})$ of multiplex quantum channel $\mc{N}$
\begin{equation}
\tilde{P}_{\LOCC}(\mc{N})\leq E_{\max,E}(\mc{N}).
\end{equation}
Here, $E_{\max,E}(\mc{N})$ denotes the max-relative entropy of entanglement of the multiplex channel $\mathcal{N}$. $E_{\max,E}(\mc{N})$ is based on the max-relative entropy \cite{Dat09}, see Eqs.~\eqref{eqn:DivergenceForm} and~\eqref{eqn:DivergenceForm2}. Moreover, in the case of finite-dimensional Hilbert spaces (associated with subsystems of honest parties) in Theorem 4 of Ref.~\cite{DBWH19}, we show another strong converse upper bound in terms of regularized relative entropy of entanglement 
\begin{equation}
\tilde{P}_{\LOCC}(\mathcal{N})\leq E^{\infty}_{E}(\mathcal{N}).
\end{equation}
Additionally, we observe that if $\mc{N}_{\vv{A'}\vv{B}\to\vv{A}\vv{C}}$ is teleportation-simulable~\cite{TGW14,Pirandola2017,Takeoka_2016,WTB16,LP17,Pirandola_2018}, the upper bounds on $P_{\LOCC}(\mathcal{N})$ are reduced to the relative entropy of entanglement of the resource state $\theta_{\vv{LA}\vv{R}\vv{C}}$ (see Theorems 5-7 and Corollary 5 in Ref.~\cite{DBWH19}). 
\begin{align}
    &\hat{P}_{\LOCC}(\mc{N})\leq E^{\infty}_{GE}(:\vv{LA}:\vv{R}:\vv{C}:)_{\theta}, \\
    &\tilde{P}_{\LOCC}(\mc{N})\leq E_E(:\vv{LA}:\vv{R}:\vv{C}:)_{\theta}, \label{eqn:WeakStrong}
\end{align}
where $E^{\infty}_{GE}$ is the regularized relative entropy of genuine entanglement and $E_E$ is relative entropy of entanglement. Finally, the derived upper bounds on the secret-key agreement capacities are also upper bounds on channel capacities, where the goal is to distill GHZ states, because GHZ state is an example of a multipartite private state \cite{AugusiakH2008-multi}. 


The second main result of Ref.~\cite{DBWH19} is the application of the upper bounds described above to the specific setups. Firstly, we identify protocols like MDI-QKD~\cite{braunstein2012side,LCQ+12,FYCC15,PhysRevA.93.022325,OLLP19}, and quantum key repeaters~\cite{bauml2015limitations,CM17,christandl2017private} are in fact special cases of LOCC-assisted SKA protocols realized with some particular multiplex quantum channel. First, we discuss that there are multiple ways in which multiplex quantum channel can describe quantum key repeater depending on the specific repeater protocol in use. In the simplest situation two honest parties Alice and Bob, send subsystems of their locally prepared maximally entangled (singlet) states  $\Phi^+_{AR_A}$ and $\Phi^+_{BR_B}$ via quantum channels  ${\mathcal N}_1^{A\to C_A}$, ${\mathcal N}_2^{B\to C_B}$ to Charlie who is assumed to be cooperative but not trusted. In the next step, Charlie performs a joint measurement ${\mathcal M}_{C_AC_B\to  XY}$ on $C_AC_B$ subsystems, in this way realizing a step in entanglement swapping protocol \cite{Zukowski_1993}, and reports the outcome to Bob who performs unitary on his reference system $R_B$. This procedure creates entanglement between Alice and Bob, which can be used for cryptographic purposes. The net multiplex quantum channel describing the quantum repeater is therefore
\begin{equation}
{\mc N}_{AB\to XY}^{\text{repeater}}\coloneqq {\mc M}_{C_AC_B\to  XY}\circ {\mc N}_1^{A\to C_A}\otimes{\mc N}_2^{B\to C_B}.
\end{equation}
We further discuss the possibility in which the channels of Alice and Bob are noisy, and the scheme requires to incorporate many rounds combined with error correction for entanglement distillation~\cite{PhysRevLett.81.5932,Dur_1999,munro2015inside}. The schemes which incorporate many relay stations and allow for distributing entanglement at arbitrarily large distances are also taken into account~\cite{Dur_1999,munro2015inside} (we refer to these as chain schemes). The upper bounds on the rates at which the key can be distributed in key repeaters setting were widely studied in the literature~\cite{bauml2015limitations,CM17,christandl2017private}. Using Theorem~4 in Ref.~\cite{DBWH19} (or alternatively using results in Refs.~\cite{DBW17,BDW18,D18thesis}), we obtain upper bounds on setups considered in Refs.~\cite{bauml2015limitations,christandl2017private} that involve only one-way classical communication from Charlie to Alice and Bob. Our upper bounds are given by $\min\{ E_{\max,E}(\mc{N}^{\text{repeater (chain)}}),E^{\infty}_{E}(\mc{N}^{\text{repeater (chain)}}\}$. The new bounds in Ref.~\cite{DBWH19} are capable of taking into account imperfect measurements performed by Charlie or imperfect error correction, which makes it more useful in the description of practical implementations. Furthermore, in some cases of practical interest, our upper bounds are comparable and perform better than results in  Refs.~\cite{bauml2015limitations,christandl2017private}.

\begin{figure}[t]
\centering
 \includegraphics[trim=1.0cm 8.0cm 4.0cm 0.0cm,width=12.5cm,clip]{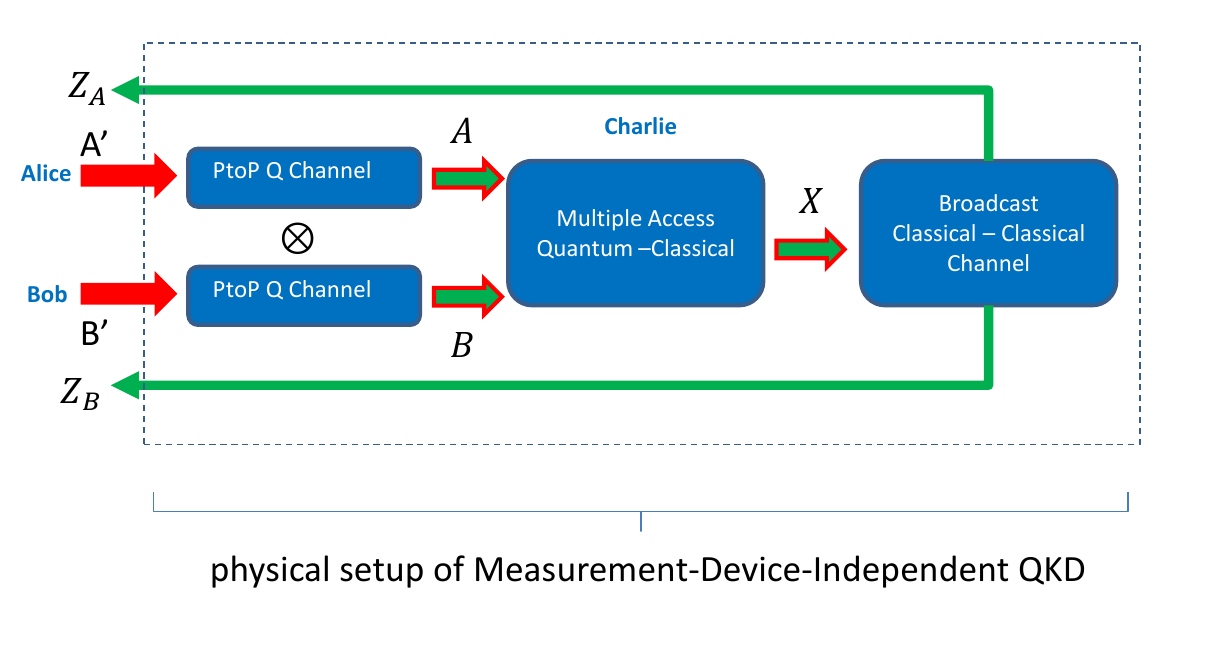}
        \caption{Graphical illustration of quantum classical multiplex channel $\mc{N}^{\text{MDI}}_{A'B'\to Z_AZ_B}$ in Ref.~\cite{DBWH19}. The channel $\mc{N}^{\text{MDI}}_{A'B'\to Z_AZ_B}$ is a composition of three types of elementary multiplex channels, i.e.,  a pair of point-to-point channels from Alice to Charlie and from Bob to Charlie composed with quantum measurement (multiple access quantum to classical channel) in the hands of Charlie followed by classical broadcast channel back to Alice and Bob. The red arrows represent the inputs to the channel, and the green arrows represent the outputs. The green arrows with red boundaries are the outputs of the multiplex channel which are also inputs to the consecutive channels. }\label{fig:MDIscheme} 
\end{figure}

\begin{figure}[t]
\centering
 \includegraphics[trim=4.5cm 7.5cm 4.5cm 6.5cm,width=6cm]{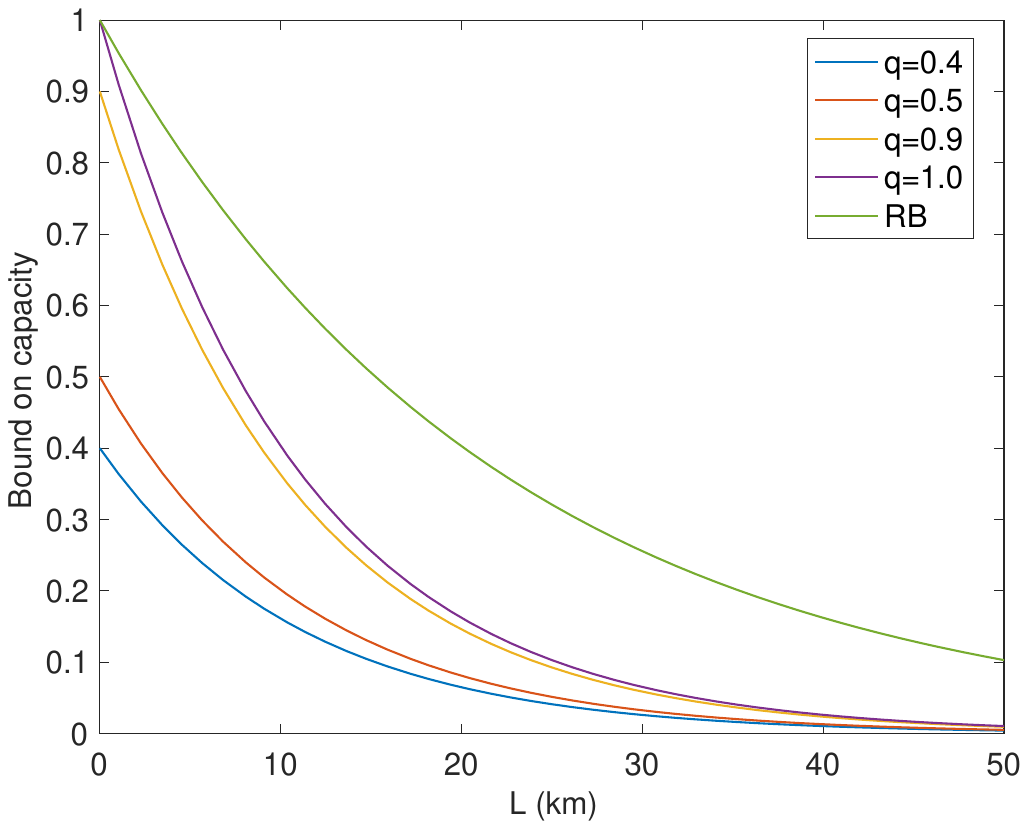}
        \caption{Plot in Ref.~\cite{DBWH19} depicting the rate-distance tradeoff comparison between different upper bound on MDI-QKD scheme for considered photon-based prototype. The values of parameters chosen are $\eta_1=\eta_2=\exp(-\alpha L)$, and $\alpha=\frac{1}{22 \textnormal{km}}$. The result of Ref.~\cite{DBWH19}, here given by~\eqref{eq:dbwhh}, are blue, red, yellow, and purple curves (plotted for different values of $q$), and the repeaterless bound (RB) is here green curve.}\label{fig:MDI} 
\end{figure}

Let us now describe the results in Ref.~\cite{DBWH19} regarding the MDI-QKD protocol (see Fig. \ref{fig:MDIscheme}). The MDI-QKD protocol is a form of QKD in which the honest parties, Alice and Bob, trust the quantum state they are supplied with but do not trust the detectors (measurements) \cite{braunstein2012side,LCQ+12}. The MDI-QKD scheme is important because it solves the problem of imperfect detectors, which can be attacked~\cite{Makarov2009}. For this reason, MDI-QKD protocol has drawn enormous theoretical and experimental attention over the last few years~\cite{pirandola2015high,FYCC15,LYDS18,MZZ18,Tamaki_2018,LL18,cui2019twin,curty2019simple,LWW+19,MPR+19,PLG+19,LoPan_review}. In the typical MDI-QKD setup~\cite{braunstein2012side,LCQ+12}, the honest parties prepare quantum states, which are sent to the relay station via quantum channels ${\mc N}^1_{A'\to A}$ and ${\mc N}^2_{B'\to B}$. The relay station where a joint measurement $\mc{M}_{AB\to x}$ is performed might be in control of eavesdropping Eve that communicates the outcome of the measurement to Alice and Bob via classical broadcast channel $\mc{B}_{X\to Z_AZ_B}$. 
\begin{equation}\label{eq:MDIchannel}
\mc{N}^{\text{MDI}}_{A'B'\to Z_AZ_B}\coloneqq\mc{B}_{X\to Z_AZ_B}\circ\mc{M}_{AB\to X}\circ{\mc N}^1_{A'\to A}\otimes{\mc N}^2_{B'\to B},
\end{equation}
where $Z_A$ and $Z_B$ are classical registers of Alice and Bob, respectively. In the virtue of Theorem 3 and Theorem 4 of Ref.~\cite{DBWH19} (alternatively, using results in Ref.~\cite{DBW17,BDW18,D18thesis}), we obtain an upper bound on the achievable key rate in terms of $E_{\max,\E}(\mc{N}^{\text{MDI}}_{A'B'\to Z_AZ_B})$ and $E^\infty_{\E}(\mc{N}^{\text{MDI}}_{A'B'\to Z_AZ_B})$ in case of general systems and finite-dimensional systems respectively. Moreover, we discuss in detail a photon-based prototype of MDI-QKD protocol in the form of the dual-rail scheme in which qubits are encoded in two orthogonal modes of a single photon~\cite{RP10}. For the noise model present in the quantum channel, we chose the pure-loss bosonic channel with transmissivity $\eta$~\cite{DKD18}. We observe that in the considered case, the action of the pure-loss bosonic channel is identical to the erasure channel~\cite{GBP97}, and therefore the former channel is tele-covaraint as the latter one is. We further assume that the central relay station operated by Charlie (or Eve) performs a perfect measurement in the Bell's basis with probability $q$ and fails with probability $1-q$ of what the honest parties are informed. Finally, using Theorem 7 in Ref.~\cite{DBWH19} we obtain the following capacity for the considered bipartite prototype of MDI-QKD protocol  $\mathcal{N}^{\text{MDI},\{\mathcal{E}_i\}_{i=1}^2}$
\begin{equation}\label{eq:dbwhh}
    \wt{P}_{\LOCC}(\mathcal{N}^{\text{MDI},\{\mathcal{E}_i\}_{i=1}^2})=q\eta_1\eta_2.
\end{equation}
The equality in Eq. \eqref{eq:dbwhh} above is because $q\eta_1\eta_2$ is known to be an achievable rate for the given setup (what follows from results in Refs.~\cite{GES16,Pirandola2017,WTB16}). The direct application of results from Refs.\cite{GES16,Pirandola2017} gives a repeaterless upper bound (RB) on the MDI-QKD capacity to be $\min\{\eta_1,\eta_2\}$ \cite{CM17,pirandola2019capacities} what is never less than $q\eta_1\eta_2$. See Fig.~\ref{fig:MDI} for a comparison between the performance of the RB upper bound and bound derived in Ref.~\cite{DBWH19}.

\newpage
Our next main result concerns upper bounds on the rates of conference key distillation from quantum states rather than the secret-key-agreement capacity of quantum channels. First, we state the security condition for the output of $(n,K,\varepsilon)$ LOCC-assisted conference key agreement protocol
\begin{equation}
    F(\mathcal{L}_{\vv{A^{\otimes n}}\to \vv{SK}}(\rho^{\otimes n}_{\vv{A}}), \gamma_{\vv{KS}})\geq 1-\varepsilon,
\end{equation}
where $\mathcal{L}_{\vv{A^{\otimes n}}\to \vv{SK}}$ is the LOCC (operation) channel performed by $M$ honest parties. In Theorem~8 of Ref.~\cite{DBWH19} we show that the one-shot secret-key distillation rate $K_{\textnormal{D}}^{(1,\varepsilon)}$ from a single copy of multipartite quantum state is upper bounded by 
\begin{equation}\label{eqn:advanatge1}
        K_{\textnormal{D}}^{(1,\varepsilon)}(\rho)\leq E^\varepsilon_{h,\GE}(:\vv{A}:)_{\rho} \coloneqq \inf_{\sigma\in\BS(:\vv{A}:)} D^\varepsilon_h(\rho\Vert\sigma).
\end{equation} 
Further, in Proposition 2 of Ref.~\cite{DBWH19} we show that the secret key rate $K_{\textnormal{D}}$ \cite{AugusiakH2008-multi} obtained directly from the definition $K_{\textnormal{D}}(\rho)=\inf_{\varepsilon>0} \limsup_{n \to \infty}\frac 1n K_{\textnormal{D}}^{(n,\varepsilon)}(\rho^{\otimes n})$ is upper bounded with the regularized relative entropy of the genuine entanglement
\begin{equation}
K_D(\rho_{\vv{A}})\leq E^\infty_{GE}(\rho_{\vv{A}}).
\end{equation}
The above result generalizes Theorem 9 in Ref.~\cite{keyhuge}. Moreover, we observe that (see Corollary 6 in Ref.~\cite{DBWH19}) $K_D(\rho_{\vv{A}})=0$ provided $\rho_{\vv{A}}$ is a tensor-stable biseparable state~(see Sec. \ref{subsec:Entanglement} for details). In this place, we discuss that already in tripartite setting ($M=3$), there exist two nonequivalent classes of genuinely entangled states, i.e., the $\Phi_M^\mathrm{GHZ}$ class and $\Phi_M^\mathrm{W}$ class of states~\cite{nonEquivWandGHZ,bennett2000exact,Horodecki2009,amico2008entanglement,HSD15,spee2017entangled}. Despite the fact, that both classes contain the states that are pure entangled they can not be transformed between each other at unit rate~\cite{FL-2007,FL-2008,Hoi-Kwong2010,vrana2015asymptotic,vrana2019distillation,streltsov2020rates}.  Since $\Phi_M^\mathrm{GHZ}$ has a role of perfect implementation of CKA protocols, the task of distillation of  $\Phi_3^\mathrm{GHZ}$  from $\Phi_3^\mathrm{W}$ has been studied \cite{SVW,FL-2007,KT-2010,CCL-2011,vrana2015asymptotic,vrana2019distillation}. In particular, it is known that a single $\Phi_3^\mathrm{W}$ state can not be transformed into $\Phi_3^\mathrm{GHZ}$ state even in a probabilistic manner \cite{SVW}. On the other hand, a lower bound on the asymptotic conversion rate is known to be $\approx 0.643$ (see Theorem~2 in Ref.~\cite{SVW}). However, as we exemplify, the $\Phi_3^\mathrm{W}$ to $\Phi_3^\mathrm{GHZ}$ conversion in the one-shot regime is still possible. Namely, two $\Phi_3^\mathrm{W}$ states can be transformed into a single $\Phi_3^\mathrm{GHZ}$ state with probability arbitrarily close to $\frac 23$. Since, distillation of $\Phi_M^\mathrm{GHZ}$ is only an instance of conference key distillation strategy \cite{Cabello2000,Scarani2001,Chen2005,AugusiakH2008-multi,Augusiak2009W,Grasselli2019,vrana2015asymptotic,vrana2019distillation}, and more general scenario incorporates distillation of private states, i.e., twisted $\Phi_M^\mathrm{GHZ}$ states  \cite{Pankowski2008,Horodecki2009,AugusiakH2008-multi,Pankowski2010,bauml2017fundamental} Eq.~\eqref{eqn:advanatge1} (Theorem 8 in Ref.~\cite{DBWH19}) constitutes an upper bound on $\Phi_M^\mathrm{GHZ}$ distillation rate. In this way, the lower bound of $\frac 23$ can be compared with the upper bound in Eq.~\eqref{eqn:advanatge1}. In order to obtain non-trivial upper bounds on CKA distillation rates from single or multiple copies of noiseless, dephased, and depolarized $\Phi_3^\mathrm{GHZ}$ and  $\Phi_3^\mathrm{W}$ we devise two families of biseparable states 
\begin{align}
    &\pi_\textnormal{GHZ}^{n,M}:= \frac{1}{M}\sum_{i=1}^M \left ( \mathcal{S}_{1,i} \left( \frac{I}{2}\otimes \Phi_{M-1}^\mathrm{GHZ}\right)\right)^{\otimes n}, \label{eqn:bisepGHZ}\\
    &\pi_\textnormal{W}^{n,M}:= \frac{1}{M}\sum_{i=1}^M \left ( \mathcal{S}_{1,i} \left( \op{0}\otimes \Phi_{M-1}^\mathrm{W}\right)\right)^{\otimes n},\label{eqn:bisepW}
\end{align}
\begin{figure}[t]
\centering
 \includegraphics[trim=2.0cm 7.5cm 2.0cm 6.5cm,width=7.5cm,clip]{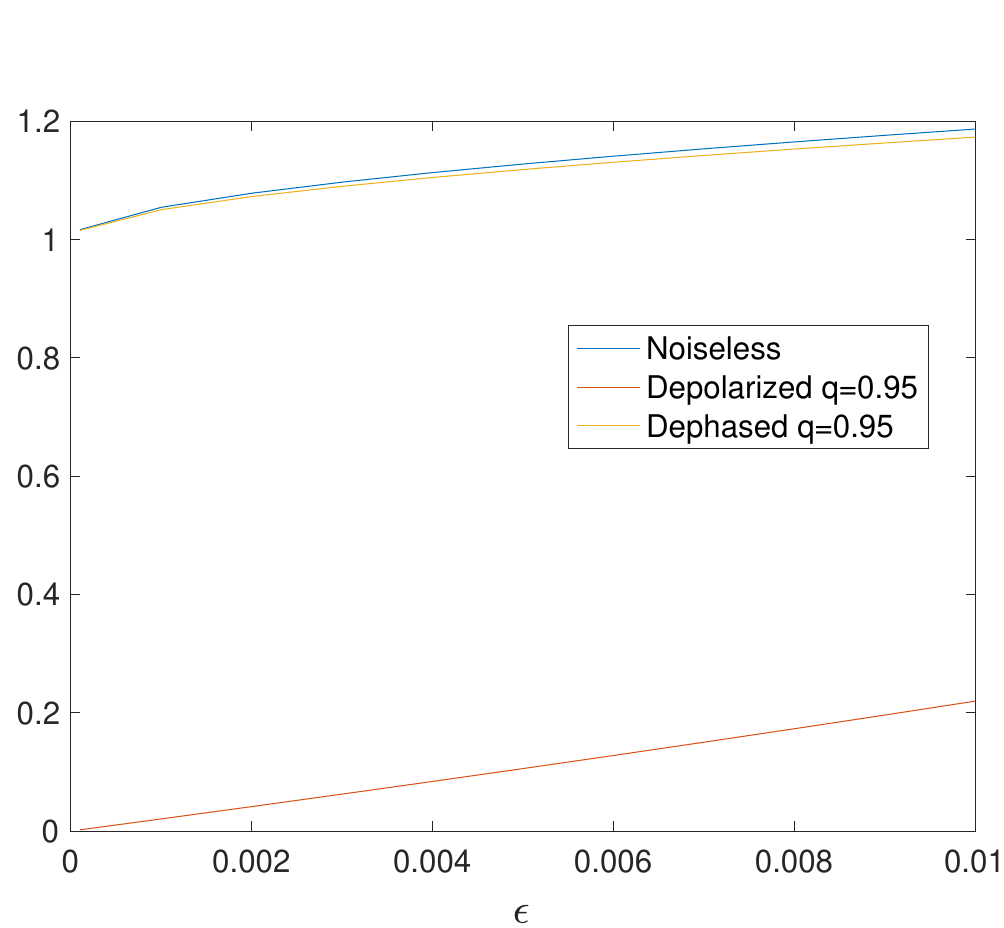}
 \includegraphics[trim=2.0cm 7.5cm 2.0cm 6.5cm,width=7.5cm,clip]{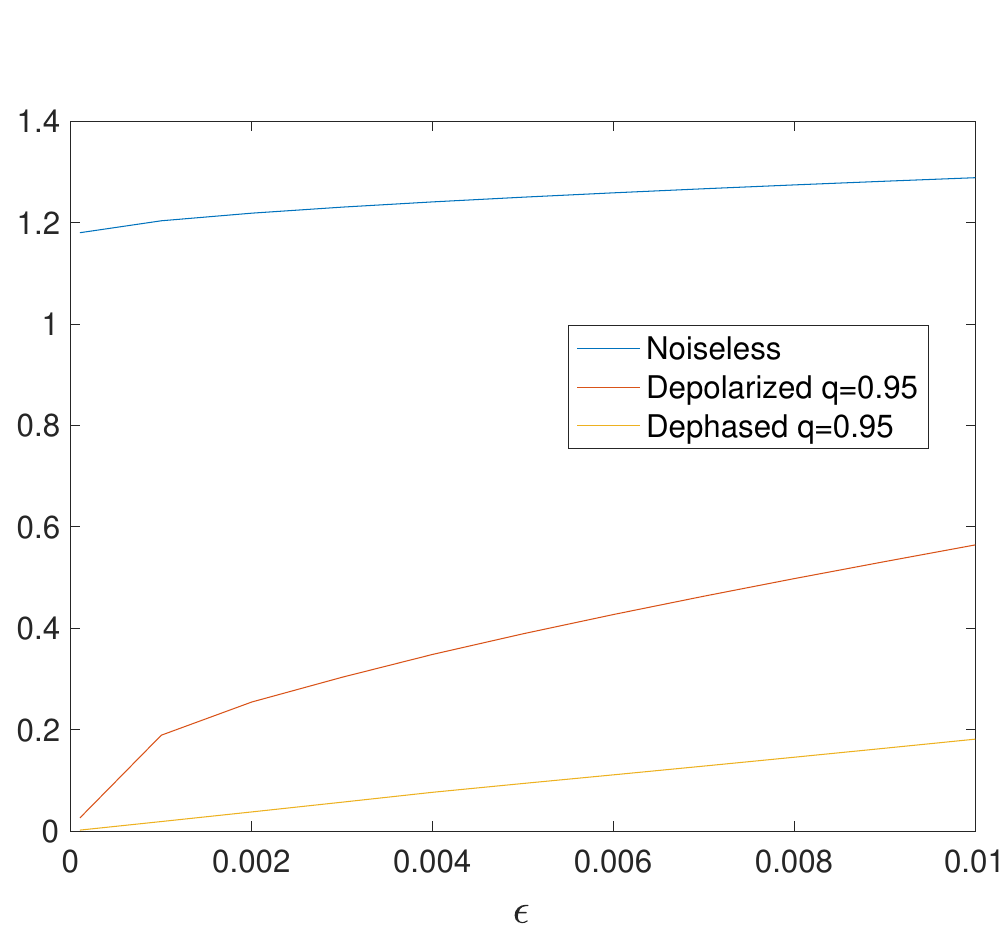}
        \caption{Plots in Ref.~\cite{DBWH19} depicting upper bounds in Theorem~8 (see Eq.~\eqref{eqn:advanatge1}) on one-shot conference key distillation rates from a single copy of $\Phi_3^\mathrm{GHZ}$ state (left) and two copies of $\Phi_3^\mathrm{W}$ state (right). The dephasing ($\mathcal{D}^q_{deph.}(\omega)=q\omega+ (1-q)  \sigma_z  \omega  \sigma_z$, where $\sigma_z$ is the Pauli Z operator) and depolarizing ($\mathcal{D}^q_{depol.}(\omega)=q \omega + (1-q)\frac{\mathbbm{1}}{2}$) noise act separately on each qubit.}\label{fig:OSUB} 
\end{figure}
here the operator $\mathcal{S}_{1,i}$ swaps the qubit of the first system (party) with the qubit of the $i$-th system. The number-of-copies-sensitive definition of biseparable states $\pi_\textnormal{GHZ}^{n,M}$ and $\pi_\textnormal{W}^{n,M}$ is due to non-closure of the set of biseparable states under the tensor product. Interestingly, we point out that the state $\pi_\textnormal{W}^{1,3}$ devised by us in Ref.~~\cite{DBWH19} is closer in the Hilbert-Schmidt norm to $\Phi_{3}^\mathrm{W}$ than the state in Ref.~\cite{VDM-2002} that was found to be the closest (in Ref.~\cite{VDM-2002} alternative definition of biseparability was employed). See Fig.~\ref{fig:OSUB} for the performance of the upper bound in Eq.~~\eqref{eqn:advanatge1} (Theorem~8 in Ref.~\cite{DBWH19}) employing biseparable states in Eqs. \eqref{eqn:bisepGHZ} and \eqref{eqn:bisepW}. Finally, in Proposition~3 and Corollary~7 of Ref.~\cite{DBWH19}, we show a method of constructing a family of (in general) nonequivalent upper bounds on the asymptotic key rate $K_D$. The method is based on Proposition~2 in Ref.~\cite{DBWH19} and the observation that $M-1$-partite key is no less than $M$-partite key if two parties unite. This fact is a consequence of the strict inclusion of the set of $M$-partite LOCC within the set of $M-1$-partite LOCC. In this way, we obtain 
\begin{align}
    &K_D\left( \Phi_{3}^\mathrm{W}\right) \le 
    K_D\left( \Phi_{2+1}^\mathrm{W}\right) \le E_\mathrm{GE}^{\infty}\left(\Phi_{2+1}^\mathrm{W}\right) = h_2\left(\frac{1}{3}\right)\approx 0.9183~ \mathrm{bit},
\end{align}
where $h_2(x)$ is the binary entropy function, and $\Phi_{2+1}^\mathrm{W}$ denotes $\Phi_{3}^\mathrm{W}$ state in which arbitrary two parties united, identical to one of Bell's states. The above upper bound can be compared with the lower bound of $\approx 0.643$ bit obtained in Ref.~\cite{SVW}.

Additionally, as another contribution of Ref.~\cite{DBWH19}, we provide lower bounds on the secret-key-agreement rates of multiplex quantum channels that can be achieved with the assistance of cppp operations. The multipartite protocols we devise are based on the Devetak-Winter protocol \cite{DevetakWinter-hash}. The Devetak-Winter protocol is a point-to-point protocol that incorporates one-way classical communication from Alice to Bob \cite{DevetakWinter-hash}. In this way, we generalize the lower bound on secret-key-agreement rates for multipartite quantum states derived in Ref.~\cite{AugusiakH2008-multi} along with the lower bound for point-to-point quantum channels shown in Ref.~\cite{pirandola2009direct} to a network of bidirectional quantum channels. The first lower bound we derive is a direct extension of the result obtained for quantum states in Ref.~\cite{AugusiakH2008-multi}. Here, the idea is to choose the so-called distributing party that performs the (directed) Devetak-Winter protocol with each of the other parties. In this case, the achievable rate is given by the worst-case Devetak-Winter protocol rate between the distributing party and any other party. As an enhancement of the lower bound, we maximize over possible choices of the distributing party. Our second protocol considers a chain of parties in which each party performs the Devetak-Winter protocol with the subsequent party. The achievable rate is given then by the lowest Devetak-Winter protocol rate between all links in the chain. In this case, we optimize with respect to all possible permutations of the parties in the chain. Furthermore, we consider a bidirectional quantum network, i.e., a network in which each node is connected to all of its neighbors via a product of point-to-point quantum channels in opposite directions. Our technique is based on some facts from the graph theory~\cite{Wil96book}. In particular, one can represent a bidirectional quantum network as a weighted, directed multigraph in which nodes represent parties, and each edge, having some weight, represents a product bidirectional channel. In order to find the best lower bound, in Ref.~\cite{DBWH19}, for each spanning tree, we attribute its lowest rate of Devetak-Winter protocol~\cite{Devetak2003} achieved among any pair of parties connected by an edge in the tree. We further maximize this rate over all spanning trees. Finally, we exemplify that the last approach is advantageous with respect to the two described before.


To summarize, by providing universal limitations on the rates at which quantum conference key can be distributed over multiplex quantum network, we created a unified approach for the treatment of a diverse class of secure communication setups. In this way, we contributed to a better understanding of the limitations of the aforementioned class of protocols and, consequently, to a better understanding of the whole CKA-DD security paradigm. The derived upper bounds provide benchmarks on the present experimental setups and forthcoming quantum Internet of the future.  
\newpage
~~
\newpage

\newpage
\section{Fundamental Limitations on the Device-Independent Quantum Conference Key Agreement [C]}\label{sec:C_Fundamenal}

\begin{figure}[t]
    \center{\includegraphics[trim=0cm 0cm 0cm 0cm,width=10.5cm]{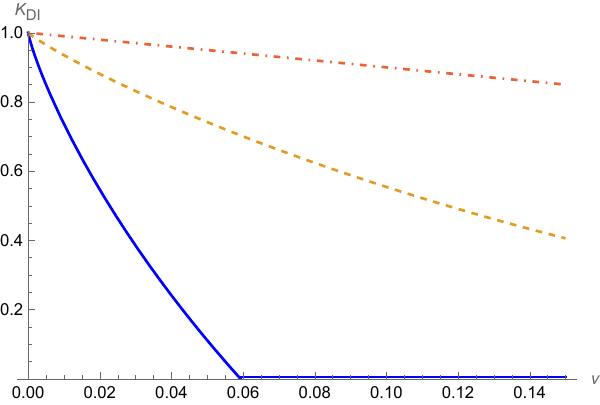}}
    \caption{Modified plot from Ref.~\cite{HWD22} of upper bounds introduced in Ref.~\cite{HWD22} and lower bounds on the DI-CKA scenario in Ref.~\cite{RMW18}. Here, $\nu$ is the parameter of the noise in the depolarizing channel ($\mathcal{D}_\nu^{depol.}(\rho)=(1-\nu) \rho + \nu\frac{\mathbbm{1}}{2}$) that acts on each qubit. The yellow dashed curve represents an upper bound (not fully optimized) on the upper bound $\frac{1}{N-1}I(N(A)\downarrow E)$ from Eq.~\eqref{eqn:UBclassicalE}  (Corollary 2 in ~\cite{HWD22}) with eavesdropper post-processing channel in Eq.~\eqref{eqn:attack1}. The red dash-dotted curve is the upper bound obtained in Corollary 6 of Ref.~\cite{HWD22} based on the relative entropy of entanglement bound ($1-\nu$). The solid blue line represents a lower bound, i.e., the rate of the protocol in Ref.~\cite{RMW18}.}
    \label{fig:attack}
\end{figure}

In Ref.~\cite{HWD22}, we initiate the study of the upper bounds in the device-independent quantum conference key agreement (DI-CKA) scenario. We firstly provide a multipartite generalization of the cc-squashed entanglement introduced in Ref.~\cite{AFL21} and further developed in Ref.~\cite{KHD21}. With a little abuse of notation, we call the new measure reduced c-squashed entanglement~$E_{sq,dev}^{c}$. We show that  $E_{sq,dev}^{c}$ upper bounds the device-independent conference key rate $K_{DI,dev}^{iid,{\hat{\textbf x}}}$ achieved by (standard) protocols (see, e.g., \cite{AFFRV18}) that use a single input $\hat{\textbf{x}}$ to generate the key. This relation holds in the independent and identically distributed setting as we aim at the upper bounds on the rate of DI-CKA (see Sec.~\ref{subsubsec:Security}). Here, the subscript dev refers to the fact that the adversary mimics the full statistics of the honestly implemented device. We generalize further the scenario to the case in which the eavesdropper has to mimic only some relevant parameters of the device. The latter case is denoted with subscript par (e.g., $K_{DI,par}^{iid,{\hat{\textbf x}}}$). To achieve our goal, we generalize the upper bounds via intrinsic information studied in Ref.~\cite{Christandl12} to the case in which the system of the adversary can be infinite-dimensional. In that way, we closed a possible loophole in the proofs of Corollary 3 and Corollary 4 of Ref. \cite{KHD21}. Our upper bounds are then compared with the lower bound on the DI-CKA secret key rate given in Ref.~\cite{RMW18}  (see Fig.~\ref{fig:attack}). We also provide a non-trivial upper bound on the device-independent key rate in the parallel measurement scenario in which all parties simultaneously set all values of inputs $x_i$. Moreover, we generalize reduced bipartite entanglement measures (see Ref.~\cite{KHD21}) to the multipartite case. In this way, we show that the reduced regularized relative entropy of genuine entanglement~\cite{DBWH19} upper bounds the DI-CKA rate for multipartite quantum states. We also discuss the issue of genuine nonlocality~\cite{Bell-nonlocality} and genuine entanglement (see Secs.~\ref{subsec:Entanglement} and \ref{subsec:Nonlocality}) in the context of the DI-CKA~\cite{Horodecki2009}. Finally, we show a strict gap between the rate of the quantum device-independent conference key $K_{DI}$ and quantum device-dependent conference key rate $K_{DD}$. This result is achieved by providing a method to lift the gap in bipartite case to the multipartite case. In this manner, we construct multipartite states that exhibit a strict gap using bipartite states for which the gap was proven in Ref.~\cite{CFH21}.


Building a quantumly secured Internet would provide worldwide information-technologically secure communication~\cite{DM03,Wehner_2018}. Quantum repeaters \cite{Dur_1999,Muralidharan_2016,ZXC+18} give hope that this formidable task will be achieved in the not-too-distant future. Unfortunately, the level of quantum security proposed in the seminal paper of C.~H.~Bennett and G.~Brassard~\cite{BB84} appears to be insufficient. This insufficiency is because, for example, a malicious eavesdropper can change the inner workings of a quantum device on the way of the device from the trusted manufacturer to the honest users, making the cryptographic device totally insecure~\cite{Pironio2009}. Indeed, attacks on the quantum cryptographic devices and quantum internet are already considered~\cite{Becker2013,Makarov2009,Sakarya_2020,Satoh_2021}. Fortunately, the idea of device-independent cryptography overcomes this obstacle~\cite{E91,Pironio2009} and importantly seems to be feasible in practise~\cite{Harald_exp,Renner_exp,JWP_exp}. Simultaneously, the limitations of the device-independent approach in terms of the upper bounds on the distillable key were considered~\cite{Kaur-Wilde,CFH21,Farkas_2021,KHD21}. However, present studies focus on the point-to-point device-independent secure communication. In Ref.~\cite{HWD22}, we initiate the study of upper bounds in the device-independent conference key agreement (DI-CKA)~\cite{Murta2020,RMW18}, where the task is to distribute the same secure key between $N>2$ honest parties, security of which does not depend on the inner working of the devices of the honest parties. The key is used later for one-time-pad encryption. A protocol achieving the DI-CKA task has been shown in~\cite{RMW18}. Our considerations set upper bounds on the performance of any DI-CKA generating protocol, including the mentioned one, in the network setting.

In the considered setup, $N$ trusted spatially separated allies intend to extract secret-key against a quantum adversary. Aiming at the upper bounds it is enough to consider that the honest parties share $n$ identical devices. The honest implementation of the device, which consists of state and measurement, is denoted $(\rho,\mathcal{M})$. However, the adversary can replace the honest device, provided by the manufacturer, with a different device $(\sigma,\mathcal{N})$, such that it yields the same input-output statistics as the honest one. The (coarse-grained) statistics tested by the honest parties are usually the level of violation of some multipartite Bell inequality $\omega(\rho, \mathcal{M})$ (see Refs.~~\cite{Mer90,Ard92,BK93,SS02,ZB02,WW02,YCLO12,HSD15,Luo21})), and the quantum bit error rate (QBER) $P_{err}(\rho,\mathcal{M})$. The QBER is the probability that the outputs of the honest parties' device are not equal to each other, given that the input for the key generating round has been chosen. Still, in some cases, the honest parties would like to test the whole statistics of the $(\rho,\mathcal{M})$ device. The honest measurement is described as a family of tensor product of POVMs for each of the honest party $\mathcal{M}\equiv \{M^{x_1}_{a_1}\otimes M^{x_2}_{a_2}\otimes\ldots\otimes M^{x_N}_{a_N}\}_{\textbf{a}|\textbf{x}}$, where $ \textbf{x}\coloneqq (x_1,x_2,\ldots , x_N)$ and $\textbf{a}\coloneqq (a_1,a_2,\ldots ,a_N)$ correspond to the settings of inputs and outputs for some number $N\in\mathbb{N}$ of the honest parties (denoted $N(A)\equiv A_1 \dots A_N$). The set $\{a_i\}_{i=1}^N$ denotes a finite set of measurement outcomes for measurement choices~$x_i$. In that way, the device yields a probability distribution
\begin{align}
    p({\textbf a}|{\textbf x})=\Tr[M^{x_1}_{a_1}\otimes M^{x_2}_{a_2}\otimes\ldots\otimes M^{x_N}_{a_N} \rho_{N(A)}], 
\end{align}
for a measurement $\mathcal{M}$ on a $N$-partite state $\rho_{N(A)}$. Provided the statistics $\{p({\textbf a}\vert {\textbf x})\}_{{\textbf a}|{\textbf x}}$ obtained from devices $(\rho,\mathcal{M})$ and $(\sigma,\mathcal{N})$ are identical, we write $(\sigma,\mathcal{N})= (\rho,\mathcal{M})$. Consequently, $(\sigma,\mathcal{N})= (\rho,\mathcal{M})$ implies $\omega(\sigma,\mathcal{N})= \omega(\rho,\mathcal{M})$ and $P_{err}(\sigma,\mathcal{N})= P_{err}(\rho,\mathcal{M})$. Furthermore, if the statistic $p$ and $p^\prime$ corresponding to the devices  $(\sigma,\mathcal{N})$ and $ (\rho,\mathcal{M})$ satisfy
\begin{equation}
    d(p,p')=\sup_{{\textbf x}}\norm{p(\cdot\vert {\textbf x})-p'(\cdot \vert {\textbf x})}_1\leq \varepsilon,
\end{equation}
\newpage
we say that $(\sigma,\mathcal{N})$ and $ (\rho,\mathcal{M})$ are $\varepsilon$-close to each other and denote the $\varepsilon$-proximity by  $(\sigma,\mathcal{N}) \approx_\varepsilon (\rho,\mathcal{M})$. In that way, we consider the relations 
\begin{align}
    (\rho,\mathcal{M}) &\approx_\varepsilon  (\sigma,\mathcal{N}) \label{eq:k-1},\\
    \omega(\rho,\mathcal{M})& \approx_{\varepsilon} \omega(\sigma,\mathcal{N})\label{eq:k-2},\\
    P_{err}(\rho,\mathcal{M})& \approx_{\varepsilon} P_{err}(\sigma,\mathcal{M})\label{eq:k-3},
\end{align}
that are used in the definitions of the DI-CKA key rate. Finally, for the sake of upper bounds, it is enough to consider the scenario that contains independent and identically distributed (iid) copies of a device. This simplification is because the iid scenario automatically constitutes an upper bound on the general scenario \cite{CFH21}. In this, we work with two different notions of DI-CKA, iid maximal distillation rates for a device $(\rho,\mathcal{M})$, i.e., dev and par key rates (see Sec.~\ref{subsubsec:Security})
\begin{align}
    &K^{iid}_{DI,dev}(\rho,\mathcal{M})\coloneqq \inf_{\varepsilon>0}\limsup_{n\to \infty} \sup_{\hat{\mathcal{P}}} \inf_{\eqref{eq:k-1}} \kappa^\varepsilon_{n} \left(\hat{\mathcal{P}}\left((\sigma,\mathcal{N})^{\otimes n}\right)\right),\\
    & K^{iid}_{DI,par}(\rho,\mathcal{M}) \coloneqq \inf_{\varepsilon>0}\limsup_{n\to \infty} \sup_{\hat{\mathcal{P}}} \inf_{\eqref{eq:k-2},\eqref{eq:k-3}} \kappa^\varepsilon_{n}\left(\hat{\mathcal{P}}\left((\sigma,\mathcal{N})^{\otimes n}\right)\right),
\end{align}
where $\hat{\mathcal{P}}$ is a protocol that consists of classical local operations and public (classical) communication (CLOPC), $\kappa_n^\varepsilon$ is the $\varepsilon$-secure key rate of a protocol $\hat{\mathcal{P}}$ achieved for any fixed value of security parameter $\varepsilon>0$,  number of copies $n$, and measurements $\mathcal{N}$. Because, Eq.~\eqref{eq:k-1} implies Eq.~\eqref{eq:k-2} and Eq.~\eqref{eq:k-3} we automatically have $K^{iid}_{DI,dev}(\rho,\mathcal{M}) \ge K^{iid}_{DI,par}(\rho,\mathcal{M})$. Furthermore, we observe that DI-CKA maximal key distillation rates are upper bounded by the device-dependent distillation key rates (DD-CKA) and also by the entanglement measures that upper bound the latter. 

In order to obtain our main result, we first generalize the notion of the cc-squashed entanglement~\cite{AFL21,KHD21} to the multipartite setting based on the notion of multipartite squashed entanglement \cite{multisquash}. W call the newly introduced quantity the c-squashed entanglement that reads 
\begin{align}
    E^{c}_{sq}(\rho_{N(A)},{\mathrm M}):= \inf_{\Lambda:~{E\rightarrow E'}}I(A_1:\ldots  :A_N|E')_{{\mathrm M}_{N(A)}\otimes\Lambda \psi^{\rho}_{N(A) E}}\label{eqn:def:CC_primal_channel}.
\end{align}
Here, ${\mathrm M}_{N(A)}\equiv{\mathrm M}_{A_1},\ldots , {\mathrm M}_{A_N}$ is an $N$-tuple of positive-operator-valued measures (POVMs) and state $\ket{\psi^{\rho}_{N(A)E}}$ is a purification of $\rho_{N(A)}$. Furthermore,
\begin{align}\label{eq:i_multipartite}
    I(A_1:\ldots  :A_N|E)_{\rho_{N(A)}}=\sum_{i=1}^{N} S(A_i|E)_{\rho_{N(A)}} - S(A_1,\ldots  ,A_N|E)_{\rho_{N(A)}},
\end{align}
where $S(A_i|E)_{\rho_{N(A)}}=S(A_iE)_{\rho_{N(A)}}-S(E)_{\rho_{N(A)}}$ is the conditional entropy, with $S(AB)\coloneqq -\Tr[\rho_{AB}\log_2  \rho_{AB}]$ and $S(A)\coloneqq -\Tr[\rho_A\log_2 \rho_A]$ being von Neumann entropies. Having the new object defined, we provide the first upper bound (Theorem 1 of Ref. \cite{HWD22})
\begin{align}\label{eqn:InfiniteEve}
K_{DD}({\mathrm M}_{N(A)}\otimes \id_E\psi^\rho_{N(A)E}) \leq \frac{1} {N-1}E^{c}_{sq}(\rho_{N(A)},{\mathrm M_{N(A)}}).
\end{align}
Here, $K_{DD}$ is the quantum device-dependent conference key agreement (DD-CKA) rate, and $\id_E$ is the identity channel (quantum operation). The above upper bound is a multipartite generalization (and improvement) of Theorem~$5$ in Ref.~\cite{KHD21}. The improvement is because in  Ref.~\cite{HWD22}, we additionally closed the problem of an eavesdropper holding an infinite-dimensional system. This is achieved directly by allowing the dimension of Eve's system in Eq. \eqref{eqn:InfiniteEve} to be infinite dimensional which is natural in the case of nonlocality. We then move to our main goal, i.e., to study the upper bounds in the DI-CKA scenario. In this case we concentrate on the standard protocols~\cite{AFFRV18} which use a single tuple of measurements $(\hat{x}_1,\ldots  ,\hat{x}_N)\equiv \hat{\textbf{x}}$ applied to ${\cal M}$ of a device $(\rho_{N(A)},{\cal M})$ yielding ${\cal M}(\hat{\textbf{x}})$ (Theorem 2 and Theorem 3 in Ref.~\cite{HWD22})
\begin{align}
    &K_{DI,dev}^{iid,\hat{\textbf{x}}}(\rho_{N(A)},{\cal M}) \le E_{sq,dev}^{c}(\rho_{N(A)},{\cal M}(\hat{\textbf{x}})) \coloneqq \frac{1}{N-1}\inf_{{(\sigma_{N(A)},{\cal L})\equiv(\rho_{N(A)},{\cal M})}}E_{sq}^{c}(\sigma_{N(A)},{\cal L}(\hat{\textbf{x}})), \label{eqn:DevBound}\\
    &K_{DI,par}^{iid,\hat{\textbf{x}}}(\rho_{N(A)},{\cal M})  \le E_{sq,par}^{c}(\rho_{N(A)},{\cal M}(\hat{{\textbf x}})) \coloneqq  \frac{1}{N-1}\inf_{\underset{P_{err}(\sigma_{N(A)},{\cal L})= P_{err}(\rho_{N(A)},{\cal M})}{\omega(\sigma_{N(A)},{\cal L})=\omega(\rho_{N(A)},{\cal M})}}E_{sq}^{c}(\sigma_{N(A)},{\cal L}(\hat{\textbf{x}})).
    \label{eqn:ParBound}
\end{align}
We refer to the quantities in the right hand sides of the Eqns. \eqref{eqn:DevBound} and \eqref{eqn:ParBound} above, as to the reduced c-squashed entanglement, ``dev'' and ``par'' respectively. We then observe that in the case $N=2$, the upper bound in Eq.~\eqref{eqn:ParBound} recovers the results from Ref.~\cite{KHD21}. Moreover, we notice that in the definition of the reduced c-squashed entanglement $E_{sq,par(dev)}^{c}$, one can obtain a weaker upper bound by considering solely classical extensions to Eve's system ~\cite{multisquash}. In that case a for a single tuple of measurements $\hat{\textbf{x}}$ the bound reads (Corollary 2 in Ref.~\cite{HWD22})
\begin{align}
    K_{DI,dev}^{iid,\hat{{\textbf x}}}(\rho_{N(A)},{\cal M})&  \leq \inf_{(\sigma_{N(A)},{\cal L})=(\rho_{N(A)},{\cal M})} \frac{1}{N-1}I(N(A)\downarrow E)_{P(A_1:\ldots  :A_N|E)} \nonumber  \\
    &\equiv \inf_{(\sigma_{N(A)},{\cal L})=(\rho_{N(A)},{\cal M})} \inf_{\Lambda_{E\rightarrow F}} \frac{1}{N-1}I(N(A)| F)_{P(A_1:\ldots  :A_N|\Lambda(E))}, \label{eqn:UBclassicalE}
\end{align}
where measurements ${\cal L}(\hat{\textbf{x}})$ performed on the purification of $\sigma_{N(A)}$ yield the distribution $P(A_1:\ldots  :A_N|E)$. The above is especially useful for numerical investigation. Furthermore, we notice that the reduced c-squshed entanglement $E^{c}_{sq}$ can be defined for multiple measurements in analogy to results in Ref.~\cite{KHD21}
\begin{align}
    E^{c}_{\sq}(\rho_{N(A)},\mathcal{M},p({\textbf x})) \coloneqq \sum_{{\textbf x}}p({\textbf x})E^{c}_{sq}(\rho_{N(A)},\mathrm{M}_{{\textbf x}}).
\end{align}
From the above, we construct an upper bound on the DI-CKA rate for the protocols in which parties broadcast their inputs (see Proposition 2 in Ref. \cite{HWD22})
\begin{align}
  K^{iid,broad}_{DI,dev}(\rho_{N(A)},\mathcal{M},p({\textbf x})) &\le  E^{c}_{sq, dev}(\rho_{N(A)},\mathcal{M},p({\textbf x})) \nonumber \\
&\coloneqq \frac{1}{N-1} \inf_{(\sigma_{N(A)},\mathcal{N})=(\rho_{N(A)},\mathcal{M})}E^{c}_{sq}(\sigma_{N(A)},\mathcal{N},p({\textbf x})).
\end{align}

The investigation on the upper bounds constructed with the reduced c-squashed entanglement contains two significant technical results. Firstly we show that $E_{sq,par}^{c}$ is a convex function with respect to mixtures of quantum states that constitute the device (see Proposition 1 in Ref. \cite{HWD22}), i.e., 
\begin{align}
    E_{sq,par}^{c}(\bar{\rho},{\cal M}({ \hat{\textbf x}})) \leq p_1E_{sq,par}^{c}(\rho_1,{\cal M}({\hat{\textbf x}})) +
    p_2 E_{sq,par}^{c}(\rho_2,{\cal M}({\hat{\textbf x}})),
\end{align}
where $\bar{\rho}=p_1\rho_1+ p_2 \rho_2$ and $p_1+p_2=1$ with $0\leq p_1 \leq 1$. The above results are useful because of the possible application of the convexification technique~\cite{WDH2021} when studying numerical upper bounds. Secondly, we do not restrict the eavesdropper system to be finite-dimensional (see Lemma 6 in Ref.~\cite{HWD22}). This improvement closes a possible loophole in the proofs of Corollary 3 and Corollary 4 of Ref. \cite{KHD21}. 


Our second main result is the numerical investigation that compares in the tripartite ($N=3$) setting the upper bound derived by us with the rate of the parity Clauser-Horne-Shimony-Holt (parity CHSH) based protocol considered in Ref.~\cite{RMW18}. We restrict ourselves to the case with a classical eavesdropper in which the channels acting on Eve's system have only classical outputs. In this way, we exemplify the upper bound in Eq.~\eqref{eqn:UBclassicalE} (see Corollary 2 in Ref.~\cite{HWD22}). To compare our upper bounds with the known lower bound, we consider the case of tripartite Greenberger-Horne-Zeilinger (GHZ) state $\ketbra{\Phi^{\GHZ}_{\vv{3}}}{\Phi^{\GHZ}_{\vv{3}}}$, on which depolarizing noise $\mathcal{D}_\nu^{depol.}$ parameterized with $\nu$ acts locally on each from the three qubits~\cite{RMW18}. For such a setting of the DI-CKA protocol in Ref.~\cite{RMW18}, the statistics of the honest device reads 
\begin{align}
    &P_\nu (a,b_1,b_2|x,y_1,y_2)
    =\Tr \left[M_{a|x}\otimes M_{b_1|y_1} \otimes M_{b_2|y_2} \mathcal{D}_\nu^{\otimes 3} (|\Phi^{\GHZ}_{\vv{3}}\>\<\Phi^{\GHZ}_{\vv{3}}|_{AB_1B_2}) \right] \nonumber\\
    &=(1-\nu)^3 P_\mathrm{GHZ} (a,b_1,b_2|x,y_1,y_2) 
    + (1-(1-\nu)^3) P_\nu^\mathrm{L}(a,b_1,b_2|x,y_1,y_2),
\end{align}
where the ranges of inputs and outputs are $x \in \{0,1\}$, $y_1 \in \{0,1,2\}$, $y_2 \in \{0,1\}$, and $a,b_1, b_2 \in \{0,1\}$. The setting $(x,y_1,y_2)=(0,2,0)$ in the key-generating round is associated with measurements of $\sigma_z$ observable. Furthermore, $P_\mathrm{GHZ}$ arises from local measurements on the GHZ state, and $P_\nu^\mathrm{L}$ arises from the same measurements (for $\sigma_z$ observable) on the fully separable state 
\begin{align}
    \chi_\nu:& = \frac{1}{1-(1-\nu)^3
    } \left((1-\nu)^2\nu\kappa_{AB_1}\otimes \frac{{\mathbbm{1}}_{B_2}}{2} +
    (1-\nu)^2\nu\kappa_{AB_2}\otimes \frac{{\mathbbm{1}}_{B_1}}{2}\right.\nonumber\\
    &\left.+(1-\nu)^2\nu\kappa_{B_1B_2}\otimes \frac{{\mathbbm{1}}_{A}}{2}
        +(3-2\nu)\nu^2\frac{\mathbbm{1}_{AB_1B_2}}{2}\right), 
\end{align}
where $\kappa_{X_1X_2}=\frac{1}{2}\left(|00\>\<00|_{X_1X_2}+|11\>\<11|_{X_1X_2}\right)$.  The eavesdropper's strategy is based on the fact that $P_\nu^\mathrm{L}$ can be expressed as a convex combination of deterministic behaviors. Therefore, Eve prepares a convex combination attack~\cite{AcinGM-bellqkd,acin-2006-8,Farkas_2021}
\begin{align}
    P_\nu^\mathrm{CC} (a,b_1,b_2,e|x,y_1,y_2) 
    &= (1-\nu)^3 P_\mathrm{GHZ} (a,b_1,b_2|x,y_1,y_2) \delta_{e,?} \nonumber\\
    &+ [1-(1-\nu)^3] P_\nu^\mathrm{L}(a,b_1,b_2|x,y_1,y_2) \delta_{e,(a,b_1,b_2)},
\end{align}
where the label $e=?$ means that Eve is not correlated to the nonlocal part of the honest parties, i.e., Alice, Bob1, and Bob2, and $\delta_{e,(a,b_1,b_2)}$ means that Eve is correlated to all event associated with the local device. The above attack does not need to be optimal since it uses a particular decomposition of $P_\nu$. In order to find a better strategy, Eve prepares a channel $E \to F$ that decomposes $P_\nu^\mathrm{L}$ and does not diminish the weight of local behaviors $1-(1-\nu)^3$. In analogy to the attack given in bipartite case in Ref.~\cite{Farkas_2021}, we consider only the distribution coming from a key-generating measurement (see above)
\begin{align}
    &P_\nu^\mathrm{ATTACK} (a,b_1,b_2,f|020)=\Lambda_{E\to F} ~P_\nu^\mathrm{CC} (a,b_1,b_2,e|020) \nonumber\\
    &= (1-\nu)^3 P_\mathrm{GHZ} (a,b_1,b_2|020) \delta_{f,?} \nonumber\\
    &+ (1-(1-\nu)^3) P_\nu^\mathrm{L}(a,b_1,b_2|020)  \left(\delta_{a,b_1,b_2}\delta_{f,a}+(1-\delta_{a,b_1,b_2})\delta_{f,?} \right),
    \label{eqn:attack1}
\end{align}
where $\delta_{a,b_1,b_2}$ is $1$ when $a=b_1=b_2$ and $0$ otherwise. The above attack strategy is thus a tripartite generalization of attack proposed in Ref.~\cite{Farkas_2021} in which the eavesdropper aims to be correlated only to the $a=b_1=b_2$ events associated with $P_\nu^\mathrm{L}$ behavior. By applying the above attack, we plot the upper bound on $K_{DI,dev}^{iid,\hat{{\textbf x}}}(\rho_{N(A)},{\cal M})$ in Fig.~\ref{fig:attack} using Eg.~\eqref{eqn:UBclassicalE} (see Corollary 2 in Ref.~\cite{HWD22}).

\newpage
For the third main result, we show that there exists a gap between device-dependent and device-independent conference key agreement rates. To obtain the result, we first generalize the reduced device-dependent conference key rate to the multipartite setting. The reduced device-dependent key agreement rate of state $\rho_{N(A)}$ is given by (see also Sec.~\ref{subsubsec:Security})
\begin{align}
    K^{\downarrow}(\rho_{N(A)})\coloneqq \sup_{\cal M}\inf_{(\sigma_{N(A)},{\cal L})=(\rho_{N(A)},{\cal M})} K_{DD}(\sigma_{N(A)}),
\end{align}
what yields (see Theorem 6 in Ref.~\cite{HWD22}) as an analog of Theorem $6$ in Ref.~\cite{CFH21}
\begin{align}
    K_{DI}(\rho_{N(A)})\equiv \sup_{\cal M} K_{DI}(\rho_{N(A)},{\cal M}) \leq K^{\downarrow}(\rho_{N(A)}).
\end{align}
To show the existence of the strict gap between $K_{DI}$ and $K_{DD}$ in the multipartite setting (Theorem 7 in Ref.~\cite{HWD22}), we use the fact that there exist bipartite states $\rho_{AB}$ for which $K^{\downarrow}(\rho_{AB})< K_{DD}(\rho_{AB})$ as described in Ref.~\cite{CFH21}. Using these states we construct multipartite states $\rho_{N(A)}$ with the analogous property, i.e., $K_{DI}(\rho_{N(A)}) < K_{DD}(\rho_{N(A)})$. Our construction constitutes a path pointed by a chain of states $\rho_{A_k^1A_k^2}$ shared between spatially separated $N$ parties $A_k$, $k \in \{1,\dots ,N-1\}$. Here, by the means of notation $A_1^1\equiv A_1$ and $A_1^2A_2^1\equiv A_2$,\ldots  , $A_{N-1}^2\equiv A_N$. In this way, we show in Theorem 7 of  Ref.~\cite{HWD22} that a multipartite state 
\begin{align}
    \widetilde{\rho}_{N(A)}\coloneqq 
    \rho_{A_1^1A_1^2}\otimes \rho_{A_2^1A_2^2}\otimes\rho_{A_3^1A_3^2}\otimes\ldots  \otimes\rho_{A_{N-1}^1A_{N-1}^2},
\end{align}
where $\rho_{A_k^1A_k^2}=\rho_{AB}$, $k \in \{1,\dots ,N-1\}$, satisfies $K_{DI}(\rho_{N(A)}) < K_{DD}(\rho_{N(A)})$ as desired, provided all $\rho_{A_{k-1}^1A_{k-1}^2}$ exhibit the gap, i.e., $K_{DD}(\rho_{A_{k-1}^1A_{k-1}^2}) - K^{\downarrow}(\rho_{A_{k-1}^1A_{k-1}^2})\geq c >0$. Indeed, such states do exist \cite{CFH21}, and so do~$\widetilde{\rho}_{N(A)}$. We remark here that the proof of Theorem 7 in  Ref.~\cite{HWD22} can be relaxed to the case in which at least one state in the chain exhibits a bipartite gap and the theorem still holds.

Moreover, we discuss the topic of genuine nonlocality in analogy to the notion of genuine entanglement (see Secs.~\ref{subsec:Entanglement} and \ref{subsec:Nonlocality}). We first define genuine quantum nonlocality by saying which devices are local and all that are not local are genuine nonlocal. Specifically, we assume that local are mixtures of devices that (i) admit quantum realization and (ii) are product in at least one partition into subsystems.
In literature, there are various definitions of locality in multipartite case. Our class of local devices falls between the so-called TOBL and NSBL devices in Ref.~\cite{TOBL} (cf. Ref.~\cite{Svetlichny1,Bell-nonlocality}). In this way, we first say that behavior $P(\textbf{a}|\textbf{x})$ is local in a cut  $(A_{i_1}\ldots A_{i_k}):(A_{i_{k+1}}\ldots  A_{i_N})$ for some $k\in\{1,\dots,N\}$, if and only if it can be written as a product of two behaviors of the systems correspondingly to the cut. Here,  $(i_1,\dots,i_N)$ is an arbitrary permutation of indices $(1,\dots,N)$. The set of all behaviors that are a product in some cut and have quantum realization is denoted by $\mathrm{LQ}$, which means locally quantum (see Sec.~\ref{subsec:Nonlocality}). Furthermore, any distribution that is not locally quantum can be treated as nonlocal, although other definitions can be found in literature~\cite{Bell-nonlocality}.  Consequently, we say that the behavior $P(\textbf{a}|\textbf{x})$ is genuinely nonlocal if and only if it is not a mixture of behaviors that are local in at least one cut. As discussed, any behavior used by honest parties to obtain a conference key in a single-shot (single run) must exhibit genuine nonlocality. We consider a single-shot device-independent key distillation rate obtained by LOPC post-processing of a distribution obtained from some behavior $P({\textbf a}|{\textbf x})$ when all inputs ${\textbf x}$ of all parties are set simultaneously
\begin{align}
    K^\mathrm{single-shot}_{DI,dev}(\rho_{N(A)},\mc{M},\varepsilon)\coloneqq  \sup_{\hat{\mc{P}}} \inf_{\eqref{eq:k-1}} \kappa^\varepsilon_{n} \left(\hat{\mc{P}}(\sigma_{N(A)},\mc{N})\right).
\end{align}
Here, $\kappa_n^\varepsilon$ is the quantum key rate achieved for any measurements $\mc{N}$ and security parameter~$\varepsilon$. In Theorem~8 of Ref.~\cite{HWD22} we show that if $K^{single-shot}_{DI,dev}(\rho_{N(A)},\mc{M},\varepsilon)>0$ then behavior $P(\textbf{a}|\textbf{x})\equiv (\rho_{N(A)},\mc{M})$ is not $\mathrm{LQ}$. Furthermore, in Theorem~9 of Ref.~\cite{HWD22} we show the following upper bounds on DI-CKA rates
\begin{align}
     K_{DI,dev}(\rho_{N(A)}, \mathcal{M})   & \leq  \inf_{(\sigma_{N(A)},\mc{L})= (\rho,\mc{M})} E^{\infty}_{GE} (\sigma_{N(A)}),\\
       K_{DI,par}(\rho_{N(A)}, \mathcal{M})   & \leq  \inf_{\underset{P_{err}(\sigma_{N(A)},{\cal L})= P_{err}(\rho_{N(A)},{\cal M})}{\omega(\sigma_{N(A)},{\cal L})=\omega(\rho_{N(A)},{\cal M})}} E^{\infty}_{GE}(\sigma_{N(A)}).
\end{align}
Here, $E^{\infty}_{GE} (\varsigma)$ is the regularized relative entropy of genuine entanglement~\cite{DBWH19} for a state~$\varsigma_{A_1A_2\ldots  A_N}$, with
\begin{align}
        E^{\infty}_{GE} (\varsigma)=\inf_{\varphi\in{\mathrm{BS}(N(A^{\otimes n}))}}\lim_{n\to\infty}\frac{1}{n}D(\varsigma^{\otimes n}\Vert \varphi),
\end{align}
where $D(\rho\Vert\sigma)=\Tr\rho \log_2 \rho - \Tr \rho \log_2 \sigma$ (if $\supp{\rho}\subseteq\supp{\sigma}$ and $\infty$ otherwise~\cite{Ume62}) is the relative entropy between two states $\rho$ and $\sigma$, and~$\mathrm{BS}$ denotes the set of biseparable states (see Sec.~\ref{subsec:Entanglement} for details) \cite{DBWH19}. 

To conclude, we have introduced several of upper bounds on the quantum secure conference key. Some of the introduced upper bounds generalize the relative entropy-based upper bound in Ref.~\cite{KHD21}, and the reduced c-squashed entanglement provides another upper bounds as the generalization of results in Ref.~\cite{Farkas_2021}. Moreover, we have shown constructive proof that a fundamental gap between the device-dependent and the device-independent key is inherited from the bipartite setting (see Ref.~\cite{AFL21,CFH21}) and holds in a multipartite case. We remark here, that Ref.~\cite{HWD22} is the first published manuscript considering upper bounds on DI-CKA key rates.\\

\emph{Note: We prove analogous upper bounds based on a different definition of the multipartite squashed entanglement. We refer to these bounds as dual bounds, as described in Sec. IV of Ref.~\cite{HWD22}. After publishing the paper, it was pointed out by M.E.~Shirokov that the aforementioned definitions are equivalent due to Theorem 7 of Ref.~\cite{DSW18}. Therefore, our dual upper bounds are equivalent as well. The erratum to the article in Ref.~\cite{HWD22} has been published in Ref.~\cite{HWD22_Erratum}.}

\newpage
\section{Limitations on Device-Independent Key Secure Against Nonsignaling Adversary via the Squashed Nonlocality [D]}\label{sec:D_Squashed}

In the recently published paper, \cite{WDH2021}, we initiate a systematic study of upper bounds on the secret key rate in the non-signaling device-independent secret key agreement (NSDI) scenario~\cite{Kent,Kent-Colbeck,Scarani2006,acin-2006-8,AcinGM-bellqkd,masanes-2009-102,hanggi-2009,masanes-2006}. We begin, by showing that the security definition, that we adopt, based on the so-called non-signaling norm, is equivalent to two notions of security already present in the literature in Refs.~\cite{masanes-2006,masanes-2009-102,Masanes2011} and Refs.~\cite{hanggi-2009,hanggi-2009b,Renner-Hanggi,Hanggi-phd}. In order to prove the above equivalence, we derive the explicit form of the non-signaling norm for the so-called $\textbf{c}$-d states. Our main result is the so-called squashing procedure. We show any secrecy quantifier that is an upper bound on the secret key rate in the SKA scenario~\cite{Maurer93,Intrinsic-Maurer,MauWol97c-intr,reduced-intrinsic,lit3} subjected to the squashing procedure becomes an upper bound on the secret key rate in the NSDI scenario (see Sec.~\ref{subsubsec:Security}). The squashing procedure is based on two elements, i.e., on the notion of the non-signaling complete extension~\cite{CE} and suitable optimization over measurements performed by the honest parties and the eavesdropper. Lifting the formalism by means of the squashing procedure was possible as we rephrased the definition of the secret key rate in the SKA scenario in the form known from quantum key distribution scenarios (QKD). Next, we concentrate on one of the squashed upper bounds, i.e., the non-signaling squashed intrinsic information hereafter called the squashed nonlocality. We prove that the squashed nonlocality is a novel unfaithful measure of nonlocality to what it owes its name. Moreover, we develop a convexification technique that allows to combine different upper bounds into a single tighter one. Finally, we perform a numerical investigation in which we compare our upper bound via the squashed nonlocality with the lower bounds given by the rates of H{\"a}nggi-Renner-Wolf~\cite{hanggi-2009} and by Ac{\'{i}}n-Massar-Pironio~\cite{acin-2006-8} protocols.


The NSDI scenario has the most relaxed assumptions from all security paradigms considered in this thesis. Here, the eavesdropper is constrained only with the no-signaling conditions that make it more powerful than the classical or quantum adversary. Similarly, the honest parties can possibly share supra-quantum correlations constrained only with the no-signaling conditions as well. In a nutshell, the no-signaling conditions restrict the input-output statistics of the devices shared between the parties in a manner that excludes the possibility of instantaneous communication~\cite{InfoProcess}. The possible advantage of the NSDI scenario over other cryptographic paradigms (e.g., SKA, QDD, QDI) is that it assures security even if a new theory of physics is established that would replace the Quantum Theory (QT) and allow for stronger violation of Bell inequalities than QT allows, provided the new theory is non-signaling one as well.

In our considerations, we assume the presence of two honest parties, Alice and Bob, and malicious Eve being the eavesdropper. The object shared by them is a tripartite non-signaling device $P(ABE|XYZ)$ (tripartite normalized, conditional probability distribution), with $X$, $Y$ and $Z$ being inputs, $A$, $B$ and $E$ being outputs, of Alice, Bob and Eve respectively. On the device, the honest parties perform direct measurements $\mathcal{M}_{x,y}^F$ (MD) $(X, Y)$ and post-process their output data $(A, B)$ with some local operations and public communication (LOPC) in order to produce the secure-key. We refer to this class of operations used for the distillation of the secret-key as to MDLOPC class (see Sec.~\ref{subsubsec:Security}). The device is assumed to be provided by Eve, who can listen to public communication and is correlated with the subsystem shared by Alice and Bob as strongly as possible under the considered assumptions. This situation constitutes the worst-case scenario for the honest parties.

In the context of the NSDI scenario, mainly lower bounds on the secret key rate have been studied so far \cite{Kent,AcinGM-bellqkd, acin-2006-8, Scarani2006, masanes-2009-102, hanggi-2009, Kent-Colbeck, masanes-2006}. To our best knowledge, the only upper bound considered so far was in Ref.~\cite{acin-2006-8}. This situation stands in contrast with the SKA \cite{CsisarKorner_key_agreement,Maurer93}, QDD \cite{BB84,E91,B92,Gisin-crypto,AcinBBBMM2004-key}, and QDI \cite{E91,Bell-nonlocality,Mayers-Yao,acin-2007-98, Masanes2011, lit12} scenarios in which both lower and upper bounds are known \cite{CsisarKorner_key_agreement,Maurer93,DevetakWinter-hash-prl, RGKinfo_sec_proof_short,DevetakWinter-hash,Christandl12,pptkey,AugusiakH2008-multi,keyhuge,Christandl_2002,Christandl12,multisquash,Wil16,TGW14,Takeoka_2014,Pirandola2017,Kaur-Wilde,Eneetthesis,CFH21,AFL21,Farkas_2021,KHD21,HWD22}. Most importantly, for the NSDI scenario, it was shown that in the presence of a collective eavesdropper's attack, a non-zero key rate can be obtained \cite{hanggi-2009,masanes-2009-102,masanes-2006} under the fully no-signaling constraints. By the fully no-signaling constraints, one means that none of the $2N+1$ subsystems of the device ($N$ for each of the honest parties and $1$ for the eavesdropper) can signal to each other. This assumption is vital because if the subsystems of an honest party can signal between each other, or the device has a memory and can perform the so-called forward signaling between the runs, then no hash function that can achieve privacy amplification against the non-signaling eavesdropper is known and the standard ones fail the security requirements \cite{hanggi-2009b,Rotem-Sha,Salwey-Wolf}. The fully non-signaling assumption can be achieved in the so-called parallel measurement model in which measurements on $2N$ subsystems are performed simultaneously.

Addressing the problem of upper bounds, it is enough to consider that the device shared between the honest parties consists of $N$ independent and identically distributed (iid) copies of a non-signaling device $P(AB|XY)$. The total system of the honest parties and the eavesdropper is then a tripartite probability distribution $Q(ABE|XYZ)$, that has $P^{\otimes N}(AB|XY)$ as its marginal after discarding (tracing out) Eve's subsystem. As we assume the worst-case scenario, we allow the eavesdropper to have access to all statistical ensembles of the device of the honest parties. This power is achieved with Eve having access to the non-signaling complete extension of the device $P^{\otimes N}(AB|XY)$ denoted $\mathcal{E}(P^{\otimes N})(ABE|XYZ)$ that is a non-signaling analog of quantum purification \cite{CE} (for more details see Sec. \ref{sec:E_CE}). Moreover, unlike the honest parties, the malicious Eve can perform a generalized measurement $M_z^{G}$ on her subsystem, allowing for additional local randomness on her input. 

\newpage
In the QDD and QDI scenarios, some upper bounds are based on the entanglement measure called ``squashed entanglement'' \cite{Christandl_2002,Christandl12,multisquash,Wil16,TGW14,Takeoka_2014,CFH21,AFL21,Farkas_2021}. A welcome feature measure of it is that it is an additive function of quantum states, as opposed to the relative entropy of entanglement \cite{keyhuge,pptkey,AEJPVM2001,Pirandola_2018,Pirandola_2020}, which allows for avoiding the need for its regularization. Although the nonlocality analog of relative entropy has been constructed~\cite{vanDamGrunwaldGill,Grudka_contextuality}, no analog of the squashed entanglement has been studied until recently \cite{Kaur-Wilde}. In our approach, which is guided by the analogies between quantum entanglement and nonlocality, we introduce (independently of Ref.~\cite{Kaur-Wilde}) a novel nonlocality measure ``squashed nonlocality". It is formulated differently but equivalent to the measure implicitly introduced in Red. \cite{acin-2006-8}. We base our formulation on the non-signaling complete extension that is an analog of quantum purification employed in the definition of the squashed entanglement. In this way, our findings contribute to more than just the development of the NSDI scenario but additionally to the fundamental topic of nonlocality.

As mentioned, our approach bases on the analogies between different cryptographic paradigms. Firstly, we introduce a novel security criterion for the NSDI key distillation protocols $\Lambda$. The criterion is based on the non-signaling complete extension and the non-signaling norm (cf. Ref.~\cite{ChristandlToner}) that plays a role of an operational distance measure between non-signaling devices. The non-signaling norm reads
\begin{align}
    \norm{P-P^\prime}_\mathrm{NS} := \sup_{g\in \mathcal{G}} \frac 12 \norm{g(P)-g(P^\prime)}_1,
\end{align}
where $\mathcal{G}$ is a certain (broad) subset (specified in Ref. \cite{WDH2021}) of linear operations $\mathcal{L}$ that map devices into probability distributions. In this way, a sequence of MDLOPC operations $\Lambda=\{\Lambda_N\}$ (the MDLOPC key distillation protocol) performed by the honest parties on $N$ iid copies of a  device $P(AB|XY)$ is said to be secure if the following condition holds (see Sec.~\ref{subsubsec:Security})
\begin{align}
    &\norm{P_\mathrm{out} - P^{(d_N)}_\mathrm{ideal}}_\mathrm{NS}  \le \varepsilon_N \stackrel{N \to \infty}{\longrightarrow} 0,\label{eqn:NSnorm}\\
    &P_\mathrm{ideal}^{(d_N)} (s_A,s_B,q,e|z) := 
	\frac{\delta_{s_A,s_B}}{|S_A|} \sum_{s_A^\prime,s_B^\prime} P_\mathrm{out} (s_A^\prime,s_B^\prime,q,e|z), \label{eqn:IdealNS}
\end{align}
where $P_\mathrm{out}= \Lambda_N \left(\mathcal{E}\left(P^{\otimes N}\right)\right)$, is the non-signaling complete extension of the $N$ iid copies of the $P(AB|XY)$ distribution, and $P^{(d_N)}_\mathrm{ideal}$ is the secure (perfect) state in which Eve is uncorrelated with the honest parties. In Eq.~\eqref{eqn:IdealNS} $s_{A}$ and $s_B$ stand for the output of the honest parties, i.e., the secret-key, $q$ represents communication, $e$ and $z$ are input and output of Eve respectively, finally $d_N$ stands for adequate dimension. Security based on the non-signaling norm constitutes an analogy to the QDD scenario, where security is based on the trace distance between quantum states. The non-signaling norm in Eq.~\eqref{eqn:NSnorm} measures the distance between the so-called classical-device (\textbf{c}-d) states shared at the end of a MDLOPC protocol. In \textbf{c}-d states the classical part is shared by Alice and Bob while Eve still holds a device. Furthermore, by showing an explicit form of the non-signaling norm in the case of \textbf{c}-d states (see Proposition 2 in Ref.~\cite{WDH2021}) we prove in Theorem 3 of Ref.~\cite{WDH2021} that the security criterion adapted by us (see Eq.~\eqref{eqn:NSnorm}) is equivalent to the two notions of security already present in the literature \cite{masanes-2006,masanes-2009-102,Masanes2011} and \cite{hanggi-2009,hanggi-2009b,Renner-Hanggi,Hanggi-phd}, respectively. The explicit form of the non-signaling norm reads 
\begin{align}\label{eq:NS-norm-simplified}
    \left| \left| P^1_{A,E|Z}-P^2_{A,E|Z}\right| \right|_\mathrm{NS} =
			\frac 12 \sum_{a} \sup_{  \mathcal{M}^F_{z}} \sum_{e}
			\left|    \mathcal{M}^F_{z} \left (P^1_{A,E|Z}    \right)\left(a,e\right) - \mathcal{M}^F_{z} \left (P^2_{A,E|Z}  \right)\left(a,e\right)\right|,
\end{align}
where $a \in A$ corresponds to multi-variable of outputs of the \textbf{c} part in the \textbf{c}-d distribution, and $\mathcal{M}^F_{z}$ is a direct measurement performed on input $Z$. By showing the equivalence between the security criteria based on the non-signaling norm and the one in Refs.~\cite{hanggi-2009,hanggi-2009b,Renner-Hanggi,Hanggi-phd} (based on the so-called distinguisher) we argue that our notion of security is restricted composable \cite{Ben-Or-Mayers-compos,Ben-OrHLMO05,Can01}, i.e., it is composable provided the device is not reused (at a risk of the leakage of key generated from the first use, during second use of the device~\cite{Kent-Colbeck}). Finally, the security definition based on the non-signaling complete extension guarantees that the memory of the eavesdropper is finite and minimal without compromising the eavesdropping power of Eve \cite{CE}. 

\begin{figure}[t]
    \centering
    \includegraphics[width=0.65\linewidth]{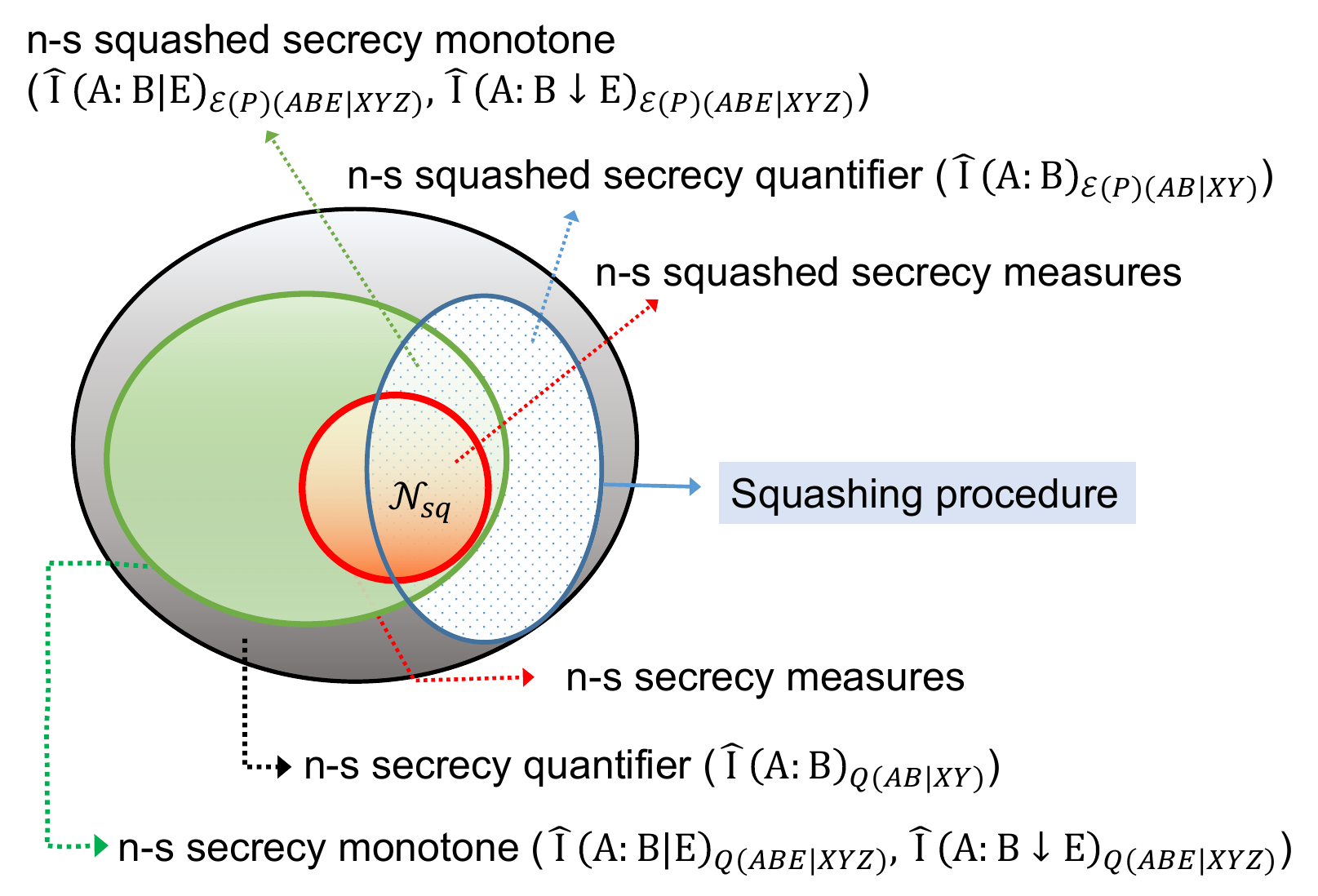}
    \caption{\label{fig:NSDI_hierarchy} The relative hierarchy of non-signaling secrecy quantifiers in Ref.~\cite{WDH2021}. The grey region represents all non-signaling secrecy quantifiers constructed from secrecy quantifiers known from the SKA security paradigm. The green region represents non-signaling secrecy monotones (MDLOPC monotones). The red circle, with the squashed nonlocality as a distinguished example, corresponds to the nonlocality measures, i.e., MDLOPC monotones that are non-negative, additive for the tensor product of devices, and zero for the local devices. Finally, the bluish region corresponds to the non-signaling secrecy quantifies incorporating the non-signaling complete extension $\mathcal{E}(P)$ of bipartite device $P$ in their construction, whereas $Q$ stands for any tripartite extension of $P$. Between brackets, we give examples of non-signaling secrecy quantifiers belonging to each set. }  
\end{figure}
Our main goal is to provide upper bounds on the secret-key rate in the NSDI scenario. To achieve our target, we introduce the so-called squashing procedure. The squashing procedure establishes an analogy between SKA and NSDI scenarios in the following manner. Suppose $\mathrm{M}(A:B||E)$ is a real-valued, non-negative function in the domain of tripartite probability distributions $P(ABE)$, that serves as an upper bound on the secret-key rate in the SKA scenario~\cite{Maurer93,Intrinsic-Maurer,MauWol97c-intr,reduced-intrinsic,lit3}, i.e., $\mathrm{M}(A:B||E) \ge \mathrm{S}(A:B||E)$. Moreover, if $\mathrm{M}(A:B||E)$ is monotonic with respect to LOPC, then we call it a secrecy monotone. Considering the above, we construct quantifiers of secret correlations in the NSDI model. Namely, the squashing procedure ``lifts up'' secrecy quantifiers $\mathrm{M}(A:B||E)$ in the SKA scenario to give non-signaling secrecy quantifiers of the NSDI scenario $\widehat{\mathrm{M}}(A:B||E)$ via mapping tripartite non-signaling devices $R(ABE|XYZ)$ into probability distributions
\begin{align}\label{eq:squashed-fn}
    &\widehat{\mathrm{M}} \left(A:B||E\right)_{R(ABE|XYZ)} := \max_{x,y} \min_z \mathrm{M}\left(A:B||E\right)_{(\mathcal{M}^F_{x,y} \otimes \mathcal{M}^G_{z}) R(ABE|XYZ)}, \\
&(\mathcal{M}^F_{x,y} \otimes \mathcal{M}^G_{z'})R(ABE|XYZ)= 
  \sum_{z} p(z|z') R(ABE|X=x,Y=y,Z=z),
\end{align}
where by $max_{x,y}$ we mean maximization over all possible direct measurements $\mathcal{M}^F_{x,y}$, and by $\min_z$ we mean minimization over all possible general measurement $\mathcal{M}^G_{z}$, where optimization over probability distribution $p(Z|Z')$ is implicit. Additionally, if $R(ABE|XYZ)$ is the non-signaling complete extension, i,.e., $R(ABE|XYZ) \equiv \mathcal{E}(P)(ABE|XYZ)$, we call $\widehat{\mathrm{M}} \left(A:B||E\right)_{\mathcal{E}(P)(ABE|XYZ)}$ a non-signaling squashed secrecy quantifier. Furthermore, suppose $\widehat{\mathrm{M}} \left(A:B||E\right)_{R(ABE|XYZ)}$ is a secrecy (MDLOPC) monotone. In that case, we call it a non-signaling secrecy quantifier monotone, and if additionally, $R(ABE|XYZ)$ is the non-signaling complete extension, we call consistently $\widehat{\mathrm{M}} \left(A:B||E\right)_{\mathcal{E}(P)(ABE|XYZ)}$ non-signaling squashed secrecy monotone. The relation between ``lifted quantifiers'' is depicted in Fig. \ref{fig:NSDI_hierarchy}, where the inclusion between the sets (red, green, gray) is proved to be strict \cite{WDH2021}.

Having the squashing procedure defined, we present our main result concerning the upper bounds on the secret-key rate (Theorem~1 in Ref.~\cite{WDH2021}). Namely, we show that the rate of the secret-key $K_{DI}^{(iid)}(P)$ distilled with MDLOPC operations in the non-signaling device-independent iid scenario (see Sec.~\ref{subsubsec:Security} for details) is upper bounded with any non-signaling squashed secrecy quantifier constructed with the squashing procedure 
\begin{align}
    \forall_{P}~~
    K_{DI}^{(iid)}(P) \le \widehat{\mathrm{M}} \left(A:B||E\right)_{\mathcal{E}\left({P}\right)}. \label{eqn:NSDI_UB}
\end{align}
Here, $P(AB|XY)$ is a single copy of a bipartite non-signaling device, of which the honest parties share multiple iid copies. The above result, together with the presence of the squashing procedure, establishes a direct connection between SKA and NSDI cryptographic scenarios. In this way, we not only establish a link between two major security paradigms but also open a possibility for a systematic study of tighter upper bounds on the secret-key rate in the NSDI scenario (cf. Ref.~\cite{acin-2006-8}). The performance and numerical feasibility of the constructed upper bounds are thanks to the presence of the non-signaling complete extension in the definition of the non-signaling quantifiers. The non-signaling complete extension allows for explicit representation of the eavesdropper system for which the memory is finite (cf. Ref.~\cite{CE}).
\begin{figure}[t]
    \centering
    \includegraphics[trim=0cm 0cm 0cm 0cm,clip,width=0.49\linewidth]{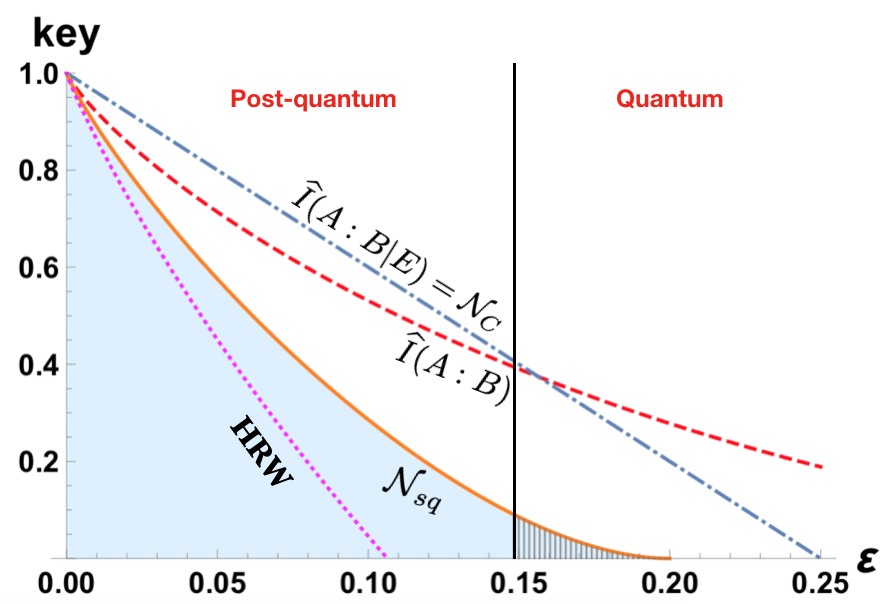}
    \includegraphics[trim=0cm 0cm 0cm 0cm,clip,width=0.49\linewidth]{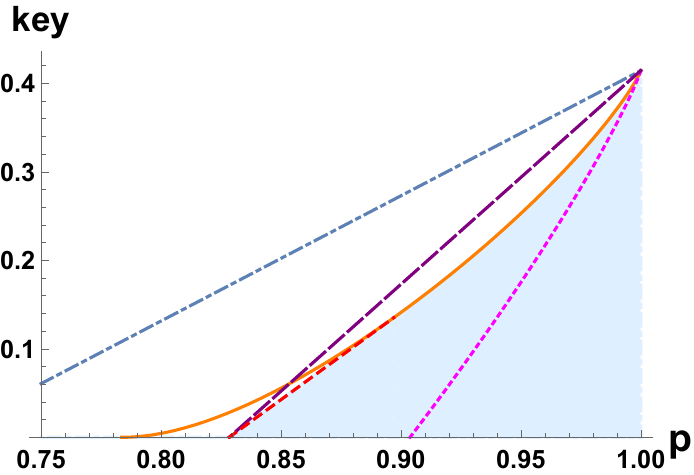}
    \caption{\label{fig:plots_NSDI_UB} Plot of the upper and lower bounds on the secret-key rate $K_{DI}^{(iid)}(P)$ in the NSDI scenario in Ref.~\cite{WDH2021}. The plot on the left corresponds to the (2,2,2,2) scenario in which the devices are probabilistic mixtures of PR and anti-PR box, i.e., $\mathrm{P}_\mathrm{iso}(\varepsilon)=(1-\varepsilon) \mathrm{PR}+\varepsilon\overline{\mathrm{PR}}$. The dashed red line corresponds to the non-signaling squashed mutual information. The straight blue line represents non-signaling squashed conditional mutual information (equal to the nonlocality cost \cite{WDH2021,Elitzur-nonloc, Brunneretal2011}). The solid orange line represents an upper bound on the squashed nonlocality obtained with the convexification technique. The dotted magenta curve (HRW) corresponds to the lower bound obtained from H{\"a}nggi, Renner, and Wolf protocol \cite{hanggi-2009}. The ``post-quantum'' region corresponds to devices that exhibit supra-quantum correlations. The plot on the right corresponds to the (3,2,2,2) scenario in which the devices are given by the $\mathrm{P}_\mathrm{AMP}(p)$ distribution (see Refs. \cite{WDH2021,acin-2006-8} for details). The solid orange line corresponds to an upper bound on the squashed nonlocality obtained with the convexification technique. The blue dash-dotted line corresponds to the non-signaling squashed conditional mutual information. The dotted magenta curve is the lower bound by Ac\'{i}n, Massar, and Pironio (AMP) in Ref. \cite{acin-2006-8}. The purple big-dashed curve is the AMP upper bound on intrinsic information of the eavesdropping strategy in Ref.~\cite{acin-2006-8}. The red dashed line is a segment of the lower convex hull of the orange and purple curves that illustrates the convexification technique. The shaded bluish region under orange and red curves represents the tightest obtained upper bound. }   
\end{figure}

The proof of general upper bound on the secret-key rate $K_{DI}^{(iid)}(P)$ in Eq. \eqref{eqn:NSDI_UB} (presented in Theorem~1 in Ref.~\cite{WDH2021}) was obtained thanks to an intermediate technical result concerning SKA security paradigm. Namely, we show (Theorem~2 in Ref.~\cite{WDH2021}) that the following definition of the secret-key rate $\mathrm{S} ( A : B||E)$ (see Sec.~\ref{subsubsec:Security}) is equivalent to the definition present in the literature \cite{CsisarKorner_key_agreement,Maurer93,MaurerWolf00CK,Christandl12}
\begin{align}\label{eq:S_equality}
	&\mathrm{S} ( A : B||E) =
	\sup_{\cal P} \limsup_{N\rightarrow \infty} \frac{\log \mathrm{dim}_\mathrm{A} \left( \mathcal{P}_N\left({P}^{\otimes N}\left(ABE\right)\right)\right)}{N},
\end{align}
with a security condition
\begin{align}
	&\norm{\mathcal{P}_N\left({P}^{\otimes N}\left(ABE\right)\right) - P_N^\mathrm{ideal}}_1 \le \delta_N \stackrel{N\to \infty}{\longrightarrow}0,\\
	\label{eqn:normAP}
    P^\mathrm{ideal}_N (S_AS_B CE^N) &:= \left(\frac{\delta_{s_A,s_B}}{|S_A|}\right) \otimes\sum_{s_A,s_B}  P^\mathrm{out}_N (S_A=s_A,S_B=s_B, CE^N), \\
    &P^\mathrm{out}_N (S_A,S_B, CE^N):=\mathcal{P}_N\left({P}^{\otimes N}\left(ABE\right)\right).
\end{align}
Here, $\mathcal{P}=\cup_{N=1}^\infty \{{\mathcal{P}}_N\}$ is a LOPC cryptographic protocol, acting on $N$ iid copies of the classical probability distribution $P(ABE)$, and  $P_N^\mathrm{ideal}$ is a distribution representing ideally secure key. Moreover, $C$ is a random variable representing public communication in the protocol. The above technical result is interesting on its own because it rephrases the definition of the secret-key rate $\mathrm{S} ( A : B||E)$ in the SKA model into the form that is characteristic for the quantum scenarios. In this way, as a byproduct, we contribute to the SKA cryptographic paradigm.

Among all non-signaling secrecy quantifiers, we pay special attention to the non-signaling squashed intrinsic information ${\cal N}_\mathrm{sq}(P)$, that we call the squashed nonlocality (cf. Ref. \cite{Kaur-Wilde} for parallel approach)
\begin{align}\label{eq:Nsq}
    {\cal N}_\mathrm{sq}(P) := 
     \max_{x,y} \min_z \mathrm{I}\left(A:B\downarrow E\right)_{(\mathcal{M}^F_{x,y} \otimes \mathcal{M}^G_{z}) \mathrm{\mathcal{E}\left(P\right)(ABE|XYZ)}}.
\end{align}
We prove that the squashed nonlocality, besides being an upper bound on the secret-key rate $K_{DI}^{(iid)}(P)$, exhibits many important properties. Namely, the squashed nonlocality is (see Proposition~1 in Ref.~\cite{WDH2021})
\begin{enumerate}
    \item \textbf{Non-negative real-valued function}, equal zero for local devices,
    \item \textbf{Monotonic} with respect to MDLOPC operations,
    \item \textbf{Convex} with respect to mixtures of non-signaling devices,
    \item \textbf{Superadditive} over joint non-signaling devices,
    \item \textbf{Additive} for product of devices,
    \item \textbf{Subextensive},
    \item \textbf{Non-faithful} (proof by inspection).
\end{enumerate}
~~\\
In consequence, we prove that the squashed nonlocality is a (novel) measure of nonlocal correlation, what makes it a nonlocality analog of the squashed entanglement being an entanglement measure. Importantly, by inspection, we prove that the squashed nonlocality is non-faithful, i.e., there exist bipartite nonlocal devices for which the squashed nonlocality is zero (see Fig. \ref{fig:plots_NSDI_UB}). We illustrate the non-faithfulness in the (2,2,2,2) setting (two binary inputs, two binary outputs) for the isotropic boxes $\mathrm{P}_\mathrm{iso}(\varepsilon)$, i.e., probabilistic mixtures of the $\mathrm{PR}$ box and anti-$\mathrm{PR}$~box~($\overline{\mathrm{PR}}$). The non-faithfulness implies that nonlocality does not imply security for the MDLOPC protocols (cf. Ref.~\cite{AcinGM-bellqkd} where a different protocol is considered). 

The intrinsic information function and, therefore, also the squashed nonlocality implicitly contains post-processing of the outputs of the eavesdropper. Therefore, in order to generate examples of the upper bounds, we perform an extensive numerical investigation that aims to find possibly optimal post-processing channels. We consider devices with statistics dependent on some set of parameters (see Fig.~\ref{fig:plots_NSDI_UB}). However, the optimal channel can be different for each set of values of the parameters. Furthermore, if one finds a post-processing channel that diminishes the upper bound curve (with respect to the identity channel), proving that it is optimal is a formidable task. This situation motivated us to develop a convexification technique that has already proved its usefulness in situations other than the NSDI scenario \cite{KHD21,HWD22}. The convexification technique states that (see Section I of Appendix in Ref.~\cite{WDH2021}) the lower convex hull of an arbitrary number of plots, each being an upper bound on the squashed nonlocality is an upper bound on $K_{DI}^{(iid)}(P)$ itself. This result is possible because the squashed nonlocality is convex with respect to mixtures of devices. In our investigation, the upper bounds on the squashed nonlocality are functions (based on the squashed nonlocality) with typically suboptimal post-processing channels. Using the convexification technique, we obtain the tightest to our knowledge upper bound (cf. Ref.~\cite{acin-2006-8}) on the NSDI secret-key in the (3,2,2,2) scenario (see Fig.~\ref{fig:NSDI_hierarchy}). We also compare our upper bound with the lower bound obtained from the protocol of H{\"a}nggi, Renner, and Wolf (HRW protocol) \cite{hanggi-2009} in order to see the proximity of the curves, and therefore the performance of the upper bound. Furthermore, the convexification technique can be easily generalized and developed for other convex non-signaling squashed secrecy quantifiers, or different scenarios \cite{KHD21,HWD22}.

To conclude, in the described paper~\cite{WDH2021}, we have developed a plethora of analogies between different security paradigms. The notion of the complete extension (see Sec.~\ref{sec:E_CE} for more details) allowed us to develop a number of analogies between the NSDI, SKA and quantum scenarios. In particular, the non-signaling complete extension allowed us to introduce a family of the non-signaling squashed secrecy quantifiers being upper bounds on the MDLOPC secret-key rate in the NSDI scenario. One of the non-signaling squashed secrecy quantifiers, called the squashed nonlocality, is proved to be a novel and nonfaithful measure of nonlocality (see Ref.~\cite{Kaur-Wilde} for parallel approach). The convexification technique developed in Ref.~\cite{WDH2021} allowed us to find the tightest upper bound in the (3,2,2,2) NSDI scenario. In this way, we contribute not only to the development of the NSDI or SKA cryptographic paradigms but also the fundamental topic of nonlocality. 


\newpage
\section{Complete Extension: the Non-Signalling Analog of Quantum Purification [E]}\label{sec:E_CE}

In the recently accepted for publication manuscript~\cite{CE}, we study a concept of the complete extension postulate (CEP) in the framework of generalized probabilistic theories~(GPTs)~\cite{InfoProcess,plavala2021general,muller2021probabilistic,lami2018non}. The CEP is meant to be a relaxation of the purification postulate (PP)~\cite{Chiribella2011}, i.e., one of the postulates in terms of which quantum theory can be reformulated. The PP is motivated by quantum purification. Considering relaxations of postulates, in terms of which quantum theory can be formulated, is a research direction motivated by the chase for beyond-quantum theory with stronger explanatory power for physical phenomena than quantum theory. On the contrary to the PP, the CEP does not require the existence, within a theory, of pure extensions for all systems and their states. Instead, the~CEP requires the existence of extension with the property of \textsc{generation}, i.e., extensions from which any other extension of the extended system can be prepared using operations available in theory. We start by discussing the limitations of the purification postulate in terms of the no-go theorem for hyperdecoherence~\cite{lee2018no}, the contradiction between the existence of fundamental superselection rules and PP~\cite{selbyReconstruction,Nakahira2020,westerbaan2022computer}, and quantum gravity in the context of theories with indefinite casual structure~\cite{hardy2005probability,Hardy_gravity,oreshkov2012quantum,chiribella2013quantum,araujo2017purification} and discrete theories. In particular, we prove a no-go theorem in which we claim that the purification postulate fails in any discrete convex theory, i.e., in a convex theory in which the number of pure states is countable. We then formally introduce the complete extension postulate (CEP) and show that \textsc{generation} property enables access to all probabilistic ensembles of the extended system via measurements on extending system (\textsc{access}), i.e., \textsc{generation} implies \textsc{access}. As we discuss, classical probability theory and superselected quantum theory both satisfy CEP. We then show that in any GPT that satisfies the CEP and the no-restriction hypothesis, the bit-commitment cryptographic task is impossible, like in quantum theory. On the other hand, consecutively, we demonstrate that the proof for the no-go theorem for hyperdecoherence of beyond-quantum theory to quantum theory that holds under the PP does not hold under the~CEP. Subsequently, we case study the theory of non-signaling behaviors and try to bypass the nonexistence of purifications therein by constructing extensions satisfying \textsc{access} and \textsc{generation} properties, i.e., the non-signaling complete extensions (NSEAs). In fact, we show that in the theory of non-signaling behaviors, the \textsc{access} and \textsc{generation} properties are equivalent. The NSEA of a generic behavior is not a pure state of the theory; still, some NSEAs are pure. As we show, NSEA of a maximally mixed binary input binary output behavior is the PR box~\cite{PR}. The above observation can be viewed as an alternative derivation of the~PR box, as it does not refer to any type of Bell inequality.  Next, we derive an upper bound on the dimension of the NSEA as a function of the dimensionality of the extended system. Importantly, we show that the dimension of NSEA is always finite. We observe that NSEA, contrary to quantum purification, does not exhibit a mirror property. Namely, in a generic case, the NSCE, $\rho^\prime_{AB}$ of the reduced state $\sigma_B$ of the extending system $B$ that is in NSCE, $\rho_{AB}$ is not the original NSCE of state $\rho_A$ of the system $A$. Finally, we conduct a numerical investigation in which we find NSEAs of particular contextual behaviors and behaviors lying on isotropic line (see Sec.~\ref{subsubsec:NSthoery}).

Whilst the quantum theory is the best-validated description of physical reality, it lacks a single agreed-on interpretation~\cite{Peres2002}, and it cannot explain some observable phenomena such as gravity. For this reason, a significant research effort has been put into finding a beyond-quantum theory with stronger explanatory power~\cite{PR,WBertramI,Bertram2018II,hardy2001quantum,zyczkowski2008quartic,smolin2006could}. The sought theory must rather be some modification of the quantum theory. However, standard axioms of quantum theory are not independent of each other~\cite{stone1932one,gleason1975measures,masanes2019measurement}, and they can not be individually modified. Therefore, recently, an interest in reformulating quantum theory in terms of mutually independent postulates occurred. This idea led to quite a number of reformulations of quantum theory in terms of information-theoretic postulates~\cite{DiakicBruckner,hardy2001quantum,Chiribella2010,Hardy2,Clifton2003,Goyal2008,MasanesMuller2011,BarnumMullerUdudec2014,Wilce16,Hhn2017,BudiyonoRohrlich2017,selbyReconstruction,STull,Wetering2019,Nakahira2020} within the framework of the so-called generalized probabilistic theories (GPTS)~\cite{plavala2021general,muller2021probabilistic,lami2018non}, \cite{InfoProcess} and references therein. One such reformulation of quantum theory introduced the purification postulate (PP)~\cite{Chiribella2011}, which gives an important insight as it distinguishes classical probability theory from quantum theory. Substantially, the purification postulate is a generalization of the notion of purification within the quantum theory to arbitrary GPTs. The PP already proved its usefulness in proving many results in the fields of pertaining to computation \cite{lee2016generalised,lee2016deriving,barnum2018oracles}, cryptography \cite{Chiribella2010,sikora2018simple}, thermodynamics \cite{chiribella2015entanglement,chiribella2017microcanonical}, and interference \cite{barnum2017ruling}. On the other hand, the PP should be modified, for several reasons, if we want to find the beyond-quantum theory to be a more fundamental description of nature than the quantum theory. Firstly, in Ref.~\cite{lee2018no}, there is a no-go theory which states that if the mentioned beyond-quantum theory is causal and reduces to quantum theory via decoherence-like mechanism (hyperdecoherence), then it does not satisfy the PP. Secondly, the~PP is contradictory with the existence of superselection rules; therefore, if one believes in those, the PP should be abandoned. Next, in Ref.~~\cite{araujo2017purification}, it was shown that a purification-like postulate fails to hold for all quantum process matrices. Therefore, there are doubts if the PP is suitable for theories that permit indefinite causal structure \cite{hardy2005probability,Hardy_gravity,oreshkov2012quantum,chiribella2013quantum}, for example, quantum gravity. Finally, in Refs.~\cite{Buniy_2005,Muller_2009, palmer2020discretisation}, ideas motivated by the pursuit of quantum gravity suggest that on the fundamental level, the quantum state space becomes discrete. We anticipate here a bit, as according to our findings, the PP can not hold in any discrete theory. All the arguments listed above suggest asking how the PP can be weakened or replaced by another postulate yielding a theory with more explanatory power. In Ref.~\cite{CE}, we propose the complete extension postulate (CEP) as a replacement for PP and study its implications in various contexts. 


We formulate the complete extension in the language of GPTs by employing the notions of extensions, ensembles, and properties of \textsc{access} and \textsc{generation}. We restrict ourselves to the convex theories, i.e., theories in which convex mixtures of states are states as well, as they seem to be natural candidates for the beyond-quantum fundamental theory. Let $s$ be a vector in the convex set of states (state space) $\Omega_A$ of a GPT system A. An ensemble for $s$, is then a set of pairs $\{(p_i,s_i)\}_{i\in I}$ such that $s_i \in \Omega_A$ and $\{p_i\}$ define a probability distribution ($p_i \in \mathbb{R}$, $p_i \geq 0$, and $\sum_i p_i =1$), satisfying the convex combination
\begin{align}
s=\sum_{i\in I} p_i s_i.
\end{align}
The set of all possible ensembles for a state $s$ is denoted $\mathbf{Ens}[s]$. Next, an extension of a state $s\in\Omega_A$ of system $A$ is a state $\sigma\in\Omega_{A\otimes E}$ of a bipartite system $A\otimes E$ for which 
\begin{align}
s = [\mathbbm{1}_A\otimes u_E](\sigma),
\end{align}
where $u_E$ is the unit effect on system $E$, and we denote the set of extensions $\sigma$, of a state $s$, as $\mathbf{Ext}[s]$. Importantly, in general, there will be extensions constructed on many different extending systems $E$. If purification exists, it is simply some $\sigma\in \mathbf{Ext}[s]$ which is pure, i.e., $\mathbf{Purif}[s] \subseteq \mathbf{Ext}[s]$. We distinguish a specific class of extensions $\mathbf{Ext}_{class}[s]$ in which the extending system is taken to be classical $\Delta_d$ for some $d\in \mathbb{N}$ (see Sec.~\ref{subsubsec:GPTsInNutshell} for details). As we show (see Proposition 9 in Ref.~\cite{CE}) that there is an isomorphism between $\mathbf{Ext}_{class}[s]$ and $\mathbf{Ens}[s]$, i.e., $\mathbf{Ext}_{class}[s] \cong \mathbf{Ens}[s]$. Third, an extension $\sigma^*\in\mathbf{Ext}[s]$ with extending system $E^*$ is said to be generating (satisfy \textsc{generation}) if and only if for every $\sigma\in\mathbf{Ext}[s]$ with arbitrary extending system $E$ there exists a transformation $T_\sigma:E^*\to E$ such that:
\begin{align}
\mathbbm{1}_A\otimes T_\sigma (\sigma^*) = \sigma.
\end{align}
Furthermore, an extension $\sigma^*\in \mathbf{Ext}[s]$ with extending system $E^*$ is an extension with access (satisfies \textsc{access} property) if and only if, for any ensemble $\sigma \in \mathbf{Ext}_{class}[s]$ with extending system $\Delta_I$ we can find a measurement $M_\sigma:E^*\to \Delta_I$ such that:
\begin{align}
\mathbbm{1}_A\otimes M_\sigma (\sigma^*) = \sigma.
\end{align}

Having the basic notions defined, in Proposition 10 and 11 of Ref.~\cite{CE}, we prove that if an extension $\sigma$ has the property of \textsc{generation}, then it necessarily has the property of \textsc{access}, but not vice versa. We discuss the above nonequivalence in terms of quantum theory and GPT, which has the same state space as the quantum theory but restricted dynamics. Finally, we formulate the complete extension postulate (CEP).\\
~~\\
\newpage
\textit{A GPT $\mathcal{G}$ satisfies the \emph{Complete Extension Postulate} (CEP) iff: for all systems $A$ and all states $s\in\Omega_A$, there exists an extension $\sigma^* \in \mathbf{Ext}[s]$ which is generating, that is, which has the \textsc{generation} property.}
~~\\
~~\\
The complete extension postulate (CEP) can be compared with the purification\\ postulate~(PP)~\cite{Chiribella2010}. \\
~~\\
\textit{A GPT $\mathcal{G}$ satisfies the Purification Postulate if and only if for all systems $A$ and all states $\omega_A\in \Omega_A$, there exists purifications ($\mathbf{Purif}[\omega_A]$) which are essentially unique.}\\
~~\\
In the above definition, essential uniqueness means that all purifications constructed on the same extending system can be related via reversible transformation on the purifying system. As we discuss, the essential uniqueness property of the PP is intuitively very closely related to the \textsc{generation} property of the CEP. However, an important difference is that essential uniqueness is a property for purifications of the same extending system, while \textsc{generation} is a property for extensions on arbitrary extending systems. Derivation of \textsc{generation} from essential uniqueness requires one more assumption and sets the relation between the CEP and the PP. As we prove in Proposition 13 of Ref.~\cite{CE} for the set of GPTs in which the product of pure states is pure, the purification postulate implies the complete extension postulate. Consequently, quantum theory satisfies CEP. Namely, purification $\ket{\psi}$ of state $\rho$ is also its complete extension, as it allows for \textsc{generation} of arbitrary other extensions as well as \textsc{access} of arbitrary ensembles. Note that the converse statement is not true, i.e., let $\omega$ be arbitrary mixed quantum state, then $\ket{\psi}\otimes \omega$ is the complete extension of $\rho$, but not its purification. Moreover, as we show, also classical probability theory satisfies CEP. Indeed, the complete extension of a probability distribution of some random variable $X$ is described by its copy. Specifically, an extension of a probability distribution $p(x)$ is any probability distribution $q(x,y)$, such that $\sum_q q(x,y)=p(x)$. In this way, $p_{ce}(x,x^\prime):=p(x)\delta_{x,x^\prime}$ is a complete extension of $p(x)$, and any other extension can be obtained by applying a stochastic map to $x^\prime$. The above results highlight the differences between the PP and the CEP, as the PP does not hold in classical probability theory. In this way, the CEP seems to be a more natural postulate to consider than the PP, as it is based on an intuitive property of \textsc{generation} and holds both in quantum and classical theory. 

As our first main result, we prove that there can not exist purification of arbitrary states in any non-signaling, convex discrete theory~\cite{Pfister-Wehner-phys-theories,CzekajHorodzie2017,PolygonTheories2011} of finite dimension. This no-go result was known previously only for the classical theory and the theory of non-signaling behaviors~\cite{Chiribella2010,Chiribella2011}. This result is not directly connected to the CEP. However, as mentioned before, the no-go theorem described below serves as an argument justifying the need to replace the PP. More precisely, a GPT $\mathcal{G}$ is said to be discrete, if and only if, for each system $T\in \mathsf{Syst}[\mathcal{G}]$ there is a discrete number of pure states, that is, where each state space, $\Omega_T$, is a polytope. The proof of Theorem 7 of Ref.~\cite{CE}, which states that in any discrete theory, there are no (non-trivial) systems that have purifications for all states, and hence, theories with purifications can not be discrete, is based on a simple observation. Namely, in Lemma 6, we observe that in any convex (non-trivial) theory, the cardinality of the set of its states is at least of power of continuum~$\mathfrak{c}$. On the other hand, the number $f$ of vertices in any discrete theory is finite, and $f <\aleph_0<\mathfrak{c}$ according to a set-theoretic fact~\cite{Kuratowski}. So there are simply not enough vertices, in theory, to purify all the mixed states of any system. The direct consequence of Theorem 7 is Proposition 8~\cite{CE}, which states that in any theory (with a countable number of system types), there must be at least one system with a continuum of pure states.

The task of bit-commitment, and its generalization to integer-commitment~\cite{May97, LC97a,sikora2018simple}, is an important two-party cryptographic primitive that can be used to build other cryptographic protocols such as coin-flipping~\cite{Blu81} and zero-knowledge proofs~\cite{GMR89}. The integer-commitment protocol for two parties, Alive and Bob, goes as follows. The task consists of two phases i) the commit phase, in which Alice chooses from the uniform random distribution an integer $j\in \{1,...,n\}$ and "commits" it to Bob by passing him a "token". Here, the "token" refers to a (sub)system prepared accordingly to the chosen integer. There are two of these tokens. At this stage, it should be impossible for Alice to change the value of $j$, as well as Bob should not be able to learn the value of $j$. In the reveal phase, ii) Alice communicates the integer $j$ to Bob and sends him a second token to verify the integer. Finally, Bob should be able to verify if the integer Alice communicated is the one she drew in the first phase. The security condition for the protocol is that Alice can not change the value of the integer $j$ after the first phase, to cheat on Bob, and Bob can not learn the value of the integer $j$ before the second phase to cheat on Alice. The cheating probabilities $p_A^*$ and $p_B^*$ of Alice and Bob, respectively, quantify to which extent the two parties can deviate the protocol from the ideal one. Namely, $p_A^*$ is the maximum probability with which Alice can reveal another integer than the one she committed to Bob. Similarly, $p_B^*$ is the maximum probability with which Bob can learn the value of $j$ before the reveal phase. It is known that (perfect realization of) bit-commitment is impossible both in classical and quantum theory~\cite{May97, LC97a}.  On the other hand, it was shown in Ref.~\cite{barnum2008nonclassicality} that any GPT which is non-classical but does not have entanglement allows for bit-commitment. It is, therefore, interesting to determine which properties of theories make the integer-commitment task impossible. In Ref.~\cite[Corollary 45]{Chiribella2010}, the impossibility of bit-commitment was proved under a set of postulates, including the purification postulate (PP). Furthermore, in Ref.~\cite{sikora2018simple} an analytical lower bound on the product of cheating probabilities $p_A^* p_B^*$, for the integer-commitment task, has been found, again relying on the PP
\begin{align}
    p_A^*\cdot p_B^* \geq \frac{\alpha}{n}>\frac{1}{2n},\label{eqn:IntCommLB}
\end{align}
where the perfect protocol would be the one for which $p_B^*=p_A^*=0$. As we show in Theorem 7 of Ref.~\cite{CE}, exactly the same lower bound holds if one replaces the PP with the CEP and assumes the no-restriction hypothesis \cite{Chiribella2010}. The proof of Theorem 7 in Ref.~\cite{CE} goes along the lines of the proof in Ref.~\cite{sikora2018simple}, with the same cheating strategies of Alice and Bob, up to the modification in which essential uniqueness is replaced with the \textsc{generation} property. The above demonstrates that the impossibility of the integer-commitment is not restricted to the theories that satisfy the PP, as the CEP suffices. On the other hand, it shows that the CEP is not an empty postulate as the lower bound in Eq.~\eqref{eqn:IntCommLB} holds. Additionally, using the CEP, we provide a unified proof applicable to both quantum and classical theory.

\begin{figure}[t]
\centering
\includegraphics[scale=0.6,angle =0]{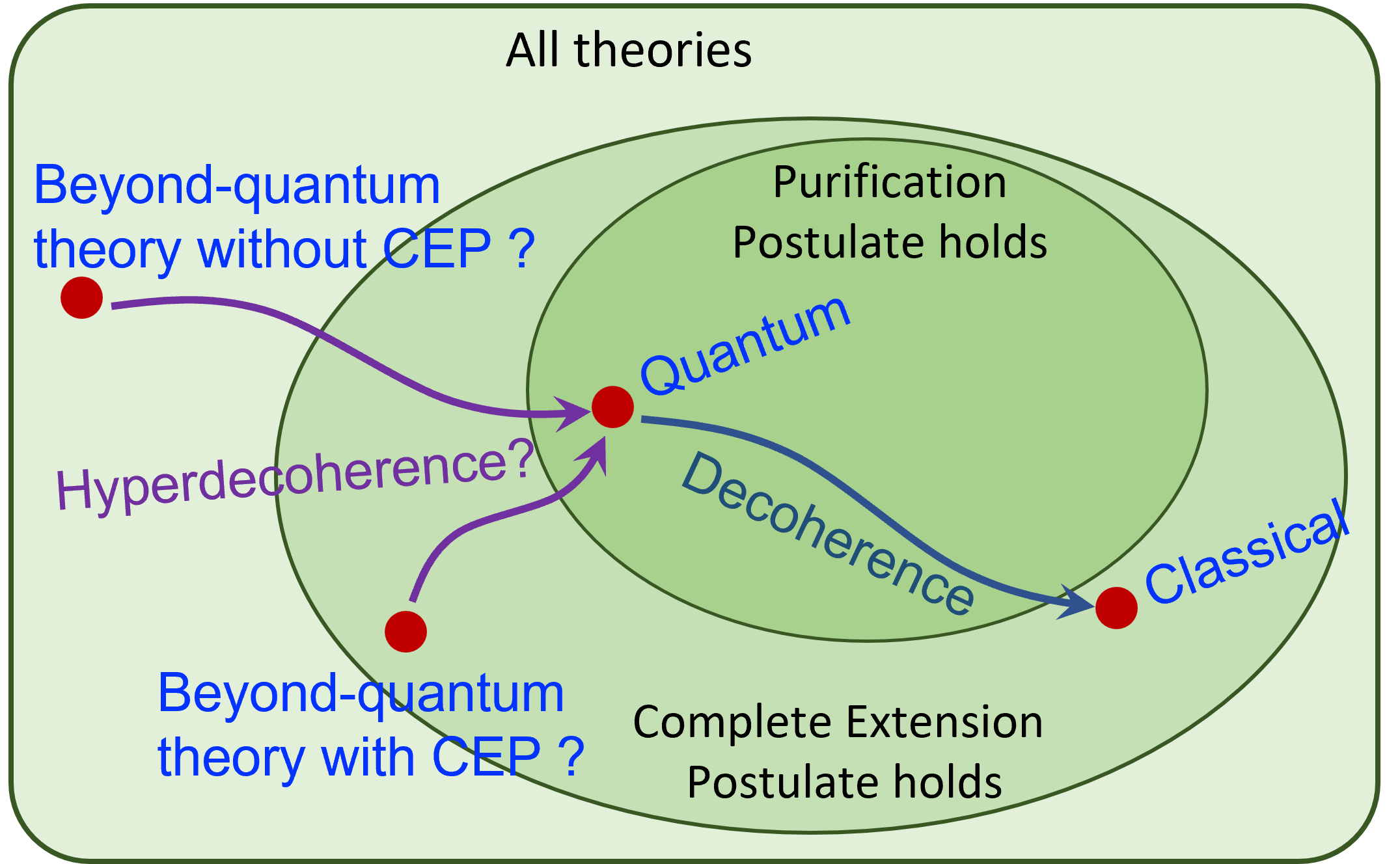}
\caption{Illustration of the relationship between the set of all theories,  theories which satisfy the CEP, and those which satisfy The PP. Hyperdecoherence mechanism (purple arrows) relates hypothetical beyond quantum theories and quantum theory in analogy to the decoherence mechanism (blue arrow), which relates quantum and classical theory). }
\label{fig:theories}
\end{figure}
Decoherence is a mechanism that shows how quantum states, measurements, and transformations become effectively classical~\cite{coecke2017two,selby2017leaks}. Note that the diagonal density operators are isomorphic to classical probability distributions. In Ref.~\cite{lee2018no}, the notion of decoherence was generalized to the concept of hyperdecoherence, a term coined in \cite{zyczkowski2008quartic}. In analogy to decoherence, hyperdecoherence (see Fig.~\ref{fig:theories}) is supposed to be a mechanism that would make underlying beyond-quantum theory effectively quantum and explain why we do not observe beyond-quantum effects in all our experimental tests (except the theory of gravity). The basic idea to describe hyperdecoherence is that for every system $A$ of beyond-quantum theory, there exists a decoherence process
\begin{align} 
    \mathbf{H}_A:A\to A,
\end{align}
which causes that system $A$ to behave effectively as a quantum system. We assume that these hyperdecoherence processes satisfy the following properties that assure consistency of the mechanism resulting in valid physical theory
\begin{enumerate} 
\item{They are unit-effect preserving, 
\begin{align} 
    u_A \circ \mathbf{H}_A = u_A,
\end{align}
in analogy to the trace-preservation condition for quantum decoherence processes.}
\item{They are idempotent, 
\begin{align} 
    \mathbf{H}_A\circ \mathbf{H}_A = \mathbf{H}_A, 
\end{align}
what reflects the fact that once the beyond-quantum features are lost, the processes of hyperdecoherence do nothing.}
\item{They must be chosen compositionally, i.e., they are such that 
\begin{align} 
    \mathbf{H}_{A\otimes B}=\mathbf{H}_A\otimes \mathbf{H}_B. 
\end{align}
In other words, if two systems hyperdecohere independently, then the joint system will behave effectively as a quantum one as well.}
\end{enumerate}
The hyperdecoherence of a beyond-quantum theory is then modeled by replacing the unit process $\mathbbm{1}_A$ with hyperdecoherence processes $\mathbf{H}_A$. The intuition behind the described model is that the hyperdecoherence happens at time scales much shorter than those we can probe experimentally and the systems hyperdecohere before we actually do something. In Ref.~\cite{lee2018no}, it was shown that any beyond-quantum theory that hyperdecoheres to the quantum theory, in the way described here, must violate either the causality principle or the PP. While abandoning the PP seems more plausible, the question arises whether the CEP can be satisfied by a beyond-quantum theory that hyperdecoheres to quantum theory. We examine the proof of the main theorem in Ref.~\cite{lee2018no} and conclude that one can not derive, using the identical proving technique, the same no-go result by replacing PP with CEP. Therefore, it remains possible that there exists a beyond-quantum theory that hyperdocoheres to quantum theory, satisfies CEP, and does not violate the causality principle.

 \begin{figure}[t]
\centering
\includegraphics[scale=0.53,angle =0]{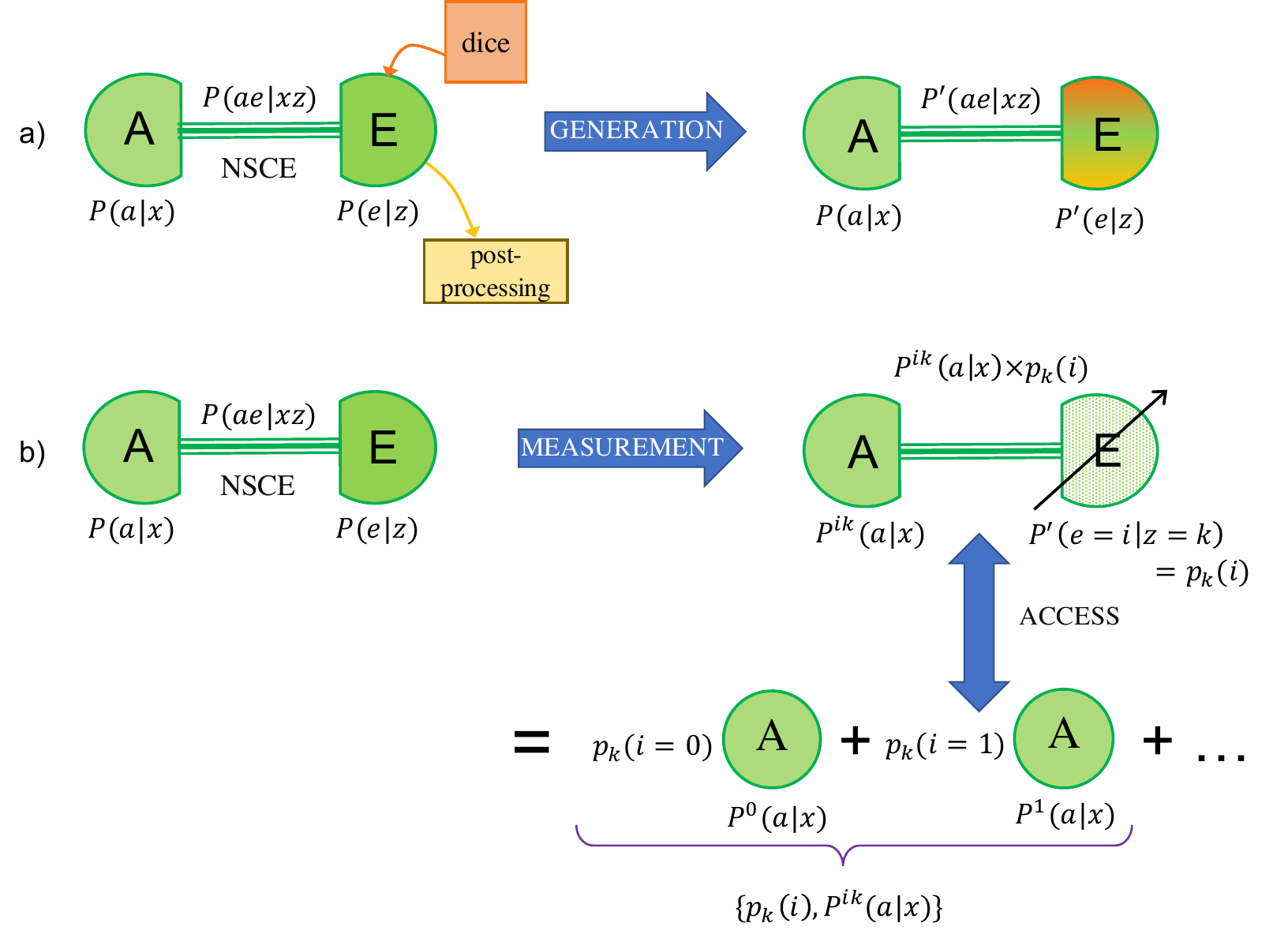}
\caption{Illustration of \textsc{access} and \textsc{generation} properties for the theory of non-signaling behaviors in Ref.~\cite{CE}. a) Using NSCE, one can generate any other extension by means of randomizing the input and post-processing of output. b) The extending system $E$ of the NSCE has access to any ensemble $\{p_k(i),P^{ik}(a|x)\}$ of the system $A$, which is generated upon measurement $z=k$ performed on the system $E$.}\label{fig:NSCE_schematic}
\end{figure}
As a case study, for the application of the CEP, we choose the theory of non-signaling behaviors~\cite{PR,InfoProcess}. The states within this theory are multipartite conditional probability distributions, e.g. $P_{\textbf{A}|\textbf{X}}$ for $\textbf{A}=A_1,\dots,A_N$ and $\textbf{X}=X_1,\dots,X_N$, that satisfy the no-signaling constraints~\cite{InfoProcess} (see Sec.~\ref{subsubsec:NSthoery} for more details). The theory of non-signaling behaviors allows for stronger correlations between subsystems than quantum theory. For example, it reaches the algebraic maximum of the Clauser-Horne-Shimony-Holt (CHSH) inequality~\cite{PR}, and it violates the Information Casuality Principle~\cite{Info-Causal}. Furthermore, the theory of non-signaling behaviors is an example of a discrete theory. Hence, it does not satisfy the PP. In our case study, we aim to bypass the lack of existence of purifications, for a generic state, with the counterpart of purifications originating from the CEP, i.e., by constructing non-signaling complete extensions (NSCEs). More precisely, for a given behavior $P_A$, we want to construct its non-signaling extension $P_{AE}$, such that it satisfies \textsc{access} and \textsc{generation} properties, with an analogy to quantum purification which satisfies these. Indeed, we achieve our goal and show that NSCEs can actually be constructed, and therefore, the theory of non-signaling behaviors satisfies the CEP. We do so by first defining the overcomplete non-signaling extension with access (ONSEA) that gives access to all pure members ensembles (PMEs) of the extended system via a choice of the measurement setting (input) on the extending system. In fact, ONSEA gives access to all ensembles of extended system, also mixed ones, as those are convex mixtures of pure members ensembles (see Theorem 21 of Ref.~\cite{CE}). Next, in Theorem 16 of Ref.~\cite{CE}, we show that \textsc{access} to  PMEs is equivalent to \textsc{access} to minimal ensembles, i.e., these ensembles that contain a set of vertices such that any of its proper subsets does not suffice to construct an ensemble of the extended system. The equivalence comes again due to dynamics available in the theory for the extending system.
This equivalence motivates us to define non-signaling extension with access (NSEA), namely\\
~~\\
\textit{Given a behavior $P_A: P_A(a|x)$, we say that a behavior $P_{AE}: P_{AE}(ae|xz)$ is 
its non-signalling extension with access extended to system $E$ if for any input choice $z = k$ an outcome  $e = i$ occurred in the extending system, there holds
\begin{align}
P_{AE}(a,e=i|x,z=k) = 
P^{i,k}_\texttt{E}(a|x)p(e=i|z=k)
\end{align}
such, that,  for each $k$,  the ensemble $\left\{\left(p(e=i|z=k),P^{i,k}_\texttt{E}(a|x)\right)\right\}_{i}$ is a minimal ensemble of the behavior $P_{A}$. Moreover, corresponding to each minimal ensemble of $P_A$, there is exactly one input $z = k$, in part of the extending system which generates it.}\\
~~\\
In Proposition 19 of Ref.~\cite{CE}, we show that for each behavior $P_A$, its NSEA $\mathcal{E}(P)_{AE}$ is unique up to local relabeling (of inputs and outputs) on the extending system $E$, what makes an analogy to essential uniqueness. In virtue of Corollary 22 of Ref.~\cite{CE} that follows from Corollary 20 and Theorem 21 of Ref.~\cite{CE}, NSEA gives access to all ensembles of the extended system, and therefore, NSEA satisfies \textsc{access}. Next, in Theorem 23 of Ref.~\cite{CE}, we show that in the theory of non-signaling behaviors, the properties of \textsc{access} and \textsc{generation} are equivalent. Finally, in Corollary 23 of Ref.~\cite{CE}, we conclude that NSEA is NSCE as it satisfies  \textsc{access} and \textsc{generation}, and therefore, the theory of non-signaling behaviors satisfies the CEP (see Fig.~\ref{fig:NSCE_schematic}). Finally, later, in Proposition 27 of Ref.~\cite{CE}, we show that NSEA (or NSCE equivalently), of arbitrary behavior $P$, has the lowest dimension amongst all non-signaling behaviors of $P$ having the property of \textsc{access}.

\begin{figure}[t]
\centering
\includegraphics[scale=0.65,angle =0]{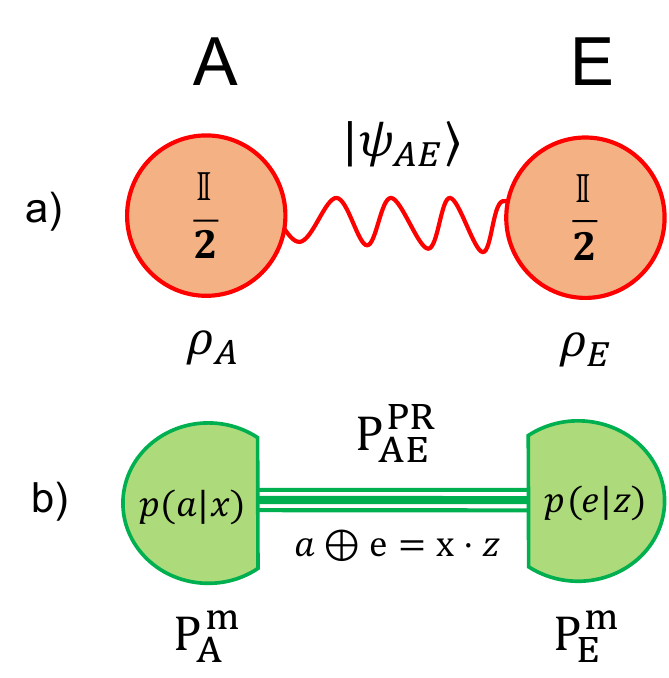}
\caption{Illustrations of purifications of system $A$ to an extended system $E$ in Ref.~\cite{CE}. a) The purification of maximally mixed state of qubit $\frac{\mathbbm{1}}{2}$ is a Bell's state which is maximally entangled. b) NSCE of a maximally mixed behavior $\mathrm{P^m}_{A}$ (see Eq. \eqref{eqn:maximally_mixed}) is the Popescu-Rohrlich box which is also pure.}\label{fig:schematic_quantum}
\end{figure}
We further give an explicit example of how NSCEs can be constructed in the theory of non-signaling behaviors. One of the behaviors for which we show the construction of NSCE is the maximally mixed behavior with a single binary input and single binary output given by 
\begin{align}
\mathrm{P_A^m}(a|x)=
\begin{array}{c|c|c}
$\diagbox[width=2em, height=2em,innerrightsep=0pt]{$a~$}{$~x$}$ & 0 & 1\\ \hline	\\[-1em]
	 0&  1/2 &  1/2  \\ \hline \\[-1em]
	 1 & 1/2 &  1/2
\end{array}
\label{eqn:maximally_mixed}
\end{align}
where $a$ is the output and $x$ is the input (measurement setting) of the behavior on system $A$. As we show, the behavior given in Eq.~\eqref{eqn:maximally_mixed} above provides an interesting example of NSCE. We start our construction by determining vertices of the polytope, i.e., state space $\mathrm{P_A^m}$ belongs to, 
\begin{align}
\mathrm{P}^0_\texttt{E} =
\begin{array}{c|c|c}
$\diagbox[width=2.0em, height=2.0em,innerrightsep=0pt]{$a$}{$x$}$ & 0 & 1\\ \hline	\\[-1em]
	 0&  1 & 1  \\ \hline \\[-1em]
	 1 & 0 & 0
\end{array}, ~~
\mathrm{P}^1_\texttt{E} =
\begin{array}{c|c|c}
$\diagbox[width=2.0em, height=2.0em,innerrightsep=0pt]{$a$}{$x$}$ & 0 & 1\\ \hline	\\[-1em]
	 0&  0 &  0  \\ \hline \\[-1em]
	 1 & 1 &  1
\end{array}, ~~
\mathrm{P}^2_\texttt{E} =
\begin{array}{c|c|c}
$\diagbox[width=2.0em, height=2.0em,innerrightsep=0pt]{$a$}{$x$}$ & 0 & 1\\ \hline	\\[-1em]
	 0&  1 &  0  \\ \hline \\[-1em]
	 1 & 0 &  1
\end{array}, ~~
\mathrm{P}^3_\texttt{E} =
\begin{array}{c|c|c}
$\diagbox[width=2.0em, height=2.0em,innerrightsep=0pt]{$a$}{$x$}$ & 0 & 1\\ \hline	\\[-1em]
	 0&  0 &  1  \\ \hline \\[-1em]
	 1 & 1 &  0
\end{array}, 
\label{eqn:extremal_boxes}
\end{align}
where the subscript $E$ corresponds to the fact that behaviors above are "extremal" points of the polytope. We now employ the fact that any point inside a convex polytope can be expressed as a convex combination of its vertices. In this way, we find all minimal ensembles of the $\mathrm{P_A^m}$ behavior
\begin{align}
\mathcal{M}_0(\mathrm{P}_A^m) &= \{(1/2, \mathrm{P_\texttt{E}^0});(1/2, \mathrm{P_\texttt{E}^1})\} 
\label{eqn:PRmin1}\\
\mathcal{M}_1(\mathrm{P}_A^m) &= \{(1/2, \mathrm{P}_\texttt{E}^2);(1/2, \mathrm{P}_\texttt{E}^3)\} 
.  \label{eqn:PRmin2}
\end{align}
Now, using the definition of NSCE (or equivalently NSEA) above, these two minimal ensembles are obtained on the extended system for two different input choices on the extending system. We associate now the choice of minimal ensembles $\mathcal{M}_0$, $\mathcal{M}_1$ in Eqs.~\eqref{eqn:PRmin1}, \eqref{eqn:PRmin2}, with input choices on the extending system $z=0$, $z=1$, respectively. In this way, we obtain the NSCE of the behavior $\mathrm{P_A^m}$
\begin{align}
		&\mathrm{P^{PR}_{AE}}\left(ae|xz\right)=
		\begin{array}{cc|cc|cc}
			&\multicolumn{1}{c}{x} & \multicolumn{2}{c}{0}	& \multicolumn{2}{c}{1} \\
			z &$\diagbox[width=2.4em, height=2.4em, innerrightsep=0pt]{$e$}{$a$}$~~  & 0	& 1	&0  & 1 \\
			\hline \\ [-0.9em]
			\multirow{2}{*}{0 }  & 0 & \frac 12	& 0	& \frac 12	& 0\\[0.3em]
			& 1	& 0& \frac{1}{2}	& 0	& \frac 12  \\ [0.2em]
			\hline  \\ [-0.9em]
			\multirow{2}{*}{1 }   &0 & \frac 12	   &  0  &   0 & \frac 12 \\ [0.3em]
			& 1 & 0	&       \frac 1 2 &  \frac 12 	& 0
		\end{array}.
\end{align}
We recognize the NSCE of $\mathrm{P_A^m}$ to be the famous Popescu-Rohrlich (PR) box \cite{PR} (see Sec.~\ref{subsubsec:NSthoery} for more details). The PR box is an extreme point in its polytope. In this point, we recognize the analogy to the purification of a maximally mixed state of a qubit, i.e., $\rho=\frac 12 \mathbbm{1}$, which purification is maximally entangled Bell's state (up to local isometry)~\cite{Horodecki2009} (see Fig.~\ref{fig:schematic_quantum}). As we conclude in Corollary 33 of Ref.~\cite{CE}, the PR box is a purification of maximally mixed behavior with a single binary input and a single binary output $\mathrm{P_A^m}$. We also discuss that observing that PR box is the purification of $\mathrm{P_A^m}$ can be seen as a derivation of PR box without referring to any notion of CHSH or the so-called Bell inequality, in contrast to Ref.~\cite{PR}. In essence, the independence of our derivation relies on the fact that we assume only a) the structure of single-partite systems and b) that the NSCE of any single-partite system is a valid state of the theory. However, we do not presuppose: i) that all non-signaling behaviors all valid states of the theory, ii) a specific composition rule, iii) the no-restriction hypothesis~\cite{Chiribella2010}. Furthermore, in section B.1 of the appendix of Ref.~\cite{PR}, we show explicit construction of NSCE for certain single-partite behavior, which is not pure. In section B.3 of the appendix of Ref.~\cite{PR}, we construct NSCE if three-cycle contextual behavior~\cite{abramsky2011sheaf,Cabello2013-noncontext} (aka Specker's triangle \cite{specker1990logik,liang2011specker}). Eventually, in section B.4 of the appendix of Ref.~\cite{PR}, we determine all minimal ensembles of bipartite behaviors lying on the isotropic line (see Sec.~\ref{subsubsec:NSthoery}) that can be readily used to construct NSCEs, and study their properties.

As our next main result, we study the dimension of the non-signaling complete extension (NSCE). By the dimension of NSCE or any other non-signaling behavior, we understand the dimension of the polytope (state space) it belongs to. Our purpose is to derive an upper bound on the dimension of the NSCE of n-partite behavior belonging to polytope $\mathcal{B}$, with $m_i$ inputs for parties, and $v_{ij}$ outputs respectively. We denote by $\tilde{\mathcal{B}}$ the polytope that contains NSCE of arbitrary behavior in $\mathcal{B}$. In Theorem 25 of Ref.~\cite{CE}, we show the following upper bound on the dimension of polytope $\tilde{\mathcal{B}}$ 
\begin{align}
         &\dim \tilde{\mathcal{B}} <\left(\dim \mathcal{B}+1\right) \nonumber\\
         &\times \left(\binom{\binom{2t -\left \lfloor{t/2}\right \rfloor - \dim \mathcal{B} }{\left \lfloor{t/2}\right \rfloor}+ \binom{3t -\left \lfloor{t/2}\right \rfloor - \left(\dim \mathcal{B}+1\right) }{t-\left \lfloor{t/2}\right \rfloor-1}}{\dim \mathrm{B}+1}\dim \mathcal{B} +1\right), \label{eqn:DimUB}
    \end{align}
     where:
     \begin{align}
         &\dim \mathcal{B} = \prod_{i=1}^{n} \left(\sum_{j=1}^{m_i}(v_{ij}-1)+1\right)-1,~~t=\prod_{i=1}^{n} \sum_{j=1}^{m_i} v_{ij}. \label{eqn:formulaPironio}
\end{align}
In the proof, we base on the formula in Eq.~\eqref{eqn:formulaPironio}~\cite{Pironio2005lift} (Theorem 1 therein) and some facts from convex geometry. To find an upper bound on $\dim \tilde{\mathcal{B}}$, we have to determine upper bounds on the number of inputs $m_{n+1}$ and outputs $v_{n+1,j}$ of the extending party. We firstly determine the maximal number of outputs, for each input generating minimal ensemble, of the extending party via Carath\' {e}odory theorem~\cite{Ziegler1995} to be $\dim \mathcal{B}+1$, i.e., $v_{n+1,j} \le \dim \mathcal{B}+1$. On the other hand, the number of inputs of the extending party is equivalent to the number of minimal ensembles. Assuming $V$ to be the number of vertices in polytope $\mathcal{B}$, and using Carath\'{e}odory theorem again, we obtain $m_{n+1} \le \binom{V}{\dim \mathcal{B}+1}$. Finally, the number of vertices $V$ can be upper bounded using McMullen's Upper Bound Theorem \cite{McMullenUpperBound,Ziegler1995,Vertices} and properties of non-signaling polytopes
\begin{align}
    V \le \binom{2t -\left \lfloor{t/2}\right \rfloor - \dim \mathcal{B} }{\left \lfloor{t/2}\right \rfloor}+ \binom{3t -\left \lfloor{t/2}\right \rfloor - \left(\dim \mathcal{B}+1\right) }{t-\left \lfloor{t/2}\right \rfloor-1}.
\end{align}
Combining the above upper bounds yields the formula in Eq.~\eqref{eqn:DimUB}. An immediate corollary from Theorem 25 (Corollary 26 in Ref.~\cite{CE}) is that the dimension of the NSCE is always finite. Investigating the dimension of the NSCE is not only interesting on its own. Namely, access to all possible ensembles of a non-signaling system has been considered operationally the worst-case extension the extending party might possess in the context of device-independent cryptography against a non-signaling adversary~\cite{Kent, Masanes, hanggi-2009, Hanggi-phd} (see also Ref.~\cite{WDH2021} and Sec.~\ref{sec:D_Squashed}), and also in the context of private randomness \cite{Colbeck_randomness, Gallego_randomness, Mironowicz_randomness, Brandao_randomness, Ramanathan_randomness}. The adversary having access to the extending system of NSCE possesses, therefore, ultimate operational power at the lowest memory cost required for having access to all ensembles of the extended system.

In summary, in Ref.~\cite{CE}, we introduce a new concept of the complete extension postulate (CEP) as a relaxation of the purification postulate (PP). We show that the CEP postulate is satisfied in classical theory, quantum theory, the theory of non-signaling behaviors, and, moreover, in any theory in which the product of pure states is pure. We have shown that the CEP does not allow for the integer-commitment cryptographic task, as well as that it does not exclude, in contrast to the purification postulate, a beyond-quantum theory that hyperdecoheres to quantum theory. In this way, the CEP sets a demarcation line between results that require the PP to hold and those for which the CEP is enough. Our case study in the theory of non-signaling behaviors presents an alternative derivation of the Popescu-Rohrlich box as the non-signaling complete extension (NSCE) of maximally mixed behavior with a single binary input and single binary output. We study the dimension of the NSCE state space. We showed that the dimension of NSCE is always finite and pointed out important implications for cryptography against the non-signaling adversary.



%

%

\chapter{Outlook}

The results described in the articles incorporated in this dissertation establish fundamental limitations of the major quantum and supra-quantum secret key distribution paradigms that are about to be the main building blocks of the future quantum-secure Internet. The vast landscape of adamant limitations on the secret key distribution tasks drawn in this dissertation is, however, a single element of the quantum revolution that is about to happen in the not-too-distant future. Still, determining the upper bounds is of no lesser importance than constructing new cryptographic protocols and tasks. This concern is because any newly created or experimentally applied secret key distribution protocol has to confront the limitations of the secret key distillation rates. In this context, the upper bounds described in this dissertation set the aforementioned type of limitations on the several selected secret key distribution scenarios of crucial importance. Importantly, any violation of the upper bounds on the secret key rate instantly implies that the produced cryptographic key can not be secure and points out theoretical shortcomings in the application of the proposed protocol or technological problems with the devices used for its implementation. Furthermore, if one finds a correct and feasible protocol for a specific cryptographic task that achieves the corresponding upper bound, then the search for a better protocol can be completed. This situation can only happen if we have the upper bounds on the considered scenario or the upper bounds we know they are tight enough. The findings described in this dissertation open a pathway for further investigations in the meaningful field of upper bounds on the secret key rates. The techniques developed here allow both, at least in some cases, for developing tighter upper bounds and extensions to other cryptographic scenarios. Moreover, apart from the application-directed perspective on the upper bounds that establish the core of this dissertation, the novel measures of nonlocality and multipartite entanglement presented here in the context of upper bounds on the secret key rate, have their reflection in the foundations of the quantum theory. Determining the properties of the proposed measures, along with the development of their frameworks, constitute tasks of no lesser importance than finding their applications in the description of the cryptographic scenarios. In the forthcoming paragraphs, we sketch possible further progress that can be done in the direction indicated in this dissertation.

Regarding the results of Ref.~\cite{Sakarya_2020} described in Sec.~\ref{sec:A_Hybrid}. The performance of the countermeasure to the rerouting attack proposed therein is benchmarked with two quantities, i.e., the gap of the scheme and the percentage of the memory overhead. The definitions of the gap of the scheme and the memory overhead include the rate of the device-dependent secret key. It is, therefore, a crucial question if the solution proposed as a countermeasure to the rerouting attack can be extended to the case of device-independent security of the secret key while preserving the reasonable gap and memory overhead of the scheme. The change of the security paradigm in this place is, however, not meaningless regarding the performance of the scheme. Private states and PPT states approximating them are known to provide a low rate of device-independent key, significantly lower than in the device-dependent key \cite{CFH21}. This drawback can dramatically diminish the gap of the scheme and significantly increase the memory overhead resulting in extra low performance of the countermeasure. Determination of the set of states yielding relatively good performance of the secure hybrid network scheme is, therefore interesting direction for further investigation. 

The upper bounds on the secret key rates derived in Ref.~\cite{DBWH19} and described here in Sec.~\ref{sec:B_Universal} are given by various types of relative entropy functions~\cite{Ume62,MDSFT13,D09,Dat09,MDSFT13,WWY14,BD10,WR12}. The framework developed in Ref.~\cite{DBWH19} allows for describing a plethora of different scenarios in a unified way. On the other hand, the squashed entanglement~\cite{Christandl_2002,Christandl12} and the c-squashed entanglement \cite{multisquash} are known to be entanglement measures that upper bound the bipartite and multipartite (conference) secret key rates respectively. It is, therefore, an interesting open question if one can develop a similar unified framework based on a multipartite generalization of the squashed entanglement rather than the one based on relative entropies originating from the generalized divergence. Furthermore, as stated in Ref.~\cite{DBWH19}, the identification of new information processing tasks and determination of bounds on the achievable rates for the classical and quantum communication protocols over a multiplex channel (see, for example, Refs.~\cite{Sha61,GKbook,BHTW10,WDW16,D18thesis,LALS19,TR19,DW19}) is undoubtedly an important future direction of research.

Speaking of the possible future developments of results in Ref.~\cite{HWD22}, described in Sec.~\ref{sec:C_Fundamenal}, the situation is a mirror reflection to the described in the previous paragraph in the case of Ref.~\cite{DBWH19}. In the Ref.~\cite{HWD22}, the derived upper bounds on device-independent conference key rate that are given in terms of the c-squashed entanglement based on the squashed and cc-squashed entanglement functions~\cite{Christandl_2002,Christandl12,multisquash,AFL21,KHD21}. It is, therefore, tempting to ask if tighter upper bounds can be derived in terms of the relative entropy functions~\cite{Ume62,MDSFT13,D09,Dat09,MDSFT13,WWY14,BD10,WR12} and in particular in terms of the reduced relative entropy of entanglement in analogy to results in Ref.~\cite{KHD21}. Furthermore, as remarked in Ref.~\cite{HWD22} our results hold in the static scenario of quantum states. A possible further investigation includes a generalization of the results in Ref.~\cite{Muralidharan_2016} to the dynamic case of quantum channels by employing and developing the formalism developed in Ref.~\cite{HWD22}. Moreover, determining whether there exists a strict gap between $K_{DI,dev}$ and $K_{DI,par}$ is an interesting question on its own. Finally, it would be interesting to determine whether Theorem~8 in Ref.~\cite{HWD22} still holds if we consider the TOBL class of devices instead of the locally quantum class.

In Ref.~\cite{WDH2021}, we developed a method for obtaining upper bounds on the device-independent secret key secure against a non-signaling adversary (NSDI scenario) by lifting the upper bounds on the secret key rate in the secret key agreement scenario (SKA). The upper bound we concentrate on the most in Ref.~\cite{WDH2021} is the new measure of nonlocality, i.e., the so-called squashed nonlocality (non-signaling squashed intrinsic information) ${\cal N}_\mathrm{sq}$ that is based on the intrinsic information function~\cite{Intrinsic-Maurer,MauWol97c-intr}. However, it is known that reduced intrinsic information \cite{lit3,reduced-intrinsic} is an example of an upper bound on SKA key rate that is, for some probability distributions, strictly lower than intrinsic information~\cite{reduced-intrinsic}. Therefore, the first step in obtaining tighter upper bounds on the NSDI key rate is to construct non-signaling squashed reduced intrinsic information and investigate its properties. Furthermore, it is an important question to answer whether the squashed nonlocality \cite{WDH2021} and intrinsic non-locality developed in Ref.~~\cite{Kaur-Wilde} are in fact the same functions. The negative answer to the aforementioned issue would yield many new questions about the relations between the squashed nonlocality and intrinsic non-locality. The generalization of the method developed in Ref.~\cite{WDH2021} to the multipartite case constitutes a new primary research direction (cf.~Ref.~\cite{PKBW21}). Eventually, the study of the properties of the squashed nonlocality and its generalization is interesting on its own. For instance, the study that proves the asymptotic continuity property of the squashed nonlocality is currently in an advanced stage of development.

Finally, the investigation conducted in Ref.~\cite{CE}, and described here in Sec.~\ref{sec:E_CE}, considers the beyond quantum theory analog of the purification postulate (PP) that holds in quantum theory, i.e., the so-called complete extension postulate (CEP). Our hitherto research raises numerous further questions not only about the consequences of the complete extension postulate (CEP) but also about its best shape. In particular, an interesting further research direction is highlighted when one considers generalized probabilistic theories (GPTs) in which GENERATION and ACCESS properties are non-equivalent. In the present shape, the complete extension postulate is defined to satisfy the GENERATION property by definition. As proved in Ref.~\cite{CE}, GENERATION implies ACCESS, but the reverse statement is not true in the general case. Therefore, replacing the GENERATION property with the ACCESS property in the definition of the complete extension postulate can supposedly bring a richer and more interesting structure of the complete extensions. The existence of a GPT in which the (complete) extensions allow access to all of the statistical ensembles of the extended system but do not allow to generate an arbitrary extension (assuming no-restriction hypothesis~\cite{Chiribella2010} holds) is an interesting question about the structure of such a theory. Furthermore, the next step to develop the framework of CEP would be to consider the possibility of studying its dynamical counterpart, i.e., a GPT analog of the Stinespring dilation theorem. Finally, there are still some open questions regarding the non-signaling complete extension (NSCE). One of them is the lack of the mirror property that holds for quantum purification but does not hold for NSCE (see Ref.~\cite{CE} for more details).

In summary, the articles included in this dissertation \cite{Sakarya_2020,DBWH19,HWD22,CE} provide many fundamental and application-directed results in the fields of quantum and supra-quantum cryptography. However, as it usually is, the insightful investigations conducted therein yield even many more interesting questions apart from those already answered and open numerous new research directions. Fortunately, the approaches developed in Refs.~\cite{Sakarya_2020,DBWH19,HWD22,CE} have structures flexible enough to immediately notice the possible generalizations and further applications of their formalism. We look forward to the new results in the research areas considered in this dissertation, as well as those connected to them.


\backmatter
\bibliographystyle{alpha}
\refstepcounter{chapter}
\bibliography{references}
\clearemptydoublepage







%

%
%
%
\end{document}